\documentclass[10pt, aps, prd, amsmath, floats, floatfix, twocolumn, notitlepage,
superscriptaddress, nofootinbib, showpacs, longbibliography]{revtex4-2}
\usepackage[T1]{fontenc}
\usepackage[utf8]{inputenc}
\usepackage{lmodern}
\usepackage{verbatim}
\usepackage{physics}
\usepackage{comment}
\usepackage{orcidlink}
\definecolor{linkcolor}{rgb}{0.6, 0.0, 0.0}
\usepackage{aas_macros}
\usepackage[all]{hypcap}
\usepackage{graphicx}
\usepackage{xspace}
\usepackage{amssymb}
\usepackage{amsmath}
\usepackage[normalem]{ulem} 
\usepackage{bm} 
\usepackage{microtype}
\usepackage[english]{babel}
\usepackage{blindtext}
\usepackage{array}
\usepackage{tabularx}
\usepackage{multirow}
\usepackage{subfigure}
\usepackage{natbib}
\usepackage{float}
\usepackage{cleveref}
\crefname{section}{Section}{Sections}
 \crefname{equation}{Eq.}{Eqs.} 
 \crefname{figure}{Fig.}{Figs.}
 \crefname{table}{Table}{Tables}
 \crefname{appendix}{Appendix}{Appendices}
 \usepackage{soul}
\allowdisplaybreaks
\definecolor{db}{rgb}{0.0, 0.0, 0.62}
\definecolor{dm}{rgb}{0.7, 0.01, 0.7}
\definecolor{dr}{rgb}{0.55, 0.0, 0.0}
\newcommand{\lt}{\left}
\newcommand{\rt}{\right}
\newcommand{\p}{\partial}

\begin{document}
\title{The Ringdown and the Tide: Fingerprints of Dark Matter Halo Profiles}
\author{Arkadip Bhowmik\orcidlink{0009-0002-4946-4917}}
\email{b.arkadip@iitg.ac.in}
\affiliation{Department of Physics, Indian Institute of Technology Guwahati, Assam-781039, India}
\author{Avijit Chowdhury\orcidlink{0000-0002-7235-5076}}
\email{avijit.chowdhury@iiap.res.in}
\affiliation{Indian Institute of Astrophysics, Block 2, 100 Feet Road, Koramangala, Bengaluru 560034, India}
\author{Sayan Chakrabarti\orcidlink{0000-0003-1332-0006}}
\email{sayan.chakrabarti@iitg.ac.in}
\affiliation{Department of Physics, Indian Institute of Technology Guwahati, Assam-781039, India}

\date{\today}
\begin{abstract}Astrophysical black holes (BH) are not isolated, but embedded in matter supplied by their host galaxies. We study how the shape of a surrounding dark matter (DM) halo modifies the ringdown and tidal response of an asymptotically flat, static, spherically symmetric BH. The halo is modelled as an anisotropic Einstein cluster with vanishing radial pressure and a generalized $(\alpha,\beta,\gamma)$ density profile, supplemented by an inner cut-off near the BH and, where required, an outer tidal truncation. We derive the axial gravitational perturbation equation and compute the quasinormal mode (QNM) spectrum using sixth-order Wentzel-Kramers-Brillouin (WKB) methods and time-domain evolutions. The halo redshifts both the oscillation frequency and the damping rate, by an amount set not only by the halo compactness but also by the profile parameters: the inner logarithmic slope $\gamma$ dominates for centrally concentrated halos, while $\alpha$ and $\beta$ give subleading but profile-dependent corrections. In the eikonal limit, the shift is governed by a single redshift integral encoding the halo mass distribution outside the light ring, explaining the close correspondence between the QNM frequencies, the light ring frequency, and the Lyapunov exponent. We also show that different combinations of compactness and profile shape can yield nearly degenerate ringdown spectra. Time-domain evolutions confirm the WKB frequencies and display the expected late-time Price law decay, with an intermediate tail controlled by the outer density falloff for slowly decaying profiles. Finally, we compute the static axial tidal Love number and show that it probes the halo with a radial weighting different from the ringdown sector. The combined ringdown and tidal response therefore provides a possible way to distinguish environmental effects from genuine deviations of the vacuum BH geometry.
\end{abstract}
\maketitle
\section{\label{Introduction}Introduction}

Since the landmark first detection of gravitational waves (GW) in 2015 from the
binary black hole (BBH) merger GW150914 by the LIGO Scientific Collaboration,
GW astronomy has rapidly evolved from the discovery phase
to an era of increasingly precise measurements \cite{LIGOScientific:2016aoc,Berti:2025hly}.  The
observations by the LIGO-Virgo-KAGRA (LVK) collaboration have now
reported a large and steadily growing population of compact binary
coalescences, with BBH mergers forming the dominant class of
sources \cite{LIGOScientific:2018mvr,LIGOScientific:2020ibl,KAGRA:2021vkt,LIGOScientific:2026tep,LIGOScientific:2026uyd}.  The first three observing runs (O1–O3) yielded 90 confident detections. The subsequent GWTC-4.0 and GWTC-5.0 releases, covering O4a and O4b respectively, increased the total number of confirmed GW events to 390~\cite{LIGOScientific:2026ifv}.
This rapid increase in both the number
and the quality of detections is changing the role of GW
observations.  They are no longer used only to establish the existence of
BBH mergers; they are now becoming precision probes of black
hole spacetimes and of the astrophysical environments in which these objects
form and evolve.

A particularly clean part of the gravitational waveform is the ringdown phase, during which the remnant BH relaxes through damped oscillations known as quasinormal modes (QNMs)~\cite{Vishveshwara:1970zz,Berti:2009kk}. These modes are fixed by the background geometry together with purely ingoing boundary conditions at the horizon and purely outgoing conditions at infinity. In vacuum general relativity, the Kerr QNM spectrum depends only on the remnant mass and spin, making ringdown a direct probe of the strong-field regime and the basis of BH spectroscopy and tests of the Kerr hypothesis \citep{Vishveshwara:1970cc,Press:1971wr,chandrasekhar1983mathematical,Kokkotas:1999bd,Berti:2009kk,Konoplya:2011qq}.

Ground-based detectors primarily probe stellar-mass BHs, whose dominant ringdown frequencies lie in the range $\sim 10$ - $10^{2}\,{\rm Hz}$~\cite{Berti:2025hly}, whereas LISA will target massive and supermassive BH mergers with masses $\sim 10^{5}$--$10^{7}M_{\odot}$ or larger and ringdown frequencies in the mHz band \cite{LISA:2017pwj,Baibhav:2020tma,Piro:2022zos,LISA:2022yao,Deng:2025qhx}. Their potentially high signal-to-noise ratios make them promising systems for precision ringdown studies. It is therefore important to identify effects that can shift the spectrum away from the vacuum Kerr or Schwarzschild prediction, including modified gravity, exotic compact objects, and additional fields~\cite{Berti:2015itd,Cardoso:2019rvt,Berti:2025hly}, as well as environmental corrections arising within general relativity because astrophysical BHs are not perfectly isolated~\cite{Berti:2025hly,Barausse:2014tra}.

Supermassive BHs are ubiquitous in galactic centres. Observations of stellar orbits around Sgr~A*~\cite{Ghez:2008ms,Gillessen:2008qv}, together with very long baseline interferometric measurements of water megamasers~\cite{1995Natur.373..127M}, provide strong evidence for compact objects with masses in the range $10^{6}$--$10^{9}M_{\odot}$~\cite{Kormendy:2013dxa}. Empirical relations such as the $M_{\rm BH}$--$\sigma$ relation further indicate a close connection between central BHs and their host galaxies \cite{Ferrarese:2000se,Gebhardt:2000fk,Kormendy:2013dxa}. Since galaxies are embedded in extended DM halos~\cite{Navarro:1995iw,Navarro:1996gj,Salucci:2018hqu}, supermassive BHs should not be regarded as perfectly isolated vacuum objects; the surrounding matter distribution may leave measurable imprints on strong-field observables~\citep{Chowdhury:2025tpt,Chowdhury:2025lrr,Macedo:2024qky,Fonseca:2025ehf, Barausse:2014tra,Miller:2025yyx,Dosopoulou:2025jth,Cardoso:2019rvt,Bertone:2024rxe}. The DM distribution near a BH is shaped by several competing processes. BH growth can steepen the central density and produce a DM spike~\cite{Gondolo:1999ef,Sadeghian:2013laa,Gnedin:2003rj}, whereas capture of low angular momentum particles depletes the density close to the horizon~\cite{Sadeghian:2013laa}. Stellar scattering, baryonic feedback, BH mergers, and long-term galactic evolution can further reshape the inner profile~\cite{Merritt:2002vj,Merritt:2003qk,Gnedin:2003rj,Ullio:2001fb}. At larger radii, the halo contributes to the galactic potential and may influence the evolution of massive BH binaries. The surrounding matter is therefore not merely a passive background, but can affect both the pre-merger dynamics and the ringdown of the remnant~\cite{Bertone:2019irm,Miller:2025yyx}.
The influence of matter environments on BH oscillations has received growing attention in recent years \cite{Eda:2013gg,Konoplya:2021ube,Konoplya:2022hbl}. Even a dilute distribution can modify the background geometry, shift the light ring properties, and alter the effective potential governing perturbations, thereby introducing environmental corrections to the QNM spectrum. Recent studies of BHs surrounded by generic matter distributions and specific DM halos have shown that the corresponding QNM frequencies can be redshifted relative to the vacuum Schwarzschild values \cite{Cardoso:2022whc,Figueiredo:2023gas,Speeney:2024mas,Pezzella:2024tkf}. Although this redshift is strongly influenced by the mass and compactness of the surrounding matter, its dependence on the detailed shape of the halo is rarely talked about. In the generalized relativistic DM halo profile parametrized by three parameters $(\alpha,\beta,\gamma)$ (see~\cref{eq:density-prof}), the inner logarithmic slope $\gamma$, the outer logarithmic slope $\beta$, and the transition sharpness $\alpha$ can redistribute matter differently and may therefore leave distinct imprints on the strong-field geometry. {\it This leads to the central question of the present work: at fixed halo compactness, does the ringdown depend only on the total surrounding mass, or can it also distinguish between different halo profiles?}

In order to answer the question, we study axial gravitational perturbations of an asymptotically flat, static, spherically symmetric BHs embedded in relativistic DM halos. The halo is modelled as an anisotropic fluid using the generalized Einstein cluster construction~\cite{Einstein:1939ms,Geralico:2012jt}, in which DM particles move on circular geodesics with random orientations. After averaging, the effective fluid has vanishing radial pressure and non-zero tangential pressure. For each density profile, the mass and redshift functions are obtained from the background field equations together with asymptotic flatness. In the axial sector, density and pressure perturbations do not enter the master equation directly~\cite{Cardoso:2022whc,Chakraborty:2024gcr}. The remaining matter perturbations are associated with the axial components of the fluid velocity and the auxiliary spatial vector. Of these, the velocity perturbation is fixed by requiring the perturbed flow to remain irrotational, whereas, the perturbation of the auxiliary vector is chosen to be consistent with conservation of the perturbed energy--momentum tensor. The linearized equations reduce to a Regge-Wheeler type equation whose effective potential depends on the halo mass function, redshift factor, and density profile. It recovers the Schwarzschild Regge-Wheeler equation when the halo is removed, while a nonzero DM distribution modifies the potential barrier and hence both the oscillation frequency and damping rate of the QNMs.

We compute the QNM spectrum in both the frequency and time-domains. In the frequency domain, the WKB approximation \cite{Iyer:1986np, Schutz:1985km, Konoplya:2003ii} provides the fundamental modes and selected higher multipoles of the halo profile, while the Gundlach-Price-Pullin~\cite{Gundlach:1993tp} characteristic scheme is used to extract the ringdown signal directly from time evolution. We also calculate the light ring frequency and its principal Lyapunov exponent \cite{Cardoso:2008bp} as a useful analytic guide, because in the eikonal limit, these quantities determine the oscillation frequency and damping rate, respectively, and therefore provide a simple interpretation of the halo-induced QNM shifts. Our results show that the axial QNM frequencies are redshifted relative to the Schwarzschild values. The shift increases with halo compactness, {\it but it also depends on the shape parameters of the density profile}. The inner logarithmic slope $\gamma$ gives the strongest effect for centrally concentrated halos, while $\beta$ and $\alpha$ produce sub-leading but non-trivial corrections. We find that {\it different combinations of compactness and profile shape can lead to nearly identical QNM shifts, producing a degeneracy in the ringdown sector}. Time-domain evolutions confirm the WKB frequencies and show the expected late-time Price-law decay, with an intermediate tail controlled by the outer density fall-off for slowly decaying profiles.

We also compute the static axial tidal Love number (TLN), which characterizes the conservative response of a compact object to an external tidal field. In four-dimensional vacuum general relativity, the static polar and axial TLNs of a Schwarzschild BH vanish identically \cite{Binnington:2009bb,Damour:2009vw,Kol:2011vg,2013arXiv1304.2228C,Gurlebeck:2015xpa,Chia:2020yla,Bhatt:2023zsy,LeTiec:2020bos,Charalambous:2021mea,Chakraborty:2025zyb,Chakraborty:2026qru,Rodriguez:2026iot,Chowdhury:2026cjv}. Nonzero TLNs can nevertheless arise when the assumptions of the vacuum Schwarzschild solution are relaxed, including in other spacetime dimensions, modified gravity, non-asymptotically flat geometries, altered near-horizon boundary conditions, or matter environments \cite{Kol:2011vg,Cardoso:2019vof,Hui:2020xxx,DeLuca:2024ufn,Bhatt:2024mvr,Chakravarti:2018vlt,Rodriguez:2023xjd,Cardoso:2018ptl,Cardoso:2017cfl,DeLuca:2022tkm,Singha:2025xah,Emparan:2017qxd,Nair:2024mya,Franzin:2024cah,Silvestrini:2025lbe,Chakraborty:2023zed,Nair:2022xfm,Pani:2015hfa,Chakraborty:2024gcr,Chakravarti:2025awj,Cardoso:2019upw,Cardoso:2021wlq,DOnofrio:2026ulh,Zhao:2026eti,Cannizzaro:2024fpz}. Focusing on the environmental contribution, we derive the axial TLNs of the $(\alpha,\beta,\gamma)$ halos to leading order in compactness and confirm the analytic results through numerical solutions, in agreement with Ref.~\cite{DOnofrio:2026ulh}. Interestingly enough, we find that the tidal response is complementary to the ringdown spectrum: QNM shifts are governed mainly by an inner-weighted redshift integral, whereas the axial TLN depends on a higher radial moment and is therefore more sensitive to the outer halo and its truncation radius. Thus, {\it halo configurations that are nearly degenerate in ringdown, can be distinguished through their tidal response}.
The physical implications of our results are threefold. First, matter-induced shifts should be taken into account in precision tests of BH spectroscopy before a deviation from the vacuum QNM spectrum is interpreted as evidence for new gravitational physics. Such a deviation may instead contain information about the environment surrounding the remnant. Second, ringdown observations may provide a new way to probe the distribution of DM in the strong-field region, which is difficult to access through conventional galactic measurements. Third, the degeneracy between halo compactness and profile shape in the ringdown spectrum can be reduced by combining QNM measurements with tidal Love numbers. The two observables probe different parts of the halo: the ringdown is more sensitive to the inner region, whereas the tidal response is weighted more strongly toward the outer halo. For ordinary stellar-mass BBH mergers, the surrounding DM environment is expected to be weak, so any constraints should be interpreted cautiously. Massive BH binaries and extreme mass-ratio inspirals are more promising, since they may remain embedded in denser galactic environments and can be observed with high precision by future space-based detectors~\cite{Amaro-Seoane:2012vvq,LISA:2017pwj,Barausse:2014tra,Coogan:2021uqv,Duque:2023seg}.
This paper is organized as follows. In
\cref{sec: BG-density}, we construct the static BH spacetime embedded in a DM halo and introduce the generalized $(\alpha,\beta,\gamma)$ density profiles considered in this work. In \cref{sec-agp}, we derive the axial gravitational perturbation equations and obtain the corresponding Regge-Wheeler-type master equation. In \cref{sec:eikonal}, we present the frequency-domain analysis of the quasinormal modes, including the WKB calculation, the eikonal interpretation in terms of the light ring frequency and Lyapunov exponent, the dependence of the spectrum on the halo parameters, and its observational implications. In \cref{sec:td}, we perform time-domain evolutions to validate the frequency-domain results and analyze the late-time behavior of the perturbations. In \cref{sec:TLN}, we investigate the static axial tidal Love numbers of BHs surrounded by DM halos and discuss their complementarity with the ringdown observables in \cref{sec:complemenarity}. We conclude in
 \cref{sec: disc} with a summary of our main results and their implications for environmental effects in GW observations. Additional analytic derivations and supplementary results on 
 \cref{sec:lightring} and \cref{sec:QNMhalo} are respectively presented in the \cref{App:eikonal,App:QNM}. Throughout this work, we use geometrized units with $G=c=1$.

\section{Background geometry and density profiles}\label{sec: BG-density}
This section reviews the construction of an asymptotically flat, static, and spherically symmetric BH spacetime embedded in a DM environment with generic density profiles~\cite{Zhao:2024bpp}.
\subsection{Background Geometry}\label{BG}

We consider a static, spherically symmetric spacetime described by the line element
\begin{align}\label{eq:metric}
\bar{ds}{}^2 = -f(r)\,dt^2 + \frac{dr^2}{1-\dfrac{2m(r)}{r}} + r^2 d\Omega^2~,
\end{align}
where $d\Omega^2$ denotes the metric on the unit two-sphere, $m(r)$ is the mass function, and  $f(r)$ determines the redshift factor.

Following Refs.~\cite{Cardoso:2021wlq,Figueiredo:2023gas,Speeney:2024mas,Chakraborty:2024gcr,Pezzella:2024tkf}, we model the DM environment using the generalized Einstein cluster formalism~\cite{Einstein:1939ms,Geralico:2012jt}. An Einstein cluster consists of collisionless particles moving on circular geodesics with all possible orientations. After averaging over orbital phases and directions, the effective stress-energy tensor takes the form
\begin{align}
\langle T^{\mu\nu} \rangle = \frac{n}{m_p} \langle P^\mu P^\nu \rangle~,
\end{align}
where $n$ is the proper number density, $m_p$ is the particle rest mass, and $P^\mu$ is the four-momentum satisfying the geodesic equation. This averaging ensures stationarity and spherical symmetry.

The resulting configuration is equivalent to an anisotropic fluid with vanishing radial pressure and non-zero tangential pressure. The spacetime therefore, satisfies Einstein’s equations,
\begin{align}\label{eq:EFE}
\bar G_{\mu\nu} = 8\pi \bar T_{\mu\nu}~,
\end{align}
with an effective stress-energy tensor
\begin{align}\label{eq:Tmunu}
    \bar T_{\mu\nu}=\lt(\bar\rho+\bar p_t\rt) \bar u_\mu \bar u_\nu + \lt(\bar p_r-\bar p_t\rt) \bar w_\mu \bar w_\nu +\bar p_t \bar g_{\mu\nu},
\end{align}
where, $\bar u^{\mu}$ is the normalized fluid four-velocity, such that $\bar u^\mu \bar u_\mu=-1$ and $\bar w^{\mu}$ is a unit space-like vector orthogonal to $\bar u^\mu$, such that $\bar w_\mu \bar w^\mu=1$ and $\bar u^\mu \bar w_\mu=0$. The quantities $\bar \rho $, $\bar p_t$ and $\bar{p_r}= 0$ represent the energy density, transverse pressure and (vanishing) radial pressure of the background DM-fluid, respectively. 

For a given density profile, the mass function is determined by the continuity equation,
\begin{align}\label{eq:mprime}
m'(r)=4\pi r^2 \bar{\rho}(r)~.
\end{align}
The metric function $f(r)$ follows from the $rr$ component of the Einstein equations,
\begin{align}\label{eq:fprime}
\frac{f'(r)}{f(r)}=\frac{2m(r)/r}{r-2m(r)}~,
\end{align}
while the  Bianchi identities determine the tangential pressure 
\begin{align}\label{eq:pt}
\bar{p}_t(r)=\frac{m(r)}{2[r-2m(r)]}\,\bar{\rho}(r)~.
\end{align}
Together, the functions $m(r)$ and $f(r)$ completely specify the spacetime geometry and its geodesic properties.

To construct the spacetime, we follow the numerical procedure of Ref.~\cite{Speeney:2024mas}. For a chosen DM density profile (see Sec.~\ref{sec:density-profile}), \cref{eq:mprime} is first integrated outward from the horizon, $r_h=2M_{\rm BH}$,  yielding the mass function $m(r)$. The metric function $f(r)$ is then obtained by integrating \cref{eq:fprime} inward from $r_\infty$(numerically $\sim 10^{10}M_{\rm BH}$), imposing asymptotic flatness through the boundary conditions
\begin{align}
m(r_\infty)&=M_{\rm ADM}=M_{\rm BH}+M_{\rm halo}~,\label{eq:M_ADM}\\
f(r_\infty)&=1-\frac{2m(r_\infty)}{r_\infty}~,
\end{align}
where $M_{\rm ADM}$ denotes the ADM mass and $M_{\rm halo}$ is the total mass of the DM halo. 
\subsection{Environmental density profiles} \label{sec:density-profile}
{
We consider a parametrised density profile~\cite{Taylor:2002zd,2017MNRAS.468.1005D}, described by 
\begin{align}\label{eq:density-prof}
\bar{\rho}(r)=\bar{\rho}_0\left(\frac{r}{a_0}\right)^{-\gamma}\left[1+\left(\frac{r}{a_0}\right)^{\alpha}
\right]^{\frac{\gamma-\beta}{\alpha}}
\end{align}
where $\bar \rho_0=2^{(\beta-\gamma)/\alpha} \bar \rho(a_0)$ is a scale factor and $a_0$ is a characteristic scale radius.  The parameter $\gamma$ controls the inner logarithmic slope of the profile, $\beta$ determines the large radius fall off, and $\alpha$ fixes the sharpness of the transition between the inner and outer regimes with larger values of $\alpha$ producing a more abrupt transition\cite{Zhao:1995cp,Taylor:2002zd}. 

For $r\gg a_0$, we note from ~\cref{eq:density-prof} that
\begin{align}
\bar\rho\approx\bar\rho_{0}\lt(\frac{r}{a_0}\rt)^{-\beta}~, 
\end{align}
whereas for $r\ll a_0$,
\begin{align}
\bar\rho\approx\bar\rho_{0}\lt(\frac{r}{a_0}\rt)^{-\gamma}~.
\end{align}
We require $\beta > \gamma \geq 0$
to ensure a monotonically decreasing density profile, where $\gamma = 0$ corresponds to a central core and $\gamma > 0$ to a central cusp. Profiles that suffer mass divergence at large distance are truncated at the tidal radius $r_t >a_0$ beyond which tidal stripping renders the spherically symmetric assumption unphysical.\footnote{The radius $r_t$ for $\beta\leq3$ is usually set to be $5a_0$, except in \cref{fig:TD}.}
For benchmarking, we also consider the Einasto density profile~\cite{1969Afz.....5..137E,Acharyya:2023rnq}, which according to recent $N$-body CDM simulations, provides an excellent fit to a wide range of DM halos. The Einasto profile is given by,
\begin{align}\label{eq:einasto}
    \bar\rho(r)=\bar\rho_e {\rm exp}\left\{-d_n \left[(r/r_e)^{1/n}-1\right]\right\}~,
\end{align}
 where $r_e$ denotes the radius of the sphere containing half of the total mass, and $\bar\rho_e$ is the mass density at $r= r_e$ with $n=6$ and $d_n=53/3$~\cite{Graham:2005xx, Prada:2005mx}. Throughout the work, we set $r_e=a_0$ for brevity.

The presence of a non-rotating BH sitting at the core of the DM distribution is expected to redistribute the DM resulting in an over-density with a sharp cut-off close to the horizon at $r=4M_{\rm BH}$ . The radius $r=4M_{\rm BH}$ corresponds to the radius of the unstable circular orbit of a marginally bound particle with angular momentum per unit mass $L=4M_{\rm BH}$. Any particle with energy $\mathcal{E}\leq 1$ and angular momentum $L\geq 4 M_{\rm BH}$ has an inner turning point at $r\geq 4M_{\rm BH}$. So, a particle reaching $r=4M_{\rm BH}$ with $\mathcal{E}=1$ is necessarily captured by the BH~\cite{Sadeghian:2013laa,Gondolo:1999ef}. Following ~\cite{Speeney:2024mas}, we model the over-density by multiplying the DM density profile in ~\cref{eq:density-prof,eq:einasto} with the cut-off factor $(1-4M_{\rm BH}/r)$,

The circular null geodesics satisfy the extremum condition
\begin{equation}
2f(r_c)=r_c f'(r_c),
\end{equation}
 which, using ~\cref{eq:fprime}, reduces to
\begin{equation}\label{eq:LR}
r_c=3m(r_c).
\end{equation}
 Since the spacetime is vacuum for $r\le 4M_{\rm BH}$, the Schwarzschild photon sphere at $r_c=3M_{\rm BH}$ always persists. Whether additional light rings, formed as stable (inner)-unstable (outer) pair~\cite{Cunha:2020azh,Ghosh:2023kge}, appear or not is set not only by the global compactness $z=M_{\rm halo}/a_0$~\cite{Fonseca:2025ehf,Chakraborty:2024gcr}, but also by how centrally the halo mass sits. The inner profile must be steep enough for $m(r)$ to exceed $r/3$ at some radius. If $m(r)<r/3$ throughout, no additional ring can form~\cite{Fonseca:2025ehf}. This bound is stronger than the dominant energy condition ($\bar\rho\geq\bar {p}_t(r)$) which yields $m(r)\le2r/5$. A halo may therefore satisfy the latter and still develop an extra pair of light rings, in the window $1/3<m(r)/r\le2/5$. However, in this narrow region, the pressure is enhanced  approaching the local density. Thus, for realistic DM halo, we restrict ourself to $m(r)<r/3$ i.e., only one lightring at $r = 3 M_{\rm BH}$~\cite{Chakraborty:2024gcr}.

Henceforth, for notational simplicity we will refer to the halo density profile represented by ~\cref{eq:density-prof} with the cut-off factor $1-4 M_{\rm BH}/r$ as the $(\alpha,\beta,\gamma)$ profile.
\begin{figure}[htbp!]\label{fig:density}
    \centering
    \includegraphics[width=\linewidth]{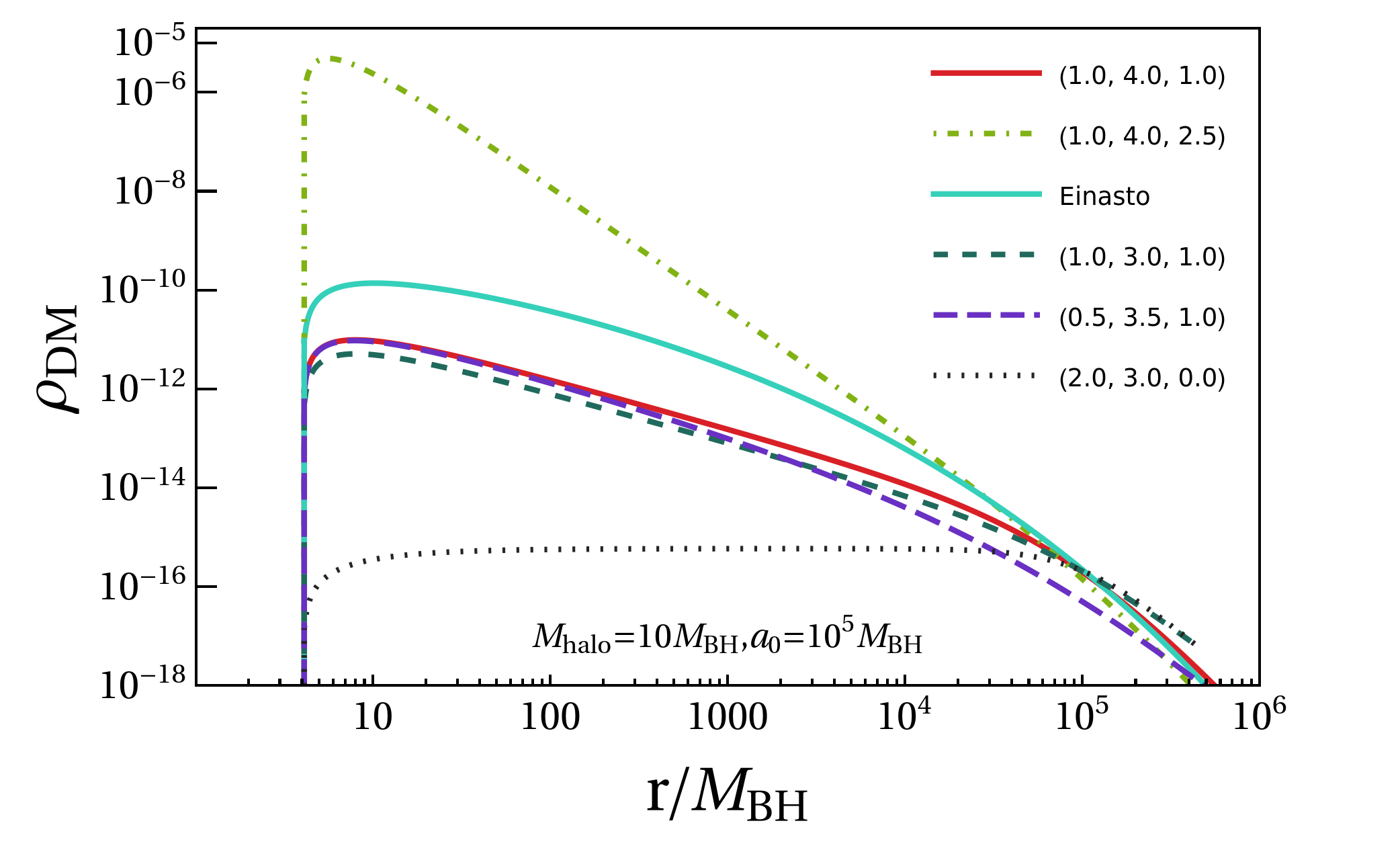}
    \caption{Plot showing the shape of the DM density profile for different values of the shape parameter ($\alpha,\beta,\gamma$) for a halo of total mass $10 M_{\rm BH}$ with $a_0=10^5 M_{\rm BH}$. In particular, we consider Hernquist ($1,4,1$), generalized Dehnen ($1,4,2.5$), NFW ($1,3,1$), Dekel-Zhao $(0.5,3.5,1)$ and King $(2,3,0)$ profiles. The Einasto density distribution with the same halo mass and compactness is also shown for comparison. 
    }
    \label{fig:placeholder}
\end{figure}
 \section{Axial Gravitational Perturbation} \label{sec-agp}
With the matter distribution and background geometry specified, we study the ringdown spectrum of an asymptotically flat, static, spherically symmetric BH in a DM halo using the BH perturbation theory.  We add small perturbations~$h_{\mu\nu}$ to the background metric $\bar{g}_{\mu\nu}$ and the matter stress tensor, such that,
 \begin{align}\label{eq:perturb}
     g_{\mu\nu}=\bar{g}_{\mu\nu}+h_{\mu\nu}~, \quad
    T_{\mu\nu}=\bar{T}_{\mu\nu}+\delta T_{\mu\nu}~. \end{align}
    The metric perturbations and the perturbation in the fluid stress tensor are related by the linearised Einstein's equation,
    \begin{align}\label{eq:pertEin}
        \delta G_{\mu\nu}=8 \pi \delta T_{\mu\nu}~.
    \end{align}
The perturbation of the fluid stress tensor is governed by the variation in the fluid energy density ($\delta \rho$),  radial pressure, $\delta p_r$, tangential pressure $\delta p_t$, as well as the perturbation of the fluid four velocity $\delta u ^\mu$ and auxiliary vector, $\delta w^\nu$ respectively.
The spherical symmetry of the background spacetime allows decomposing the metric perturbations into axial (parity-odd) and polar(parity-even) sectors depending on their behaviour under the parity transformation, $\theta\rightarrow \pi-\theta$ and $\phi\rightarrow \phi+ \pi$. Likewise, the perturbed energy-momentum tensor can be decomposed into the axial and polar components. In the present work, we focus solely on the axial perturbations and leave the polar sector for a future analysis.\\
The axial metric perturbation in the Regge-Wheeler gauge can be written as
\begin{align} \label{eq:hRG}
h_{\mu\nu}=
\begin{pmatrix}
    0 & 0 & h_0 ^{lm}(t,r) S^{l m}_\theta & h_0 ^{lm} (t,r)S^{l m}_\phi \\
    \star & 0 & h_1 ^{lm} (t,r) S^{l m}_\theta & h_1 ^{lm}(t,r) S^{l m}_\phi \\
    \star & \star & 0 & 0 \\
    \star & \star & \star & 0
    \end{pmatrix}~,
    \end{align} 
    where
    \begin{align}
        \lt(S^{l m}_\theta, S^{l m}_\phi\rt)=&\lt(-\frac{1}{\sin \theta}\p_\phi Y^{lm}(\theta, \phi),~ \sin\theta \p_\theta Y^{lm}(\theta, \phi) \rt),
    \end{align}
    are the vector spherical harmonics and $Y^{lm}(\theta, \phi)$ are scalar spherical harmonics with $l=2\ldots$, and  $-l\leq m \leq l$. The $\star$ in \cref{eq:hRG} denote the symmetric matrix components.

    Interestingly, the perturbations of the fluid energy density and pressure do not contribute in the axial sector. Thus, the axial perturbation of the fluid energy-momentum tensor is solely determined by the perturbations of the fluid four velocities, $\delta u ^\mu$ and the auxiliary vector $\delta w^\mu$ whose non-zero components are,
    \begin{align}
        \delta u^\theta &=&\frac{\sqrt{f(r)}}{4 \pi\lt(\bar \rho+\bar p_t\rt) r^2}\sum_{l,m} U_{lm}(t,r) S^{lm}_\theta~, \label{eq:uth}\\
        \delta u^\phi &=&\frac{\sqrt{f(r)}}{4 \pi\lt(\bar \rho+\bar p_t\rt) r^2 \sin^2\theta}\sum_{l,m} U_{lm}(t,r)S^{lm}_\phi~, \label{eq:uph}\\
        \delta w^\theta &=&\frac{\sqrt{1-2 m(r)/r}}{4 \pi\lt(\bar \rho+\bar p_t\rt) r^2 }\sum_{l,m} W_{lm}(t,r)S^{lm}_\theta~,
        \label{eq:wth}\\
        \delta w^\phi &=&\frac{\sqrt{1-2 m(r)/r}}{4 \pi\lt(\bar \rho+\bar p_t\rt) r^2 \sin^2\theta}\sum_{l,m} W_{lm}(t,r) S^{lm}_\phi~.
        \label{eq:wph}
    \end{align}
    Thus, we get the following non-zero components of the perturbed fluid stress tensor,
    \begin{align}
        \delta T_{t\theta}&=-\lt(\frac{f(r) U_{lm}(t,r)}{4 \pi} + h_0 (t,r)\bar \rho(r) \rt)S^{lm}_{\theta}~,
        \label{eq:Tttheta}\\
        \delta T_{t\phi}&=-\lt( \frac{f(r) U_{lm}(t,r)}{4\pi} + h_0(t,r) \bar \rho (r) \rt)S^{lm}_{\phi}~,
        \label{eq:Ttphi}\\
        \delta T_{r\theta}&=-\frac{\bar p_t W_{lm}(t,r)}{4 \pi(\bar{\rho}+\bar p_t) }S^{lm}_{\theta}~,
        \label{eq:Trtheta}\\
        \delta T_{r\phi}&=-\frac{\bar p_t W_{lm}(t,r)}{4 \pi(\bar{\rho}+\bar p_t) }S^{lm}_{\phi}~.
        \label{eq:Trphi}
    \end{align}
Using ~\cref{eq:Tttheta,eq:Ttphi,eq:Trtheta,eq:Trphi} in  \cref{eq:pertEin} in cognizance with the background equations,~\cref{eq:mprime,eq:fprime,eq:pt}, we obtain the $\theta\theta$, $r\theta$, and $t\theta$ components of the perturbed Einstein equations, which constitute the governing system of differential equations in the axial sector. The $\theta\theta$ equation relates $\dot h_0 $ with $h_1$ and $h_1'$,
\begin{align}\label{eq:h0dot}
   \dot h_0 =\frac{f(r)}{r^2} \lt[r (r-2 m(r)) h_1'+2 \left(m(r)-2 \pi  r^3 \bar\rho (r)\right) h_1\rt]~,
\end{align}
where a dot denotes derivative with respect to $t$ and a prime  denotes derivative with respect to $r$.  The spherical symmetry of the background spacetime implies degeneracy in the azimuthal index, $m$. Accordingly, we suppress the corresponding mode labels and omit the $l$ subscript for notational simplicity.
Defining, 
\begin{align}
    \psi=\frac{1}{r}\sqrt{f\lt(1- 2m(r)/r\rt)}~h_1~,
\end{align}
and substituting \cref{eq:h0dot} in the $r\theta$ equation, we get the sourced Regge-Wheeler equation as,
\begin{align}\label{eq:RWs}
 \frac{d^2\psi}{dr_*^2}  - \frac{d^2\psi}{dt^2}-&V(r)\psi=\nonumber\\
 &\frac{4 f(r) m(r) \sqrt{{f(r) (1-2 m(r)/r)}}}{r\lt(2 r-3 m(r)\rt)}W~,
\end{align}
where 
\begin{align}\label{eq:Veff}
    V(r)={f(r) \left(\frac{l (l+1)}{r^2}+\frac{4 \pi  r \rho (r)}{r-2 m(r)}-\frac{6 m(r)}{r^3}\right)}~,
\end{align}
 is the effective potential and $r_*$ is the tortoise coordinate, defined as $\frac{dr_*}{dr}=1/\sqrt{f(r)(1-2 m(r)/r)}$.
 The $t\theta$ equation, on the other hand, yields,
 \begin{align}\label{eq:rphi}
   & \lt( 1-\frac{2m(r)}{r}\rt)\lt[ h_0''- \dot h_1' -\frac{2}{r} \dot h_1\rt]-4 \pi r \lt(h_0'-\dot h_1\rt)\bar\rho(r)\nonumber\\
   & -\lt[ \frac{l(l+1)}{r^2}-\frac{4 m (r)}{r^3} +\frac{r-m(r)}{r-2m (r)} 8 \pi \rho(r)\rt] h_0\nonumber
   \\&=4 f(r) U(t,r)~.
 \end{align}
However, we still need one more equation to close the system as we have three unknown functions, namely, $h_0$, $h_1$, $U$ and $W$. To this end, we choose the fluid four velocity $u^\mu$ to be irrotational (zero vorticity)~\cite{Chakraborty:2024gcr,DOnofrio:2026ulh}.

\begin{align}\label{eq:vort}
 \omega^\mu=\epsilon^{\mu\nu\alpha\beta}u_\nu \nabla_\alpha u_\beta=0~.   
\end{align}
Equation~\ref{eq:vort} leads to the following expression for $U(t,r)$
\begin{align}\label{eq:Ulm}
    U(t,r)=-\frac{4 \pi h_0(t,r)}{f(r)}\lt[\bar p_t(r)+\bar \rho(r)\rt]~.
\end{align}
From ~\cref{eq:Ulm}, it follows that the covariant fluid-velocity perturbation vanishes, i.e. $\delta u_{\mu}=0$. 
We emphasise that the contravariant
components do not vanish: $\delta u^{\theta},\delta u^{\phi}\propto U \neq 0$ by
\cref{eq:uth,eq:uph}, so the fluid does acquire an axial velocity. What vanishes is
the combination $\delta u_\mu = \bar g_{\mu\nu}\delta u^{\nu} + h_{\mu\nu}\bar
u^{\nu}$, in which the fluid motion is exactly compensated by the frame dragging of the axial metric perturbation. Substituting \cref{eq:Ulm} in the $\theta$ component of the linearized version the conservation equation $\nabla_\mu T^{\mu\nu}$, 
\begin{align}\label{eq:cons}
\lt[
\frac{\partial}{\partial r}+\frac{B(r)}{A(r)}
\rt]
\lt[
W+4\pi h_1(\bar{p}_t+\bar{\rho})
\rt]=0,
\end{align}
where,
\begin{align}
A(r)=&2r^4m-11r^3m^2+20r^2m^3-12rm^4,\\
B(r)=&2r^3m-10r^2m^2+18rm^3-12m^4 +8\pi r^6\bar{\rho}\nonumber\\
&-40\pi r^5m\bar{\rho}
+60\pi r^4m^2\bar{\rho}
-24\pi r^3m^3\bar{\rho}.
\end{align}
It is obvious that one solution of~\cref{eq:cons} is
\begin{align}\label{eq:W}
W=-4\pi h_1(\bar{p}_t+\bar{\rho})~.
\end{align}
The choice~\cref{eq:W} is equivalent to the condition $\delta w_\mu=0$, and is therefore a statement about the perturbed matter configuration.
This in turn transform \cref{eq:RWs} to the source-free Schrodinger like form~\cite{Chakraborty:2024gcr,DOnofrio:2026ulh},
\begin{align}
\frac{d^2\psi}{d r_*^2} - \frac{d^2\psi}{d t^2}- V(r)\psi = 0,
\label{eq:mastereqn}
\end{align}
with the effective potential,~\cite{Cardoso:2021wlq, Figueiredo:2023gas, Pezzella:2024tkf}
\begin{align}\label{eq:RWpot}
V(r) = f(r) \left( \frac{l(l + 1)}{r^2} - \frac{6m(r)}{r^3} + \frac{m'(r)}{r^2} \right).
\end{align}
In the limit of vanishing halo density, Eq.~\eqref{eq:mastereqn} reduces to the standard Regge-Wheeler equation for a Schwarzschild BH~\cite{Regge:1957td}.
\begin{figure*}
    \centering
    \includegraphics[width=0.99\linewidth]{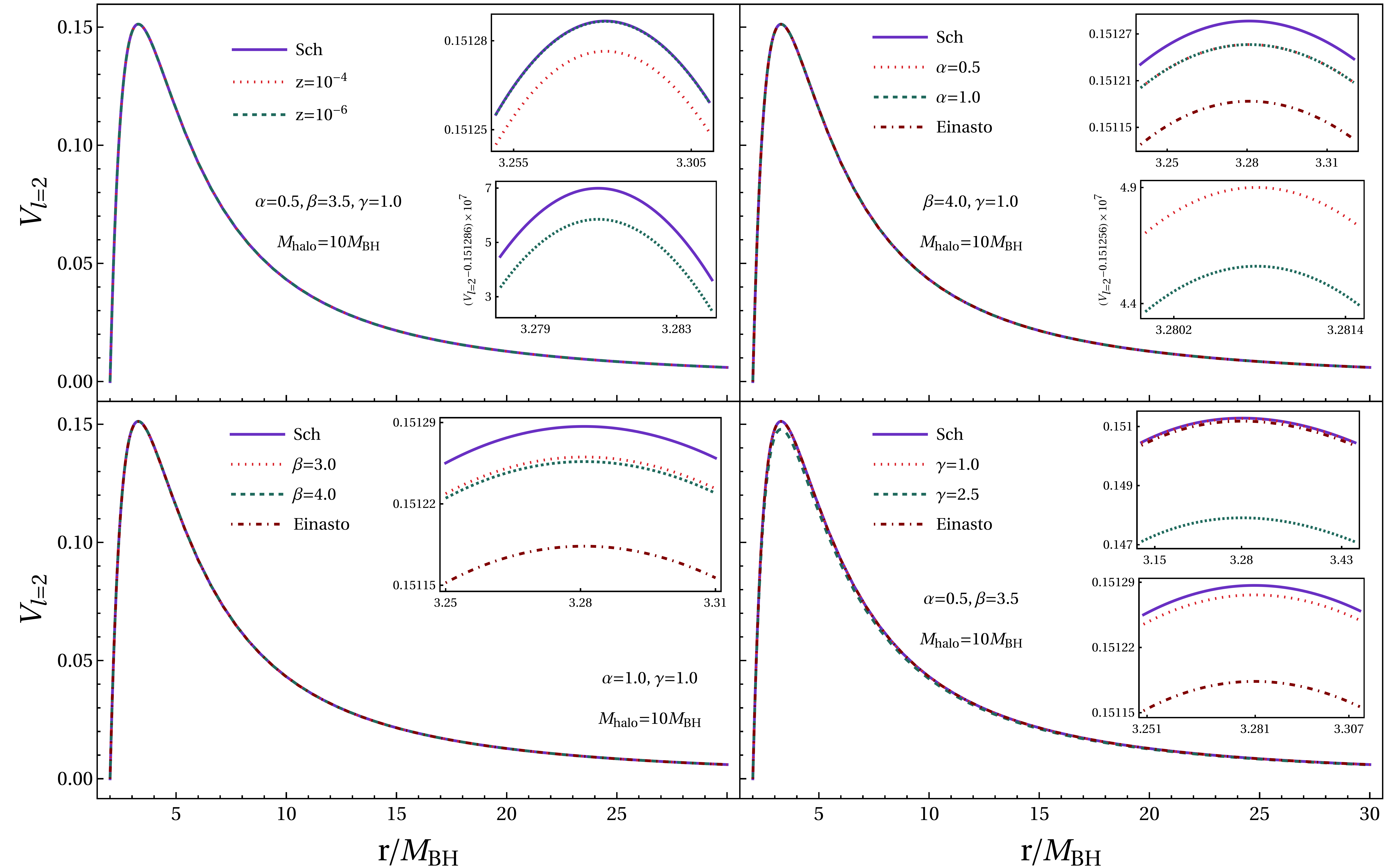}
    \caption{Effective potential in the axial sector for the $l=2$ mode for different choices of the halo parameters: (top left) compactness $z$, (top right) $\alpha$, (bottom left) $\beta$, and (bottom right) $\gamma$. In each panel, the solid blue curve denotes the Schwarzschild effective potential. The top-right and bottom panels also include the effective potential for an Einasto-type DM halo. The inset in each panel shows a magnified view of the potential near its peak. For top left plot $a_0$ is \(10^5 M_{\rm BH}\) when \(z=10^{-4}\)} and \(a_0=10^7 M_{\rm BH}\) when \(z=10^{-6}\). In remaining panels, \(a_0\) = \(10^5 M_{\rm BH}\) and \(z=10^{-4}\).
    \label{fig:V}
\end{figure*}
 From ~\cref{fig:V}, we observe that while the qualitative structure of the effective potential remains unchanged by the presence of the DM halo, quantitative deviations arise that depend on the halo compactness $(z = M_{\rm halo}/a_0)$, the inner logarithmic slope $\gamma$, the outer logarithmic slope $\beta$ and the transition parameter $\alpha$. 

\section{Ringdown Spectra: Frequency-domain analysis} \label{sec:eikonal}
Having specified the background spacetime and derived the perturbation equations, we now examine how the surrounding DM distribution modifies the ringdown spectrum of the BH. 

 We use a sixth-order WKB approximation method~\cite{Konoplya:2003ii}, which extends the original formulation of Schutz \& Will~\cite{Schutz:1985km} and Iyer\&Will~\cite{Iyer:1986np}. We start by separating the radial and temporal dependence of $\psi$ as,
 \begin{align}
     \psi(t,r)= \psi(r) \exp (- i \omega t)~,
 \end{align}
 we rewrite \cref{eq:mastereqn} in the Schrodinger-like form
\begin{align}
\frac{d^{2} \psi(r)}{d r_{*}^{2}}+Q\lt(r_{*}\rt) \psi(r)=0,
\end{align}
where 
\begin{align}
    Q\lt(r_{*}\rt)=\omega^{2}-V\lt(r_*\rt).
\end{align}
The WKB solutions are then matched across the two classical turning points of the effective potential barrier, leading to the quantization condition
\begin{align}
i\frac{Q_0}{\sqrt{2Q_0''}}-\sum_{k=2}^{6}\Lambda_k=n+\frac{1}{2}, \quad n=0,1,\cdots
\end{align}
where $Q_0$ and $Q_0''$ denote $Q$ and its second derivative evaluated at the maximum of the effective potential, and $\Lambda_k$ represent higher-order correction terms determined by higher derivatives of the potential at the same point. The fundamental mode corresponds to $n=0$, the first overtone corresponds to $n=1$ and so on. The sixth-order corrections substantially improve the accuracy of the WKB approximation and for $l>n$, achieves a accuracy comparable to Leaver's method~\cite{Leaver:1985ax}
\footnote{In~\cite{Pezzella:2024tkf}, the authors computed the QNM frequencies of a static spherically symmetric BH surrounded by a Hernquist and NFW type DM halo, with the inner cut-off radius at $r=2M_{\rm BH}$ using a generalized matrix method. However, introducing a realistic inner cutoff radius at $r=4M_{\rm BH}$ and varying the halo parameters, we observe that a large number of grid points $(N\gg17)$ are required for the QNMs to converge with accuracy comparable to that of sixth-order WKB. This greatly increases computational cost, limiting the applicability of the method in the present study.}.
\subsection{Light ring and eikonal QNMs}
\label{sec:lightring}

Before presenting the QNM spectrum, we analyze the properties of the unstable circular null geodesic (light ring). In the eikonal limit, the real and imaginary parts of the QNM frequencies are determined by the light ring angular frequency, $\Omega_c$, and the
corresponding Lyapunov exponent, $\lambda_c$, respectively. We therefore investigate the dependence of these quantities on the DM halo parameters and use the resulting behaviour to interpret the QNM spectrum discussed below.

The angular frequency of the light ring located at $r=r_c$ is
\begin{align}
   \Omega_c = \frac{\sqrt{f(r_c)}}{r_c},
\end{align}
while the principal Lyapunov exponent, which characterises the
instability timescale of the circular null orbit, is~\cite{Cardoso:2008bp}
\begin{align}\label{eq:lyapunov}
   \lambda_c = \frac{1}{\sqrt{2}}
               \sqrt{-\frac{r_c^2}{f(r_c)}
               \left(\frac{d^2}{dr_*^2}\frac{f(r)}{r^2}\right)}_{\!r=r_c}.
\end{align}
To quantify the halo-induced deviations, we define the fractional
shifts from the corresponding Schwarzschild values,
\begin{align}\label{eq:deltas}
   \delta\Omega_c = 1 - \frac{\Omega_c^{\rm DM}}{\Omega_c^{\rm Sch}},
   \qquad
   \delta\lambda_c = 1 - \frac{\lambda_c^{\rm DM}}{\lambda_c^{\rm Sch}}.
\end{align}

The last column of \cref{fig:c4c6g,fig:c4c6b,fig:c4c6a} show
$\delta\Omega_c$ and $\delta\lambda_c$ as functions of $\gamma$,
$\beta$, and $\alpha$ respectively, for two representative compactnesses
$z = M_{\rm halo}/a_0 = (10^{-6},10^{-4})$.  For a given halo model,
both $\delta\Omega_c$ and $\delta\lambda_c$ increase with halo
compactness, indicating larger deviations from the Schwarzschild
values as the halo becomes more compact, consistent with the
findings of~\cite{Pezzella:2024tkf}.

Although the DM distribution begins only at
$r = 4M_{\rm BH}$, well outside the light ring at $r_c=3M_{\rm BH}$,
the halo still redshifts $\Omega_c$ and $\lambda_c$ through its
cumulative effect on $f(r_c)$.  We show in \cref{App:eikonal} that
this redshift is exact,
\begin{align}\label{eq:frc}
   f(r_c) = \frac{1}{3}\,e^{-\mathcal{I}},
\end{align}
giving the compact form
\begin{align}\label{eq:Omega_exact}
\Omega_c = \Omega_c^{\rm Sch}\,e^{-\mathcal{I}/2},
\end{align}
where the halo integral
\begin{align}\label{eq:haloint}
   \mathcal{I} \equiv \int_{4M_{\rm BH}}^{\infty}
     \frac{2\,m_h(r)}{[r-2m(r)]\,[r-2M_{\rm BH}]}\,dr > 0
\end{align}
encodes the full non-linear effect of the halo mass distribution
$m_h(r) = m(r) - M_{\rm BH}$, so that $\delta\Omega_c =
1-e^{-\mathcal{I}/2}$.  A further exact result, also derived in
\cref{App:eikonal}, is that the local geometry at $r_c$ is
unaffected by the halo profile, so that $\lambda_c$ acquires
exactly the same redshift factor as $\Omega_c$:
\begin{align}\label{eq:lambdaOmega}
   &\lambda_c = \Omega_c = \Omega_c^{\rm Sch}\,e^{-\mathcal{I}/2},\\
  & \delta\Omega_c = \delta\lambda_c = 1 - e^{-\mathcal{I}/2},
   \label{eq:deltasEqual}
\end{align}
valid for any $(\alpha,\beta,\gamma)$ profile.  This identity is
confirmed numerically by the overlapping  of $\delta\Omega_c$ and $\delta\lambda_c$ in the
right panels of \cref{fig:c4c6g,fig:c4c6b,fig:c4c6a}. It may be noted that for  dilute halos (small $z$) with shallow inner logarithmic slope (low $\gamma$), \cref{eq:lambdaOmega} can be written in a form similar to Eqs. (19) and (20) of \cite{Pezzella:2024tkf}.

Even at fixed compactness, both $\Omega_c$ and $\lambda_c$ depend clearly on the shape of the underlying density profile, with the
strongest variation arising from the inner logarithmic parameter $\gamma$, followed
by $\beta$ and then $\alpha$, as seen in
\cref{fig:c4c6g,fig:c4c6b,fig:c4c6a}.  This ordering follows from
how each parameter redistributes halo mass relative to the radial
weight $(r-2M_{\rm BH})^{-1}$ entering $\mathcal{I}$, which is
largest near the inner cut-off. Increasing $\gamma$ moves mass
inward most efficiently and so dominates the redshift, $\beta$
acts more uniformly across all radii, and $\alpha$ only
redistributes mass over the comparatively gentle transition region near $a_0$.  \Cref{App:eikonal} derives this mechanism in detail and
identifies the regimes in which the ordering can change, including a sign reversal in the $\alpha$ dependence at small $\gamma$.

In vacuum general relativity, the QNM frequencies in the eikonal limit ($l \gg 1$) are related to the light ring parameters through
\begin{align}\label{eq:eikonal}
   \omega_c = l \Omega_c - i\!\left(n+\frac{1}{2}\right)\!|\lambda_c|,
\end{align}
where $l$ and $n$ denote the angular and overtone numbers,
respectively.  In the axial sector, where fluid modes are absent,
this correspondence is expected to remain approximately valid.
Using \cref{eq:lambdaOmega}, the eikonal frequencies take the
compact form
\begin{align}\label{eq:eikonalDM}
   \omega_c = \Omega_c^{\rm Sch}\,e^{-\mathcal{I}/2}
              \!\left[l - i\!\left(n+\frac{1}{2}\right)\right],
\end{align}
so both $\operatorname{Re}\,\omega_c$ and
$\operatorname{Im}\,\omega_c$ are redshifted by the same factor
$e^{-\mathcal{I}/2}$ relative to their Schwarzschild values.  Since
$\mathcal{I}$ depends on the full profile shape
$(\alpha,\beta,\gamma)$ and not merely on $M_{\rm halo}$, halos with identical mass and compactness but different density profiles generally produce different values of $\mathcal{I}$ and hence distinct ringdown signatures.  This same profile dependence, however, also makes different halo configurations capable of producing an identical shift, a steep inner cusp at low
compactness can mimic a shallower cusp at higher compactness,
represented by the horizontal line in the right panel of
\cref{fig:c4c6g}, whose intersections give two halo configurations
with different compactness and inner cusp but the same $\Omega_c$
and $\lambda_c$.  We return to this degeneracy in \cref{App:eikonal}.

\Cref{App:eikonal} provides a complete analytic derivation of the results summarised in this section.  We also compare in \cref{app:check} the frequencies predicted by the eikonal relation \cref{eq:eikonalDM} with those obtained from the
sixth-order WKB approximation for several representative halo models at fixed compactness.
\subsection{Effect of halo parameters on QNMs}\label{sec:QNMhalo}
To systematically quantify the influence of the halo parameters on
the ringdown spectrum, we compute the fractional deviations
(redshift) of the real and imaginary parts of the quasinormal mode
frequencies from their corresponding Schwarzschild values for a
given multipole and overtone,
\begin{align}\label{eq:frac-shift}
\delta \omega_{\rm Re/Im}
=
\frac{\omega_{\rm Re/Im}^{\rm Sch.}-\omega_{\rm Re/Im}^{\rm DM}}
{\omega_{\rm Re/Im}^{\rm Sch.}}.
\end{align}
Motivated by the dependence of the eikonal frequencies on the halo
compactness, we show in \cref{fig:zplotnew} the variation of
$\delta\omega_{\rm Re}$ and $\delta\omega_{\rm Im}$ with compactness
for several representative halo models. We find that, for a fixed
density profile, increasing halo compactness leads to progressively
larger deviations of both the QNM frequency and damping rate from
their Schwarzschild values. The fractional deviations at the leading order are linear in compactness. 
This is explicitly shown by the linear fit in ~\cref{fig:zplotnew} and is in agreement with the observations of ~\cite{Pezzella:2024tkf}. 
The fractional deviations in the real and imaginary parts of $\omega$ remain nearly identical for all halo profiles and compactnesses considered. This behaviour is a direct consequence of the common halo-redshift mechanism derived in \cref{App:QNM}, with the small residual differences reflecting higher-order corrections beyond the analytic approximation. The horizontal lines constitute the degeneracy surface, showing that the same QNM value can be obtained from DM profiles with different compactness and shape parameters.
\begin{figure}[htbp!]
    \centering
    \includegraphics[width=\linewidth]{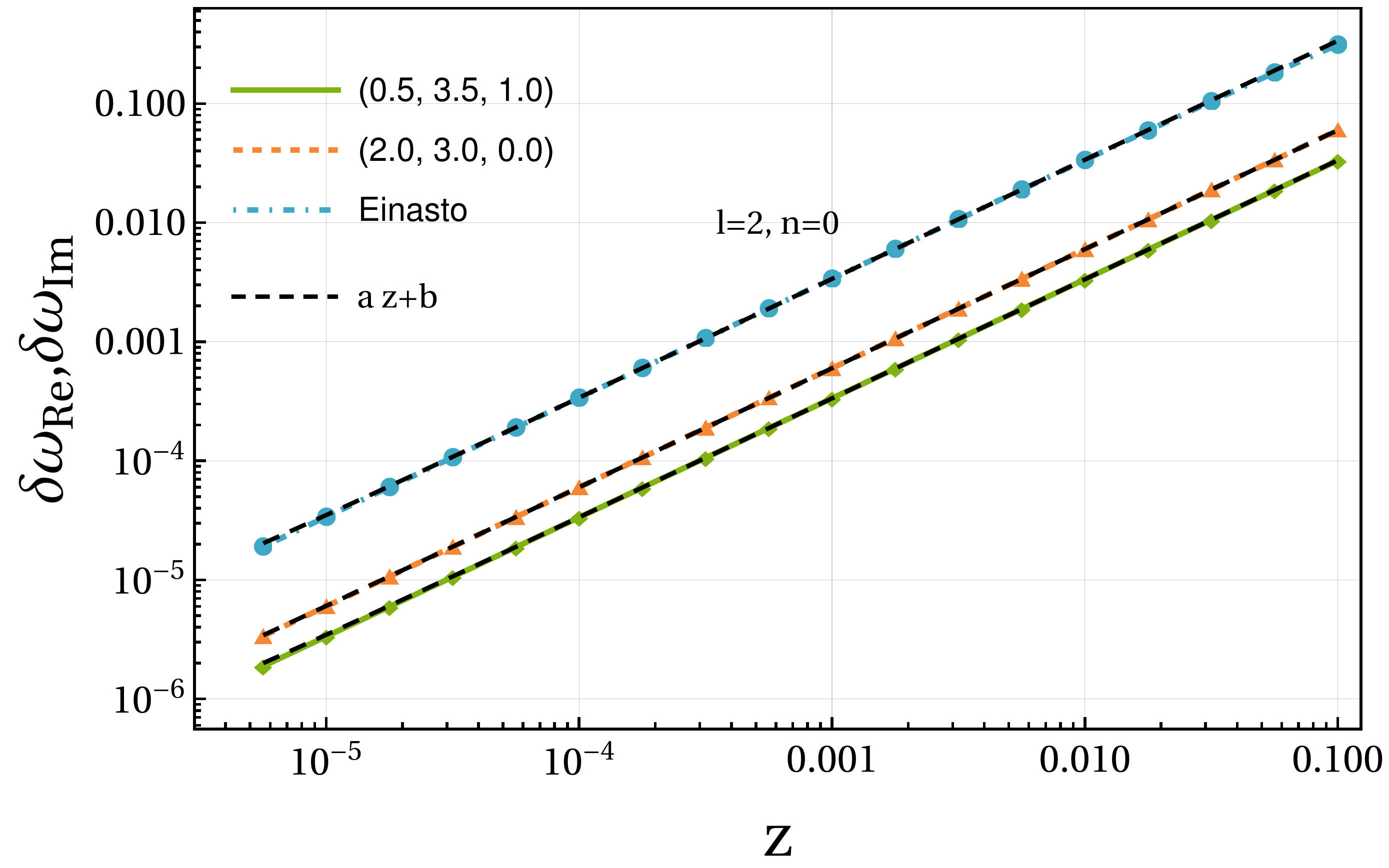}
    \caption{Fractional deviation in the real QNM frequency and damping rate from the corresponding Schwarzschild values with compactness for three halo configurations as depicted in the plot with $a_0=10^5 M_{\rm BH}$. The deviation in the real oscillation frequency is represented by distinct line styles (solid, dashed, dot-dashed), while the corresponding deviation in the damping rate is displayed as discrete points. The black dashed line represents a linear fit.}
    \label{fig:zplotnew}
\end{figure}
To highlight the dependence of the fractional redshift on the detailed shape parameters of the DM profile, we plot
$\delta\omega_{\rm Re/Im}$ for the fundamental $l=2$ mode and its first overtone against $\gamma$, $\beta$, and $\alpha$ in the first two columns of \cref{fig:c4c6g,fig:c4c6b,fig:c4c6a} respectively.
\begin{figure*}[htbp!]
    \centering
    \includegraphics[width=0.95\linewidth]{c4c6gLog.png}
    \caption{Fractional deviations from Schwarzschild values, plotted on a logarithmic scale, as functions of the halo parameter $\gamma$. In each panel, the upper and lower curves correspond to compactness values $z=10^{-4}$ and $10^{-6}$, respectively. When \(z=10^{-4}\), we took \(a_0=10^5 M_{\rm BH}\) and for \(z=10^{-6}\), we set \(a_0=10^7 M_{\rm BH}\), leading to a constant \(M_{\rm halo}=10 M_{\rm BH}\) in both of the cases. The first and second columns show the fractional deviations in the oscillation frequency and damping rate of the $l=2$ fundamental ($n=0$) and first-overtone ($n=1$) QNMs, respectively. Fractional deviation of the real oscillation frequency is represented by distinct line styles (solid, dashed), while fractional deviation in the damping rate is displayed as discrete points. The third column shows the corresponding fractional deviations in the light ring angular frequency and the principal Lyapunov exponent. Fractional deviation in the light ring angular frequency is also represented by distinct line styles (solid, dashed), while fractional deviation in the principal Lyapunov exponent is displayed as discrete points. Each row corresponds to a fixed $\alpha$ and $\beta$ as depicted in the plot.}
    \label{fig:c4c6g}
    \end{figure*}
\begin{figure*}[htbp!]
    \centering
    \includegraphics[width=0.95\linewidth]{c4c6bLog.png}
    \caption{Fractional deviations from Schwarzschild values, plotted on a logarithmic scale, as functions of the halo parameter $\beta$. In each panel, the upper and lower curves correspond to compactness values $z=10^{-4}$ and $10^{-6}$, respectively. When \(z=10^{-4}\), we took \(a_0=10^5 M_{\rm BH}\) and for \(z=10^{-6}\), we set \(a_0=10^7 M_{\rm BH}\), leading to a constant \(M_{\rm halo}=10 M_{\rm BH}\) in both of the cases. The first and second columns show the fractional deviations in the oscillation frequencies and damping rates of the $l=2$ fundamental ($n=0$) and first-overtone ($n=1$) QNMs, respectively. The third column shows the corresponding fractional deviations in the light ring angular frequency and the principal Lyapunov exponent. The inset in each panel shows the corresponding quantities on a logarithmic scale for small values of $\beta$. Each row corresponds to a fixed $\alpha$ and $\gamma$ as depicted in the plot.}
    \label{fig:c4c6b}
    \end{figure*}
    \begin{figure*}[htbp!]
    \centering
    \includegraphics[width=0.95\linewidth]{c4c6a.png}
    \caption{Fractional deviations from Schwarzschild values, plotted on a logarithmic scale, as functions of the halo parameter $\alpha$. In each panel, the upper and lower curves correspond to compactness values $z=10^{-4}$ and $10^{-6}$, respectively. When \(z=10^{-4}\), we took \(a_0=10^5 M_{\rm BH}\) and for \(z=10^{-6}\), we set \(a_0=10^7 M_{\rm BH}\), leading to a constant \(M_{\rm halo}=10 M_{\rm BH}\) in both of the cases. The first and second columns show the fractional deviations in the oscillation frequency and damping rate of the $l=2$ fundamental ($n=0$) and first-overtone ($n=1$) QNMs, respectively. The third column shows the corresponding fractional deviations in the light ring angular frequency and the principal Lyapunov exponent. Each row corresponds to a fixed $\beta$ and $\gamma$ as depicted in the plot.}
    \label{fig:c4c6a}
\end{figure*}
The redshift grows with both the inner logarithmic slope $\gamma$ and the outer logarithmic
slope $\beta$, with $\gamma$ dominating over most of the parameter
space explored. This ordering is not universal, however, for small $\gamma$ or a small gap $\beta-\gamma$, the sensitivity to $\beta$ can match or exceed that of $\gamma$, as seen in the upper panels of \cref{fig:c4c6g,fig:c4c6b}. The dependence on the transition parameter $\alpha$ is more subtle still, and changes sign with $\gamma$. Figure~\ref{fig:c4c6a} shows the redshift {increasing} with $\alpha$ at $\gamma=1.0$ (top row), but {decreasing} with $\alpha$ at $\gamma=2.5$ (middle and bottom rows), with the crossover occurring near $\gamma\approx1$.  \Cref{App:QNM} derives the radial mechanism behind each of these trends and identifies precisely where the orderings hold and where they invert.
The physical constraint $\beta>\gamma$ also bounds how large the redshift can become at fixed $\beta$ and $M_{\rm halo}$. As
$\gamma\to\beta^-$, the profile approaches a single power law and $\delta\omega$ saturates at a finite value $\delta_{\rm
sat}(\beta,M_{\rm halo})$, shared by both the eikonal and
non-eikonal quantities, rather than increasing without bound. We
derive this saturation value explicitly in \cref{App:QNM}. A finite
saturation means that, for a given $(\beta,M_{\rm halo})$, there is
a maximum redshift reachable by varying $\gamma$ alone. A measured
$\delta\omega$ above $\delta_{\rm sat}(\beta,M_{\rm halo})$ would be
inconsistent with the ($\alpha, \beta, \gamma$) family at that halo mass,
giving a direct observational constraint, while a measured shift
below $\delta_{\rm sat}$ remains consistent with a range of $\gamma$
values and feeds into the degeneracy discussed below. Because
$\delta_{\rm sat}$ depends on $M_{\rm halo}$ rather than on $a_0$
individually, this bound constrains $\gamma$ once $(\beta,M_{\rm
halo})$ are otherwise fixed or independently known, regardless of
$a_0$.
In \cref{fig:c-g}, we plot the variation of
$\delta\omega_{\rm Re/Im}$ with $\gamma$ for different compactness
at fixed $\alpha$ and $\beta$.
\begin{figure}[htbp!]
    \centering
    \includegraphics[width=\linewidth]{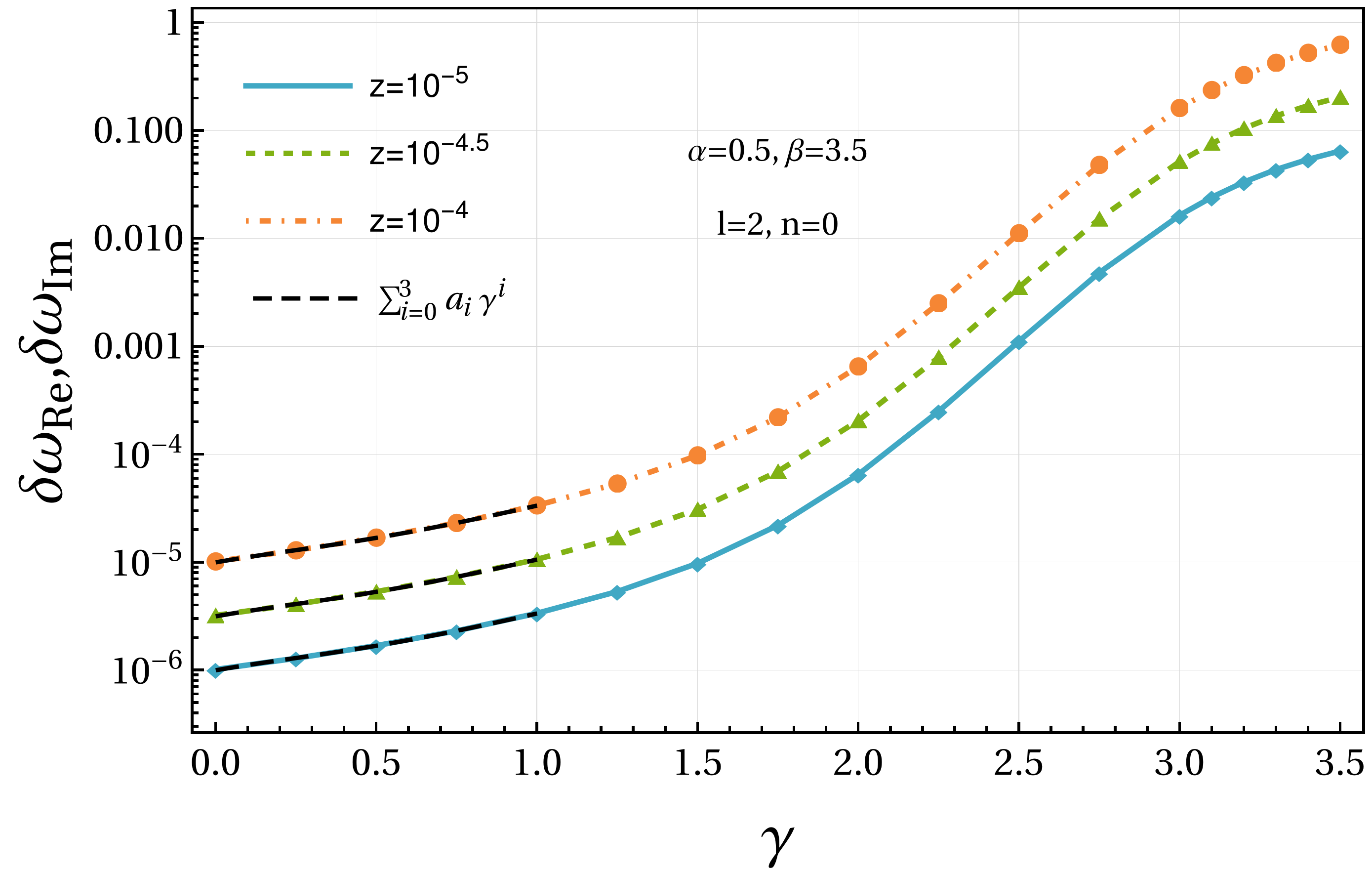}
    \caption{Degeneracy in the fractional deviations of the QNM frequencies from Schwarzschild values across different combinations of compactness ($z$) and  inner logarithmic slope ($\gamma$) with $a_0=10^5 M_{\rm BH}$. The horizontal grid lines, which intersect any two or three curves, explicitly illustrate the parametric degeneracy in which distinct physical configurations ($z,~\gamma$) yield same QNM frequencies. Fractional deviation of the real oscillation frequency is represented by distinct line styles (solid, dashed, dot-dashed), while fractional deviation in the damping rate is displayed as discrete points. The black dashed line denotes a polynomial fit to the fractional deviation for $\gamma \in [0,1]$. Here \(a_0\) has fixed value of \(10^5 M_{\rm BH}\).}
\label{fig:c-g}
\vspace{.4cm}
\includegraphics[width=\linewidth]{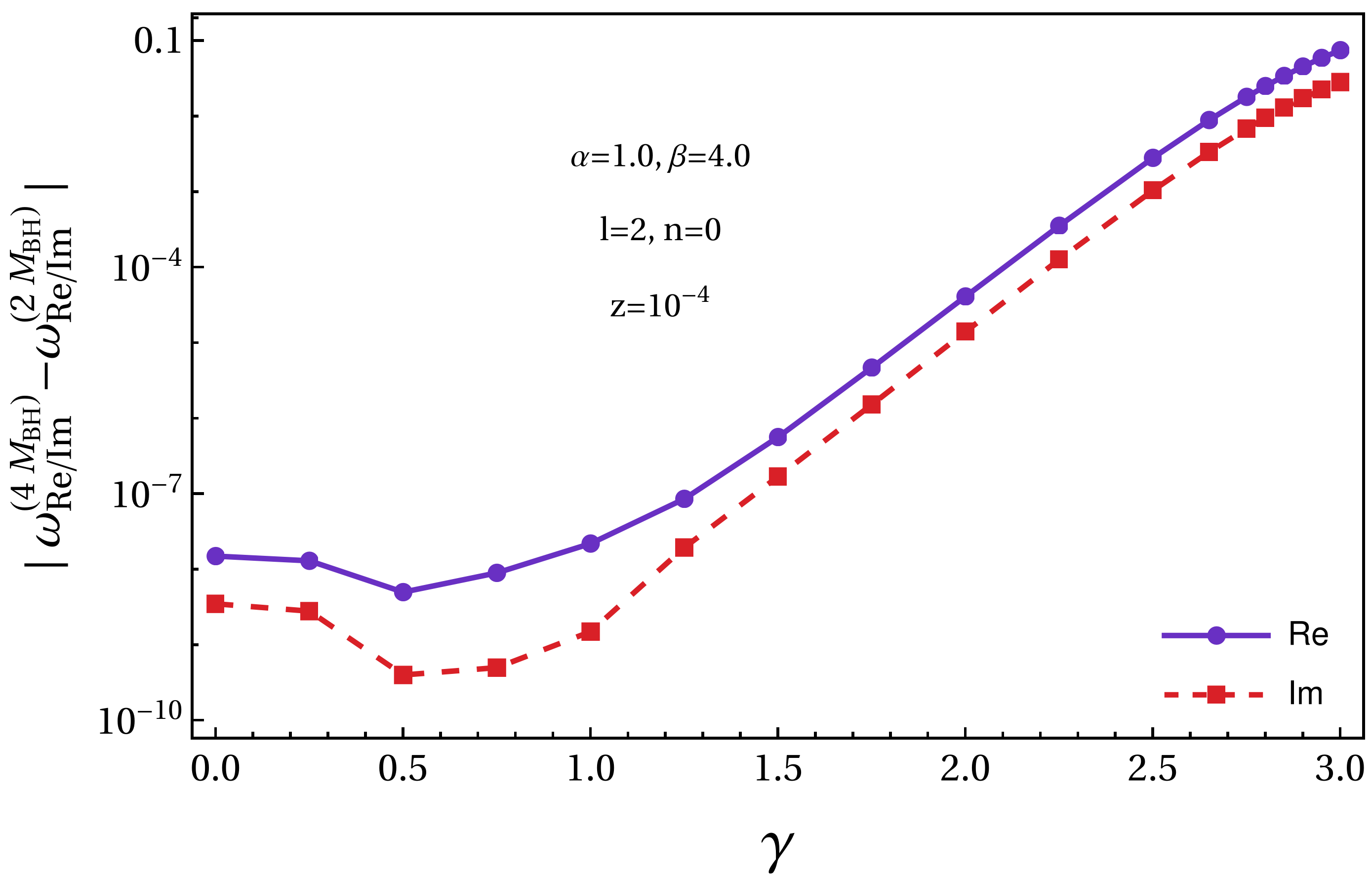}
   \caption{Difference in the fundamental ($l=2$, $n=0$) QNM oscillation frequency and damping rate computed using halo profiles with inner cut-offs at $2M_{\rm BH}$ and $4M_{\rm BH}$. The figure shows the magnitude of the differences on a logarithmic scale.}
    \label{fig:diff2M-4M}
\end{figure}

Note that unlike \cref{fig:c4c6g,fig:c4c6b,fig:c4c6a}, where the compactness
is varied by changing $a_0$ at fixed $M_{\rm halo}$, the compactness
in \cref{fig:c-g} is varied by changing $M_{\rm halo}$ at fixed $a_0$. The two choices are not equivalent away from small compactness. As $\gamma$ approaches its maximum value $\beta$, the
curves in \cref{fig:c4c6g} converge to a common value regardless of $a_0$, whereas the curves in \cref{fig:c-g} remain separated
according to $M_{\rm halo}$, since the saturated redshift depends on
$M_{\rm halo}$ but not on $a_0$ individually (see
\cref{App:QNM} for the underlying argument). For $\gamma\lesssim1$, we observe that the QNM shifts can be expressed as a polynomial, $\delta\omega\approx\sum_{i\geq0} a_i\gamma^i$, however, at larger $\gamma$ the polynomial dependence breaks down as the growth accelerates (see \cref{app:shape_QNM}). This is explicitly shown in \cref{fig:c-g} by the polynomial fit of $\delta\omega_{\rm Re/Im}$ for $\gamma\leq 1$. 

These trends closely mirror those observed for the fractional shifts in the light ring frequency and principal Lyapunov exponent, providing further support for the eikonal interpretation of the quasinormal mode spectrum.
We provide a detailed analysis in ~\cref{App:QNM}.

The strong localization of the redshift kernel near the inner cut-off naturally raises the question of how sensitive the predicted QNM shifts are to the choice of the inner truncation radius, $r_0$. Figure~\ref{fig:diff2M-4M} addresses this issue by comparing the fundamental $(\ell=2,n=0)$ oscillation frequency and damping rate obtained using  the horizon cut-off $r_0=2M_{\rm BH}$, standard in the literature~\cite{Pezzella:2024tkf} , and our value $r_0=4M_{\rm BH}$ (see also~\cite{Zhao:2023tyo}). For small $\gamma$, the difference between the two is less than $10^{-7}$. As $\gamma$ increases, the corresponding difference between $2M_{\rm BH}$ and $4M_{\rm BH}$ grows, and the difference rises to 0.1 by $\gamma=3$. The cut-off is thus a physical choice in the cuspy regime, not a convention, and $r_0=4M_{\rm BH}$ is the appropriate one, since matter crossing it is captured~\cite{Sadeghian:2013laa}.

{\subsection{Observational relevance of the halo-induced QNM shifts}
We now place the halo-induced QNM shifts in the context of present observational precision. The purpose of this comparison is not to derive a direct observational bound on the DM halo parameters, since our calculation describes axial perturbations of a static, spherically symmetric BH embedded in a prescribed matter distribution. The available observational constraints, by contrast, are obtained from binary BH merger data using waveform models for a Kerr remnant. Nevertheless, the precision achieved in current BH spectroscopy provides a useful benchmark for judging the magnitude of the environmental shifts predicted in this work.

We use as reference the recent LVK analysis of GW250114, which provides one of the strongest single-event tests of the Kerr ringdown spectrum \cite{LIGOScientific:2025wao,LIGOScientific:2025rid}. In that analysis, the oscillation frequency and damping time of the fundamental mode are parametrized as
\begin{equation}
 f_{\ell m0}=f^{\rm GR}_{\ell m0}\left(1+\delta \hat f_{\ell m0}\right),
 \qquad
 \tau_{\ell m0}=\tau^{\rm GR}_{\ell m0}\left(1+\delta \hat\tau_{\ell m0}\right).
 \label{eq:lvk_qnm_parametrization}
\end{equation}
The GR values are computed from the remnant mass and spin inferred from the full inspiral--merger--ringdown signal. For the dominant $(2,2,0)$ mode, the reported constraints are
\begin{equation}
 \delta \hat f_{220}=0.02^{+0.02}_{-0.02},
 \qquad
 \delta \hat\tau_{220}=-0.01^{+0.10}_{-0.09},
 \label{eq:gw250114_constraints}
\end{equation}
at the $90\%$ credible level. These intervals are consistent with the Kerr prediction within the present measurement uncertainty.

In the present work, we characterize the QNM spectrum in terms of the complex frequency
\begin{equation}
 \omega=\omega_{\rm Re}-i\omega_{\rm Im},
 \label{eq:complex_frequency_convention}
\end{equation}
where the imaginary part determines the damping time through
\begin{equation}
 \omega_{\rm Im}=\frac{1}{\tau}.
 \label{eq:damping_time_relation}
\end{equation}
The fractional shifts used throughout this work are defined as
\begin{equation}
 \delta\omega_{\rm Re/Im}
 =
 \frac{\omega^{\rm Sch}_{\rm Re/Im}-\omega^{\rm DM}_{\rm Re/Im}}
 {\omega^{\rm Sch}_{\rm Re/Im}}.
 \label{eq:qnm_fractional_shift_obs_section}
\end{equation}
With this convention, a positive halo-induced redshift corresponds to a lower oscillation frequency and a longer damping time. To linear order in the fractional deviation, the two conventions are related by
\begin{equation}
 \delta\omega_{\rm Re}\simeq -\delta\hat f,
 \qquad
 \delta\omega_{\rm Im}\simeq \delta\hat\tau .
 \label{eq:translation_between_conventions}
\end{equation}
We therefore use
\begin{equation}
 |\delta\omega_{\rm Re}|\sim 0.04,
 \qquad
 |\delta\omega_{\rm Im}|\sim 0.09,
 \label{eq:representative_precision}
\end{equation}
as representative reference values for the current precision with which the real and imaginary parts of the dominant QNM frequency can be tested.

Figure~\cref{fig:fewabg} compares these reference values with the fractional shifts predicted by our halo models at fixed compactness $z=10^{-4}$. The shifts grow monotonically with the inner logarithmic slope $\gamma$, showing that centrally concentrated halos produce larger changes in the ringdown spectrum. For the representative profiles shown in the figure, only sufficiently cuspy configurations lead to shifts comparable to the present benchmark precision. Standard profiles such as NFW-like or Hernquist-like profiles remain below this level for the chosen compactness, and would therefore be difficult to distinguish from the vacuum Schwarzschild prediction using a measurement of comparable precision.

We next examine the full $(\alpha,\beta,\gamma)$ parameter space. In ~\cref{fig:abgc4,fig:abgc6}, we vary the parameters, such that,
\begin{equation}
 \alpha\in[0.25,2.5],
 \qquad
 \beta\in[0.25,5],
 \qquad
 \gamma\in[0,3],
 \label{eq:parameter_ranges_obs_section}
\end{equation}
and retain only the physically allowed region satisfying $\beta\geq\gamma$. The coloured surface marks the locus where the predicted QNM shift becomes comparable to the observational reference precision. Points lying on the side of the surface with larger redshift correspond to halo configurations that would produce a potentially resolvable departure from the Schwarzschild value in an observation of similar quality. Points on the other side remain within the present measurement uncertainty.

The comparison between \cref{fig:abgc4,fig:abgc6} shows the expected compactness dependence. At $z=10^{-4}$, a larger portion of the physically allowed parameter space can produce shifts near the reference precision. At $z=10^{-6}$, the corresponding region shrinks substantially, and only very centrally concentrated profiles remain capable of producing comparable deviations. This confirms that the observational relevance of the effect is controlled jointly by the halo compactness and the profile shape, with the inner logarithmic slope $\gamma$ playing the dominant role in the parameter range considered here.

 A direct observational constraint would require waveform models that consistently include the environmental modification throughout the inspiral, merger and ringdown. The present analysis instead shows that, if future massive BH or extreme mass ratio systems are observed with comparable or better ringdown precision, sufficiently compact and cuspy DM environments could produce QNM shifts large enough to be observationally relevant. Conversely, shallow or weakly compact halos are expected to remain degenerate with the vacuum prediction at current precision.
\begin{figure}[htbp!]
    \centering
    \includegraphics[width=\linewidth]{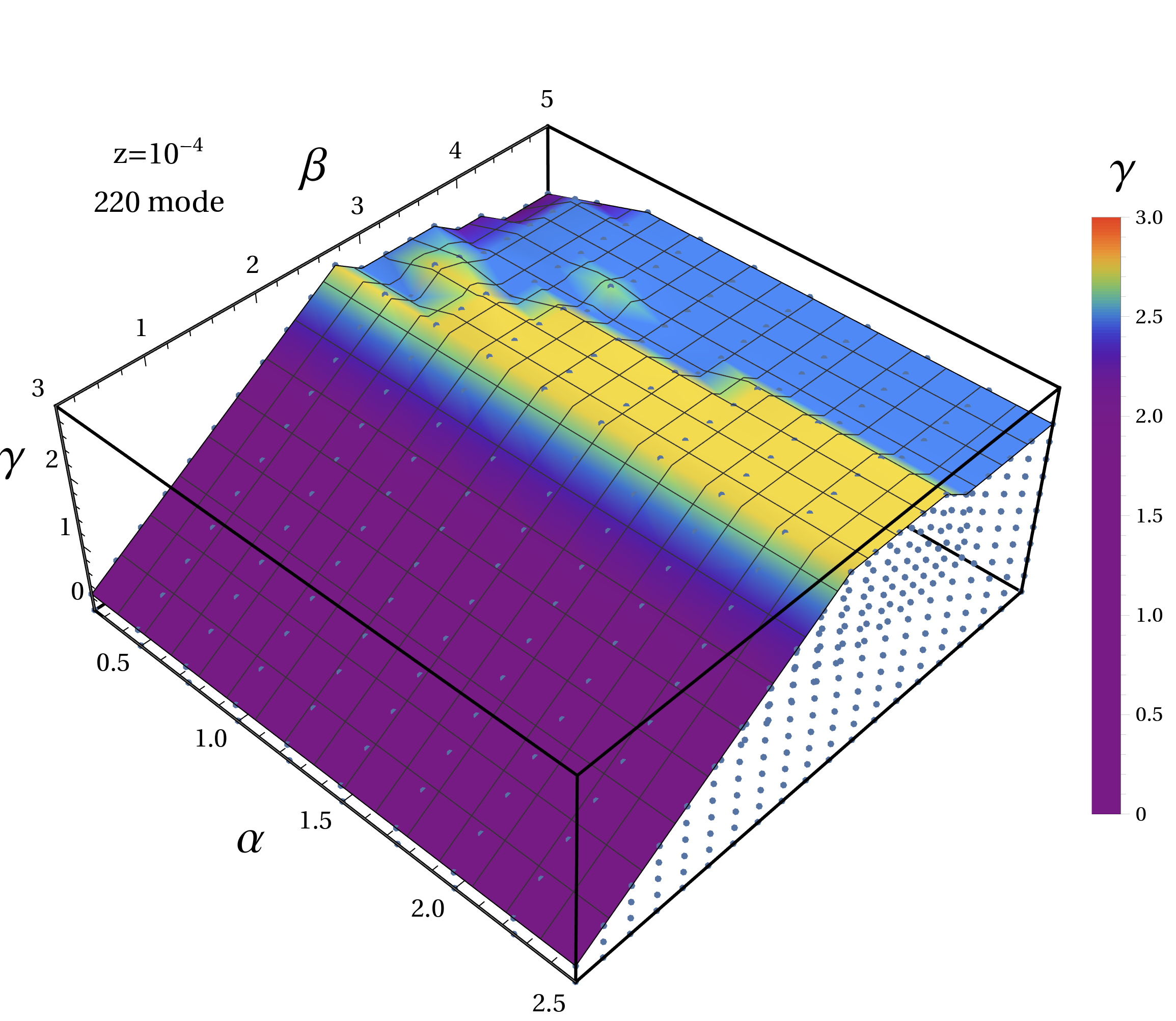}
    \caption{Three-dimensional parameter space of the generalized $(\alpha,\beta,\gamma)$ halo profile for compactness $z=10^{-4}$ and $a_0=10^5 M_{\rm BH}$. Only the physically allowed region satisfying $\beta\ge\gamma$ is shown. The coloured surface corresponds to parameter combinations for which the predicted fractional deviations in the oscillation frequency and damping rate of the $l=2$ fundamental QNM become comparable to the representative $90\%$ credible intervals inferred from GW250114.}
    \label{fig:abgc4}
\end{figure}
\begin{figure}[htbp!]
    \centering
    \includegraphics[width=1\linewidth]{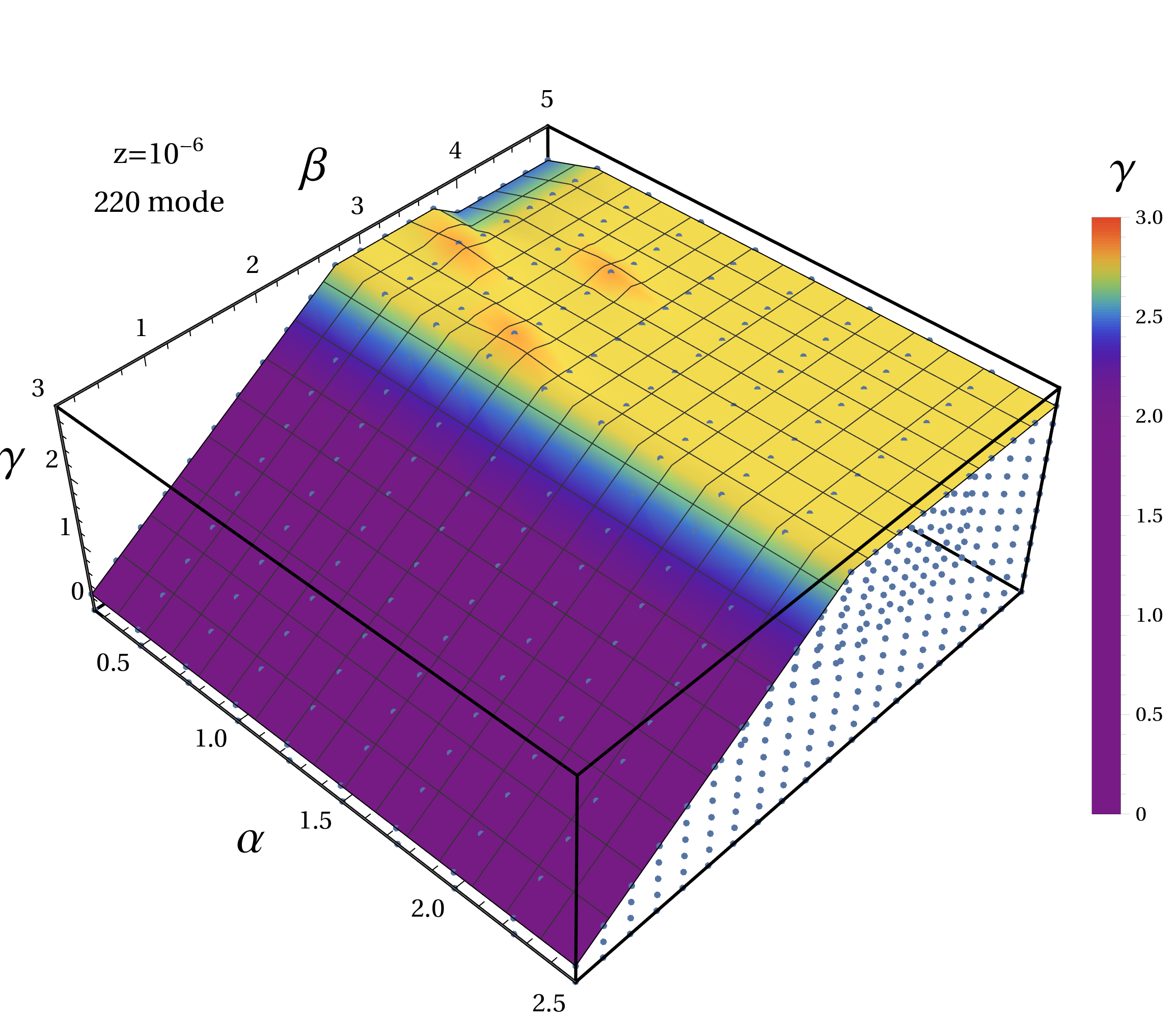}
   \caption{Three-dimensional parameter space of the generalized $(\alpha,\beta,\gamma)$ halo profile for compactness $z=10^{-6}$ and $a_0=10^7 M_{\rm BH}$. Only the physically allowed region satisfying $\beta\ge\gamma$ is shown. The coloured surface corresponds to parameter combinations for which the predicted fractional deviations in the oscillation frequency and damping rate of the $l=2$ fundamental QNM become comparable to the representative $90\%$ credible intervals inferred from GW250114.}
    \label{fig:abgc6}
\end{figure}
\begin{figure}[htbp!]
    \centering
    \includegraphics[width=\linewidth]{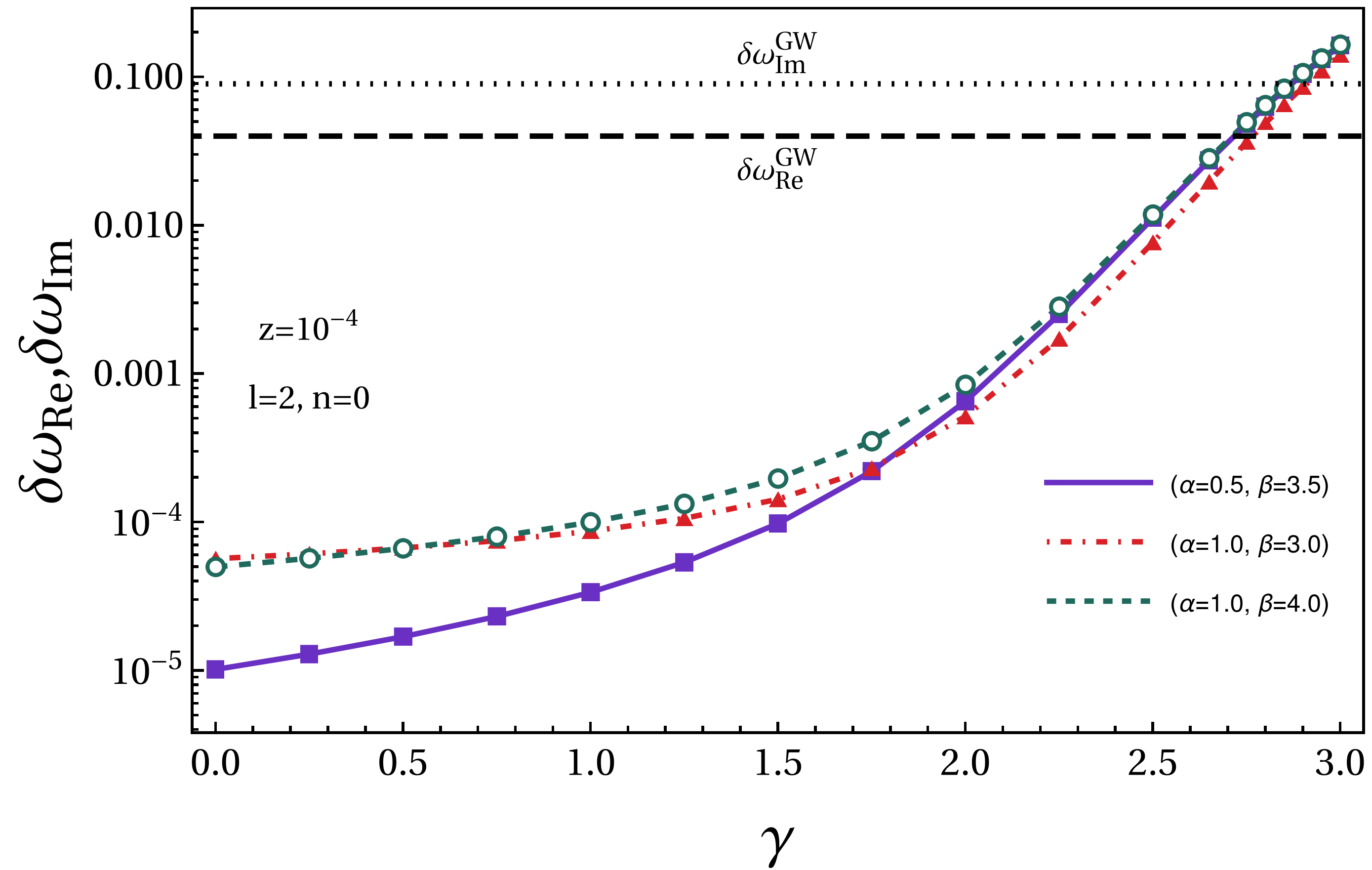}
    \caption{Comparison of the predicted fractional QNM deviations for representative halo profiles with compactness $z=10^{-4}$ and \(a_0=10^5 M_{\rm BH}\) with the current observational precision inferred from GW250114. The black dashed and dotted lines denote the representative $90\%$ credible intervals for the real and imaginary parts of the $l=2$ fundamental QNM frequency, respectively. The fractional deviations in the real part are shown by coloured curves, while those in the imaginary part are represented by the corresponding markers.}
    \label{fig:fewabg}
\end{figure}
\section{Time domain analysis}\label{sec:td}
\begin{figure}[htbp!]
    \centering
    \includegraphics[width=1\linewidth]{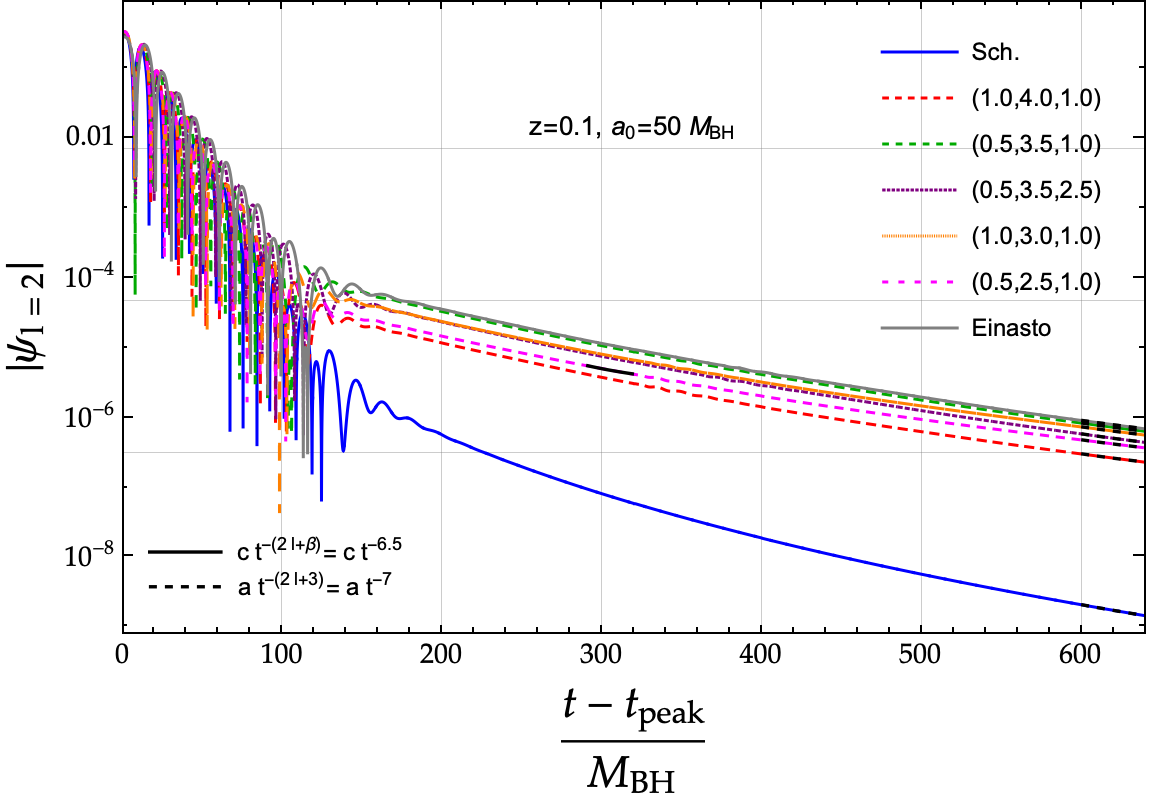}
    \caption{Evolution of the Gaussian wave packet for different DM halo configurations. The solid blue curve corresponds to the Schwarzschild BH. The time coordinate is shifted such that the peaks of the ringdown coincide. To keep the time-domain evolution computationally tractable, we use representative values of $M_{\rm halo}\sim5M_{\rm BH}$ and $a_0\sim50M_{\rm BH}$, without affecting the qualitative features of the ringdown and late-time tails. For the $(1,3,1)$ and $(0.5,2.5,1)$ halo profiles, the density is truncated at $r_t=20.5\,a_0$. The solid black line shows the intermediate-tail scaling $t^{-(2l+\beta)}=t^{-6.5}$ for the $(0.5,2.5,1)$ profile, while the dashed black line denotes the expected Price-law decay $t^{-(2l+3)}=t^{-7}$.}
    \label{fig:TD}
\end{figure}
The frequency-domain WKB results of the preceding section provide accurate quasinormal frequencies for $l>n$, but cannot access the late-time behaviour of the signal or the transient structure that connects the ringdown to the power-law tail. Thus, to complement the frequency-domain analysis and to verify the late-time behaviour predicted in \cite{Rosato:2025rtr}, we solve the master equation~\cref{eq:mastereqn} directly in the time domain using the characteristic integration scheme of Gundlach, Price, and Pullin~\cite{Gundlach:1993tp}. The evolution is initiated by a Gaussian pulse, whose scattering off the effective potential barrier generates the characteristic sequence of prompt response, quasinormal ringing, and late-time tail. The late-time tail and quasinormal ringing arise from distinct physical mechanisms. While quasinormal ringing is produced by resonant scattering off the effective potential barrier, the late-time tail originates from the backscattering of long-wavelength perturbations by the asymptotic spacetime curvature. For generic compact initial data with multipolar index $l$, the perturbation at fixed spatial position decays following Price-law~\cite{Gundlach:1993tp} at the leading order~\cite{Andersson:1996cm}
\begin{align}
    \psi\propto t^{-\lt(2l+3\rt)}
\end{align}
at asymptotically late times. This power-law behaviour arises from the branch-cut contribution of the frequency-domain Green's function and is therefore governed by the asymptotic behaviour of the effective potential rather than its near-horizon structure. In the presence of a fluid (including DM) environment around the central BH, the perturbation of the surrounding fluid excites the fluid modes which appear in the time-domain signal after the prompt-ringdown phase. Such fluid driven modes have been explicitly shown to be excited in the polar sector~\citep{Cardoso:2022whc,Spieksma:2024voy}. 
In the axial sector (with $\delta u_\mu=\delta w_\mu=0)$, however, the time evolution of the perturbation is governed by the source-free Schrödinger-like equation,~\cref{eq:mastereqn}, where the effective potential takes the asymptotic form,
\begin{align}\label{eq:vinf-largeb}
V_{\infty}(r)&\approx\frac{l(l+1)}{r^2}-2 (M_{\rm BH}+M_{\rm halo})\frac{l(l+1)+3}{r^3}\nonumber\\ 
&+4\pi C_\beta\frac{\beta+3}{\beta-3}r^{-\beta}+\mathcal{O}\lt( r^{-\rm{Min} [(\beta+\alpha),~{4}]}\rt),
\end{align}
where $C_\beta=\bar\rho_0 a_0^\beta$. Thus, for $\beta>3$ the DM contributions vanish faster than $1/r^2$ at large $r$ and consequently the late-time tail decays as $t^{-(2l+3)}$~\cite{Rosato:2025rtr}. For $\beta\leq 3$, the halo mass diverges at large $r$, thus the halo is truncated at a tidal radius $r_t>a_0$. This causes an observer at $r_t>r_{\rm obs}>a_0$, ($a_0\gg M_{\rm BH}$) to witness a shallower fall-off at intermediate time, $a_0<t<r_t$, 
\begin{align}
\psi(t)\propto t^{-(2l+\beta)}~.
\end{align}
 Beyond the truncation radius, where the spacetime becomes Schwarzschild (with an enhanced ADM mass $M_{\rm ADM}=M_{\rm BH}+M_{\rm halo}$), the standard asymptotic decay ($\psi(t)\propto t^{-(2l+3)}$) is eventually restored.

In \cref{fig:TD}, we plot the time profile for some representative DM distribution around the central spherically symmetric BH. It is useful to state the scales involved, since the parameters of \cref{fig:TD} are compressed for numerical tractability. The intermediate tail comes from backscattering off the region in which the effective potential falls as $r^{-\beta}$, extending from the scale radius to the truncation radius. Its decay rate depends only on $l$ and $\beta$, and the extent of the region only on the ratio $r_t/a_0$, so the compressed values of \cref{fig:TD} entail no loss of generality. The wider truncation (relative to $a_0$) adopted there simply makes the intermediate
scaling easier to identify. For $a_0=10^{5}M_{\rm BH}$ with the truncation $r_t=5a_0$ of Sec.~II\,B, this region lies between $10^{5}M_{\rm BH}$ and $5\times10^{5}M_{\rm BH}$, that is between $0.0048$ and $0.24$pc for $M_{\rm BH}=10^6\,M_\odot$. The two decay rates are thus complementary: the
intermediate exponent $2l+\beta$ measures the outer logarithmic slope, while the
truncation radius fixes where the Price-law behaviour is recovered. 

We verify that at late-time perturbation decays follow Price's law. However, for $(0.5,2.5,1)$-DM distribution with $\beta=2.5$ an intermediate tail proportional to $t^{-2l-\beta}=t^{-6.5}$ is observed at intermediate times.
\begin{table}
    \centering
    \begin{tabular}{|c|c|c|}
        \hline
        \textbf{DM Profiles} & \textbf{Prony-fit} & \textbf{WKB (6th Order)} \\ \hline
         1.0, 4.0, 1.0    & 0.3446328-0.0819501 i   & 0.3446369-0.0819955 i    \\ \hline
        0.5, 3.5, 1.0      & 0.3651418-0.0869178 i      & 0.3651036-0.0868649 i     \\ \hline
        0.5, 3.5, 2.5     &0.3282496-0.0781977 i    & 0.3282692-0.0781013 i     \\
        \hline
        1.0, 3.0, 1.0     & 0.3591402-0.0848168 i    & 0.3591856-0.0854569 i     \\
        \hline
        0.5, 2.5, 1.0     & 0.3656644-0.0870637 i    & 0.3656216-0.0869882 i     \\
        \hline
        Einasto     & 0.3163015-0.0752344 i    & 0.3163734-0.0752711 i     \\
        \hline
        Schwarzschild     & 0.3736769-0.0889529 i      & 0.3736194-0.0888910 i     \\ \hline
    \end{tabular}
    \caption{QNM frequencies computed from the time domain signal in ~\cref{fig:TD} and the corresponding values obtained from 6th order WKB for the $l=2$ fundamental mode.}
    \label{table:prony}
\end{table}
The QNM frequencies and damping rate for the  fundamental mode, obtained from ~\cref{fig:TD} using the Prony method~\cite{Berti:2007dg}, match the corresponding sixth-order WKB results to a high accuracy, as shown in ~\cref{table:prony}

\section{Tidal Love numbers}\label{sec:TLN}
The analysis of the preceding sections showed that the entire ringdown spectrum responds to the DM environment through a single quantity, the redshift integral $\mathcal{I}$ of \cref{eq:haloint}. Because the radial weight of $\mathcal{I}$ peaks near the inner cut-off, the QNM shift is most sensitive to the halo mass at \emph{small} radii, and hence to the inner logarithmic slope $\gamma$. In this section we compute a second, independent characterization of the same configuration, the static axial (magnetic-type) tidal Love number (TLN), and show that it is governed by a differently weighted functional of the same halo mass distribution: an integral weighted toward large radii and controlled mostly by the compactness $z$, the outer logarithmic slope $\beta$, and the truncation radius $r_t$. The two quantities therefore probe complementary regions of the halo, and respond to the shape parameters in opposite ways.

The (static) tidal Love number quantifies the conservative multipolar deformation induced in a compact object by an external static tidal field~\cite{Chakraborty:2026qru,Rodriguez:2026iot}. In vacuum general relativity in four spacetime dimensions, the static TLNs of an Schwarzschild BH vanish identically in both the polar (electric-type) and axial (magnetic-type) sectors
\cite{Binnington:2009bb,Damour:2009vw}. However, a surrounding matter environment can induce a nonzero tidal response. Consequently, TLNs can act as a a powerful probe  for surrounding environment. Here, we obtain the static axial TLNs of a Schwarzschild BH dressed with a DM halo configurations of \cref{sec: BG-density} analytically, to leading order in the halo compactness and verify the result against numerical computations Ref.~\cite{DOnofrio:2026ulh}.

\subsection{Static master 
equation}\label{sec:TLNmaster}
In the static limit, the axial sector isolates the tidal field. The $\theta\theta$ equation~\eqref{eq:h0dot} reduces to a
homogeneous first order equation for $h_1$,
\begin{align}\label{eq:h1static}
r(r-2m)\,h_1' + 2\lt(m - 2\pi r^{3}\bar\rho\rt)h_1 = 0~,
\end{align}
while the $t\theta$ equation~\eqref{eq:rphi} loses its $\dot h_1$ terms and
governs $h_0$ alone. The two sectors decouple. The nontrivial solution of
\cref{eq:h1static}, $h_1\propto r/(r-2M_{\rm BH})$ in the vacuum cavity, is
not a tidal deformation. As $h_1$ cannot affect the response, we set
$h_1=0$ with no loss of generality. This also removes the auxiliary vector from the problem: $W$ appears only in the source of \cref{eq:RWs}, so in the static limit the matter enters the perturbation equations through $U$ alone.

Substituting the irrotational condition~\eqref{eq:Ulm} into \cref{eq:rphi}
and using the background relation \cref{eq:pt}, the matter terms combine as
\begin{align}\label{eq:mattercollapse}
\frac{8\pi\bar\rho\,(r-m)}{r-2m} - 16\pi\lt(\bar\rho+\bar p_t\rt)
 = -\,8\pi\bar\rho
 = -\,\frac{2m'}{r^{2}}~,
\end{align}
so the static axial sector is governed by
\begin{align}\label{eq:staticmaster}
\lt(1-\frac{2m}{r}\rt) h_0''
 - \frac{m'}{r}\, h_0'
 - \lt[\frac{l(l+1)}{r^{2}}
        - \frac{4m}{r^{3}}
        - \frac{2m'}{r^{2}}\rt] h_0 = 0~.
\end{align}
Equation~\eqref{eq:staticmaster} coincides with the static axial equation of~\cite{DOnofrio:2026ulh}
for anisotropic fluids with vanishing radial pressure.

Two structural properties of \cref{eq:staticmaster} determine the physics of this section. First, the redshift function $f(r)$ has dropped out entirely -- the static axial perturbation is sourced by the
mass function $m(r)$ alone. The ringdown shift of
\cref{sec:QNMhalo} and the TLN thus probe two different radial weightings of the same halo mass distribution, and we quantify below how differently they respond to the shape parameters. Second, in the vacuum region $2M_{\rm BH}\le r\le 4M_{\rm BH}$, where $m=M_{\rm BH}$ and $m'=0$, the solution of \cref{eq:staticmaster} regular at the horizon is, 
\begin{align}\label{eq:vacsol}
   h_0^{(0)}(r)= C r^{l+1}\,{}_2F_1(1-l,-l-2;-2l;2M_{\rm BH}/r).
\end{align}
For $l=2$, this reduces to
\begin{align}\label{eq:vacsol2}
h_0^{(0)}(r) \propto r^{2}\lt(r - 2M_{\rm BH}\rt).
\end{align}
The large-$r$ expansion of $h_0^{(0)}$ contains no admixture of the decaying solution, implying that the axial TLN of an isolated Schwarzschild BH vanishes. Equation~\eqref{eq:vacsol} fixes the inner boundary data at
the edge of the DM-free cavity without approximation.

\subsection{Linear tidal response}\label{sec:TLNresponse}
In terms of the logarithmic derivative $y \equiv r h_0'/h_0$, \cref{eq:staticmaster} takes the Riccati form~\cite{Damour:2009vw,Ghosh:2026vig}
\begin{align}\label{eq:TLNriccati}
\lt(1-\frac{2m}{r}\rt)\lt(r y' + y^{2} - y\rt) = m' y+ l(l+1) - \frac{4m}{r} - 2m'~.
\end{align}
The vacuum limit of \cref{eq:TLNriccati} possesses two exact branches of solution, the growing branch (corresponding to~\cref{eq:vacsol})
\begin{align}\label{eq:yS}
y_1 \approx l+1 ~,
\end{align}
and a decaying branch with 
\begin{align}
    y_2 \approx -l~.
\end{align}

Horizon regularity selects \cref{eq:vacsol} in the vacuum cavity, which is pure growing solution, the halo then mixes in a small amount of the decaying solution, and that admixture is the induced response. We therefore write
\begin{align}\label{eq:expand}
y(r) = y_1(r) + \delta y(r)\,,
\end{align}
and linearize \cref{eq:TLNriccati} in $\delta y$ and in the halo, $m = M_{\rm BH} + m_h$, obtaining
\begin{align}\label{eq:dyeqn}
r\,\delta y' + \lt[\,2y_1(r)-1\,\rt]\delta y = \mathcal{S}[m_h,m_h'](r)\,,
\end{align}
where the source $\mathcal{S}$ is $\mathcal{O}(z)$ and vanishes in vacuum. Two properties of \cref{eq:dyeqn} make the expansion well posed throughout the halo region, $4M_{\rm BH} \le r \le r_t$.

First, the homogeneous part of \cref{eq:dyeqn} is stable under outward integration. Its solution is $\delta y \propto r/{[h_0^{(0)}]}^{2}$, so the response is damped wherever the coefficient $2y_1-1$ is positive. This holds throughout the exterior. Evaluating the vacuum limit of \cref{eq:TLNriccati} on the line $y = l+1$ forces
\begin{align}\label{eq:online}
r\,y'\big|_{y=l+1}
 = \frac{2M_{\rm BH}\,(l-1)(l+2)}{r-2M_{\rm BH}} > 0\,,
\end{align}
so this line can be crossed only upward. The regular solution starts at $y\to+\infty$ at the horizon and hence remains above $l+1$ for all $r$, giving $2y_1-1 \geq 2l+1$. A halo-induced deviation therefore relaxes at least as fast as $r^{-(2l+1)}$ rather than growing. The growing branch is thus the attractor of the outward flow, and the expansion is stable. 

Second, the expansion parameter is the halo compactness alone. The source in \cref{eq:dyeqn} is built from $m_h(r)/r$ and $m_h'(r)$, both $\mathcal{O}(z)$ throughout the halo, so $\delta y = \mathcal{O}(z)$
\footnote{Note that the response is small because the halo is dilute and
extended ($z = M_{\rm halo}/a_0 \ll 1$), not because it is light. Also, since $\delta y$ is measured relative to the exact vacuum solution and the cavity is vacuum, it vanishes identically there. The inner boundary condition is
simply $\delta y(4M_{\rm BH}) = 0$.}.

To carry out the linearization in the far zone $r\gg M_{\rm BH}$, where the
halo resides, we introduce the integrating factor $[h_0^{(0)}(r)]^{2}/r$, whose leading form is $r^{2l+1}$. Thus, the first-order equation becomes
\begin{align}\label{eq:TLNexact}
\frac{d}{dr}\!\lt[r^{2l+1}\,\delta y\rt]
 = (l-1)\lt[\,m_h'(r)\,r^{2l} + 2(l+2)\,m_h(r)\,r^{2l-1}\,\rt]\,.
\end{align}
At $l=2$ the source collapses to the polynomial $8\,m_h r^{3}+m_h' r^{4}$, so \cref{eq:TLNexact} is then exact to
linear order in $z$ on the \emph{full} Schwarzschild background. For
$l\geq3$ the same construction holds to leading order, with fractional
corrections $\mathcal{O}(M_{\rm BH}/a_0)$. Integrating from the cavity edge,
where $\delta y = 0$,
\begin{equation}
\label{eq:dyh}
\delta y(r_t) = \frac{(l-1)\,\mathcal{J}_l}{r_t^{2l+1}}~,
\end{equation}
\begin{align}
\mathcal{J}_l &\equiv \int_{4M_{\rm BH}}^{r_t}
\lt[\,m_h'\,r^{2l} + 2(l+2)\,m_h\,r^{2l-1}\,\rt] dr\nonumber\\
&= M_{\rm halo}\,r_t^{2l}
 + 4\!\int_{4M_{\rm BH}}^{r_t}\! m_h\,r^{2l-1}\, dr\,,
\label{eq:Jdef}
\end{align}
using $m_h(4M_{\rm BH})=0$ and $m_h(r_t)=M_{\rm halo}$ in the integration by
parts.

Equation~\eqref{eq:dyh} is not yet the tidal response. For $r > r_t$ the
spacetime is Schwarzschild with mass $M_{\rm ADM} = M_{\rm BH} + M_{\rm halo}$,
so the applied tidal field is built on $M_{\rm ADM}$ rather than
$M_{\rm BH}$, and the exterior solution reads
$h_0 \propto h_0^{(0)}(r;M_{\rm ADM}) + \kappa_l\, r^{-l}+\dots$, with
$\kappa_l$ the induced (decaying) amplitude, for which
$\delta y^{\rm ext}=-(2l+1)\,\kappa_l\,r^{-(2l+1)}$. The part of
$\delta y(r_t)$ set by the total enclosed mass $m_h(r_t)=M_{\rm halo}$, rather
than by its radial distribution, merely renormalizes the mass of the applied
field from $M_{\rm BH}$ to $M_{\rm ADM}$ and carries no information about the
response, so it must be subtracted before the TLN is read
off\footnote{Consider a spherical shell of mass added around the BH. The shell shifts
$M_{\rm BH}\to M_{\rm ADM}$ but, by spherical symmetry, induces no quadrupolar or higher deformation. The tidal response is therefore sourced not by the total halo mass but by its radial arrangement, through $\int m_h' r^{2l}dr$. The monopole piece is the difference of the two growing branches, $y_1(r;M_{\rm ADM})-y_1(r;M_{\rm BH})\to \frac{(l-1)(l+2)}{l}
\frac{M_{\rm halo}}{r}$ at large $r$ (equal to $2M_{\rm halo}/r$ for $l=2$), which removes the $M_{\rm halo}r_t^{2l}$ term of $\mathcal{J}_l$ upon integration by parts.}. Matching $y$ across $r_t$ gives
\begin{equation}
\delta y(r_t)-\frac{(l-1)(l+2)}{l}\frac{M_{\rm halo}}{r_t}
=-\frac{(2l+1)\,\kappa_l}{r_t^{2l+1}}\,,
\end{equation}
in which the monopole term cancels through the identity
$\tfrac{l+2}{l}M_{\rm halo}r_t^{2l}-\mathcal{J}_l=\tfrac{2}{l}\int m_h'\,r^{2l}\,dr$.
\begin{figure*}[!htbp]
\centering
\includegraphics[width=0.98\linewidth]{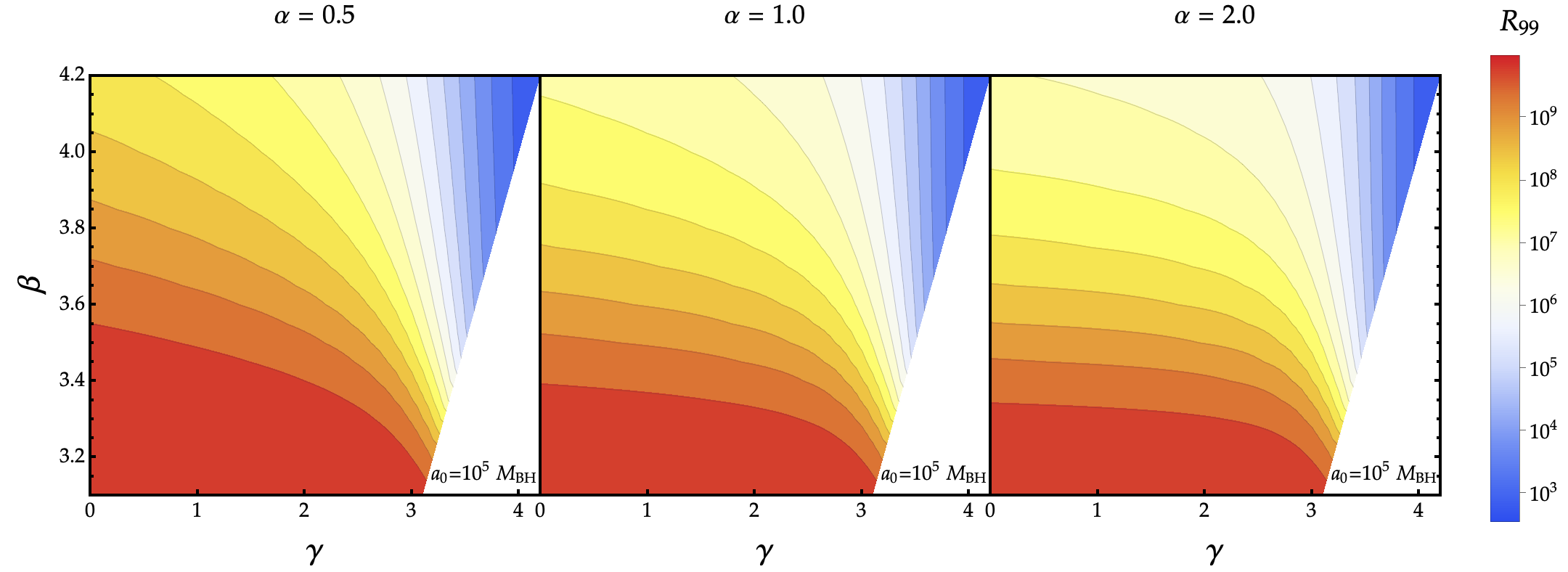}
\caption{The $99\%$ mass radius, $R_{99}$ over the $(\gamma,\beta)$ plane,
at fixed $a_0=10^{5}M_{\rm BH}$, for $\alpha=0.5,\,1,\,2$. The white wedge
is the unphysical region $\gamma\ge\beta$. $R_{99}$ is largest for shallow
profiles (small $\gamma$, $\beta\to3^+$) and smallest for steep ones,
spanning several decades across the shape parameters.}
\label{fig:R99abg}
\end{figure*}
The matching introduces two multiplicative prefactors because the interior perturbation is constructed about the Schwarzschild background with mass $M_{\rm BH}$, whereas the exterior solution is expressed in terms of the ADM mass $M_{\rm ADM}=M_{\rm BH}+M_{\rm halo}$. Each differs from unity by $O(M_{\rm halo}/r_t)\sim O(z)$. Since the induced tidal response is itself linear in the halo compactness, these prefactors modify the Love number only at $O(z^2)$ and are consistently neglected within our first-order perturbative treatment.
The response then collapses to the compact result
\begin{equation}
\kappa_l = \frac{2(l-1)}{l\,(2l+1)}\int_{4M_{\rm BH}}^{r_t} m_h'(r)\, r^{2l}\, dr \label{kappal}
\end{equation}

\begin{equation}
 = \frac{8\pi(l-1)}{l\,(2l+1)}\int_{4M_{\rm BH}}^{r_t} \bar\rho(r)\, r^{2l+2}\, dr~,
 \label{eq:kappal}
\end{equation}
which has no residual dependence on $M_{\rm BH}$. For the dominant quadrupole
$l=2$ this reduces to
\begin{align}\label{eq:kappamaster}
\kappa_2
 = \frac{1}{5}\int_{4M_{\rm BH}}^{r_t} m_h'(r)\, r^{4}\, dr
 = \frac{4\pi}{5}\int_{4M_{\rm BH}}^{r_t} \bar\rho(r)\, r^{6}\, dr\,,
\end{align}
which is exact at linear order in the halo mass on the full Schwarzschild background, the exact source cancellation of \cref{eq:TLNexact} being provably unique to $l=2$. 
\begin{figure}
\centering
\includegraphics[width=0.49\textwidth]{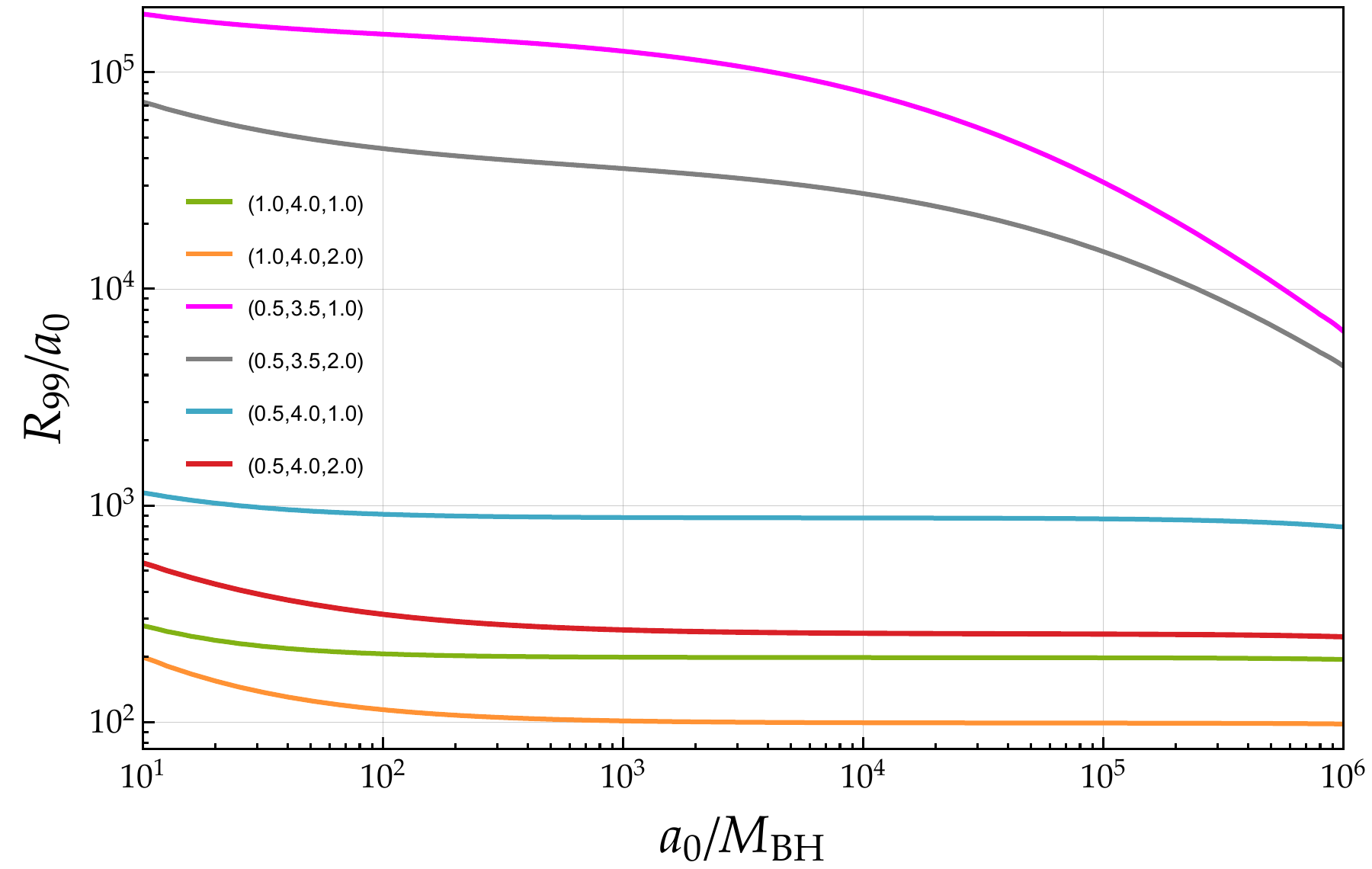}
\caption{The ratio $R_{99}/a_0$ as a function of $a_0$ for representative
convergent profiles. For $\beta=4$ the ratio is nearly independent of
$a_0$, while for $\beta=3.5$ it varies substantially, reflecting the
growing influence of the fixed inner cut-off $4M_{\rm BH}$ as the outer
logarithmic slope approaches the divergent value $\beta=3$.}
\label{fig:R99a0}
\end{figure}
Equation~\eqref{eq:kappal} vanishes identically for $l = 1$, consistent with the fact that the static axial dipole represents a shift of angular momentum rather than an induced tidal response. \footnote{It is independently confirmed by a variation-of-parameters computation~\cite{Boyce:2021ode} (see~\cite{Ghosh-etal} for details)}. 
The induced magnetic-type multipole of each mass element grows as $r^{2l}$, so the axial tidal response
is dominated by the \emph{outermost} matter in the halo, in contrast to the redshift integral $\mathcal{I}$, whose weight $(r-2M_{\rm BH})^{-1}$ favours the innermost matter. Since $m_h'\,r^{2l}\propto r^{2l+2-\beta}$, the moment in \cref{eq:kappal}
is dominated by its upper limit for every $\beta<2l+3$, and the axial TLN
of an untruncated halo is ill-defined. This is the position-space
manifestation of the logarithmic obstruction to the tidal matching
identified in Ref.~\cite{DOnofrio:2026ulh} for density profiles lacking
compact support, and it makes a finite truncation radius $r_t$ mandatory
rather than optional. We fix $r_t$ from the halo's own mass distribution.
For a convergent profile ($\beta>3$) the total halo mass is finite, and we
identify $r_t$ with the radius $R_{99}$ enclosing $99\%$ of it,
\begin{align}\label{eq:R99def}
\int_{4M_{\rm BH}}^{R_{99}} m_h'(r)\,dr = 0.99\,M_{\rm halo}
\qquad(\beta>3)\,.
\end{align}
For a divergent profile ($\beta\le3$, such as NFW) the total mass grows
without bound, $R_{99}$ is undefined, and we adopt the fixed truncation
$r_t=5a_0$ used in \cref{sec:density-profile}.
Both integrals implicit in \cref{eq:R99def} scale linearly with the density
amplitude, so $R_{99}$ is a functional of the profile shape alone,
independent of $M_{\rm halo}$ and hence of the compactness $z$. Figure~\ref{fig:R99abg}
shows $R_{99}$ across the $(\gamma,\beta)$ plane for three values of
$\alpha$: it grows as either logarithmic slope flattens, since matter is
then distributed to larger radii, and shrinks as the profile steepens.
Figure~\ref{fig:R99a0} shows the complementary dependence on $a_0$. For steep
outer logarithmic slopes $R_{99}/a_0$ is nearly constant, however, as $\beta$ approaches 3, it
acquires a pronounced $a_0$ dependence, controlled by the fixed inner cut-off $4M_{\rm BH}$ relative to the scale radius.
We render the induced multipole dimensionless following the relativistic convention of~\cite{Binnington:2009bb,DOnofrio:2026ulh}, normalizing
by the total mass $M_{\rm ADM}=M_{\rm BH}+M_{\rm halo}$,
\begin{align}\label{eq:k2def}
{k}_l^{B}\equiv-\frac{l+2}{l-1}\,\frac{\kappa_l}{M_{\rm ADM}^{2l+1}}\,,
\qquad
k_2^{B}=-\frac{4\,\kappa_2}{M_{\rm ADM}^{5}}\,,
\end{align}
which coincides with the definition of Ref.~\cite{DOnofrio:2026ulh,Chakraborty:2026qru,Cardoso:2021wlq}. Since $\kappa_l>0$, the matter-induced axial
Love number is negative. 

 A word of caution about the magnitude of $k_l^B$ is in order. The normalisation by $M_{\rm ADM}^{2l+1}$ follows the standard relativistic convention, but for a BH dressed by an extended halo it carries the geometric factor $(r_t/M_{\rm ADM})^{2l+1}$. Here $r_t$ exceeds $M_{\rm ADM}$ by several orders of magnitude, whereas for a self-gravitating body such as a neutron star the same ratio is only
a few, and this difference accounts entirely for the very large values of $|k_2^B|$ reported below. They should not be read as indicating an unusually soft or easily deformed object: the physical response $\kappa_2$ is entirely ordinary and, as shown below, strictly linear in the halo mass.

In the single-power-law limit, $\gamma\to\beta^-$
\cref{eq:kappal} evaluates in closed form,
\begin{widetext}
\begin{align}\label{eq:Qlpowerlaw}
\kappa_l=&\frac{2 (l-1)}{l(2l+1)}\,\frac{M_{\rm halo}\displaystyle\int_{4 M_{\rm BH}}^{r_t} r^{2l+1-\beta}\lt(r-4 M_{\rm BH}\rt)\,dr}{\displaystyle\int_{4 M_{\rm BH}}^{r_t} r^{1-\beta}\lt(r-4 M_{\rm BH}\rt)\,dr}\nonumber\\
=&\frac{2 (l-1)}{l(2l+1)}\,\frac{M_{\rm halo}(2-\beta)(3-\beta)}{(2l+2-\beta)(2l+3-\beta)}\,
\frac{(2l+2-\beta)\,r_t^{\,2l+3-\beta}
      -(2l+3-\beta)\,4 M_{\rm BH}\,r_t^{\,2l+2-\beta}
      +\lt(4 M_{\rm BH}\rt)^{\,2l+3-\beta}}
     {(2-\beta)\,r_t^{\,3-\beta}
      -(3-\beta)\,4 M_{\rm BH}\,r_t^{\,2-\beta}
      +\lt(4 M_{\rm BH}\rt)^{\,3-\beta}}\,,
\end{align}
\end{widetext}
depending on the halo only through $M_{\rm halo}$ and $r_t$. The scale radius $a_0$ drops out entirely, as it must: a pure power law fixes only the combination $\bar\rho_0 a_0^{\beta}$. Equation~\eqref{eq:Qlpowerlaw} further solidifies the choice of $M_{\rm ADM}$ as the normalizing length scale and provides an analytic anchor for the shape trends
discussed below.

\subsection{Dependence on compactness and halo parameters}
\label{sec:TLNparams}
\begin{figure}[!htbp]
\centering
\includegraphics[width=0.99\linewidth]{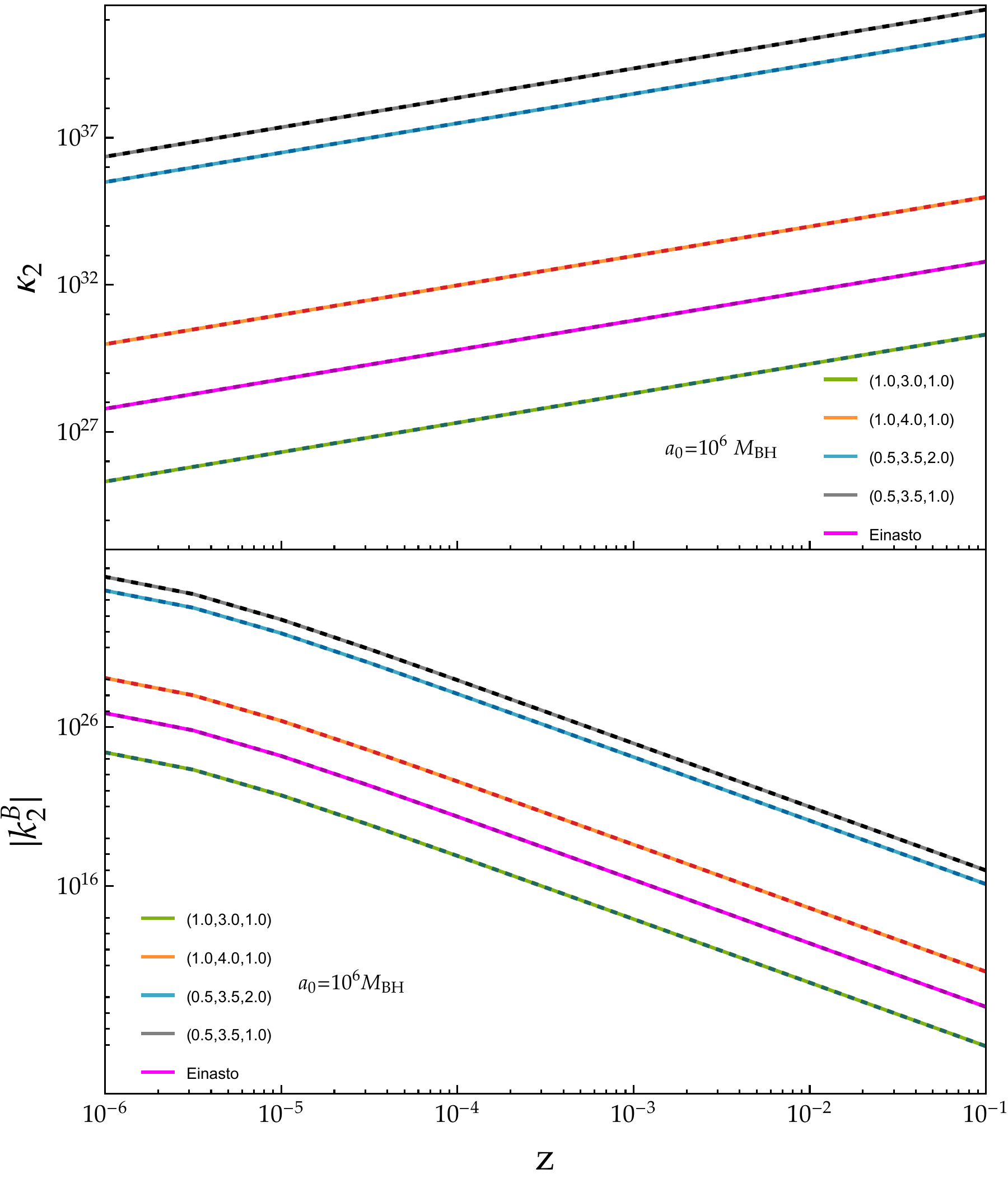}
\caption{The dimension-full response $\kappa_2$ (top) and the dimensionless Love number $\lt|k_{2}^B\rt|$(bottom) as functions of compactness $z$, at $a_0=10^{6}M_{\rm BH}$, for several halo profiles ($\beta>3$ truncated at $R_{99}$, $\beta=3$ at $5a_0$). Dashed curves are the numerical extraction,
solid curves the analytic result \cref{eq:kappal}. $\kappa_2$ grows linearly with $z$, whereas $\lt|k_2^{B}\rt|$ decreases with $z$ in accordance with \cref{eq:k2Bz}.}
\label{fig:kapk2Bz}
\end{figure}
At fixed shape and scale radius, the truncation radius $R_{99}$ (or $5a_0$) is fixed, so the response is strictly linear in the halo mass, $\kappa_l\propto M_{\rm halo}\propto z$, as shown in the upper panel of
\cref{fig:kapk2Bz}. This linear growth does not contradict the familiar decrease of tidal Love numbers with increasing compactness for self-gravitating bodies. Since the Schwarzschild BH has a vanishing axial TLN, the response originates entirely from the surrounding halo, and $z$ measures the amount of responding matter rather than the compactness of a fixed body.
The dimensionless quantity $k_2^{B}$ exhibits the opposite trend because its normalization depends on the total gravitating mass. Since
\begin{align}\label{eq:k2Bz}
k_2^{B}\propto
\frac{M_{\rm halo}}
{\lt(M_{\rm BH}+M_{\rm halo}\rt)^5},
\end{align}
it decreases monotonically with $z$ over the astrophysically relevant regime $M_{\rm halo}\gtrsim M_{\rm BH}$, as shown in the lower panel of \cref{fig:kapk2Bz}. Unlike the physical response $\kappa_2$, this behaviour is entirely a consequence of the conventional normalization by $M_{\rm ADM}^{2l+1}$ and is therefore consistent with the familiar trend that dimensionless TLNs decrease as the gravitating mass increases. 

\begin{figure*}[htbp!]
\centering
\includegraphics[width=0.49\linewidth]{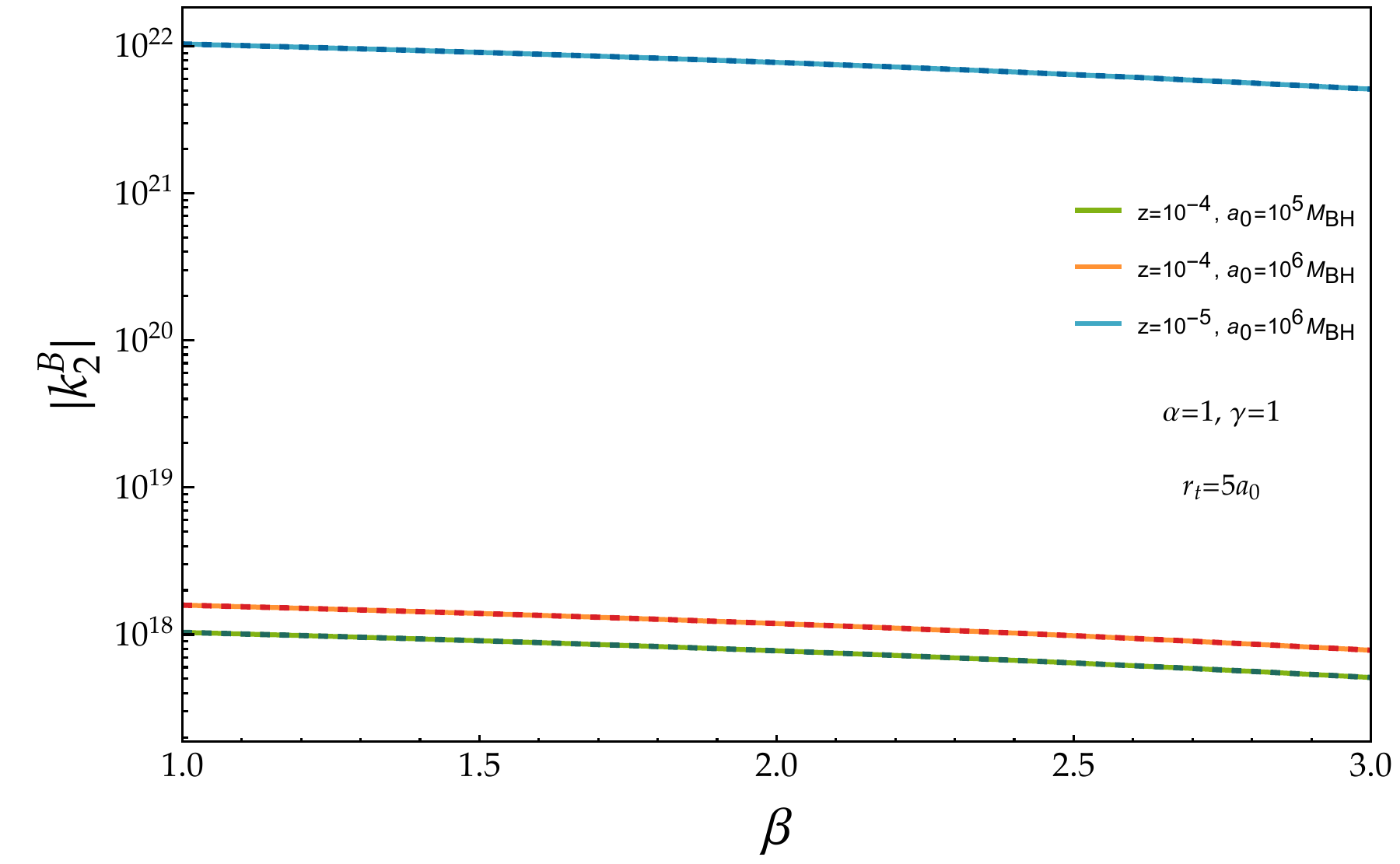}
\includegraphics[width=0.49\linewidth]{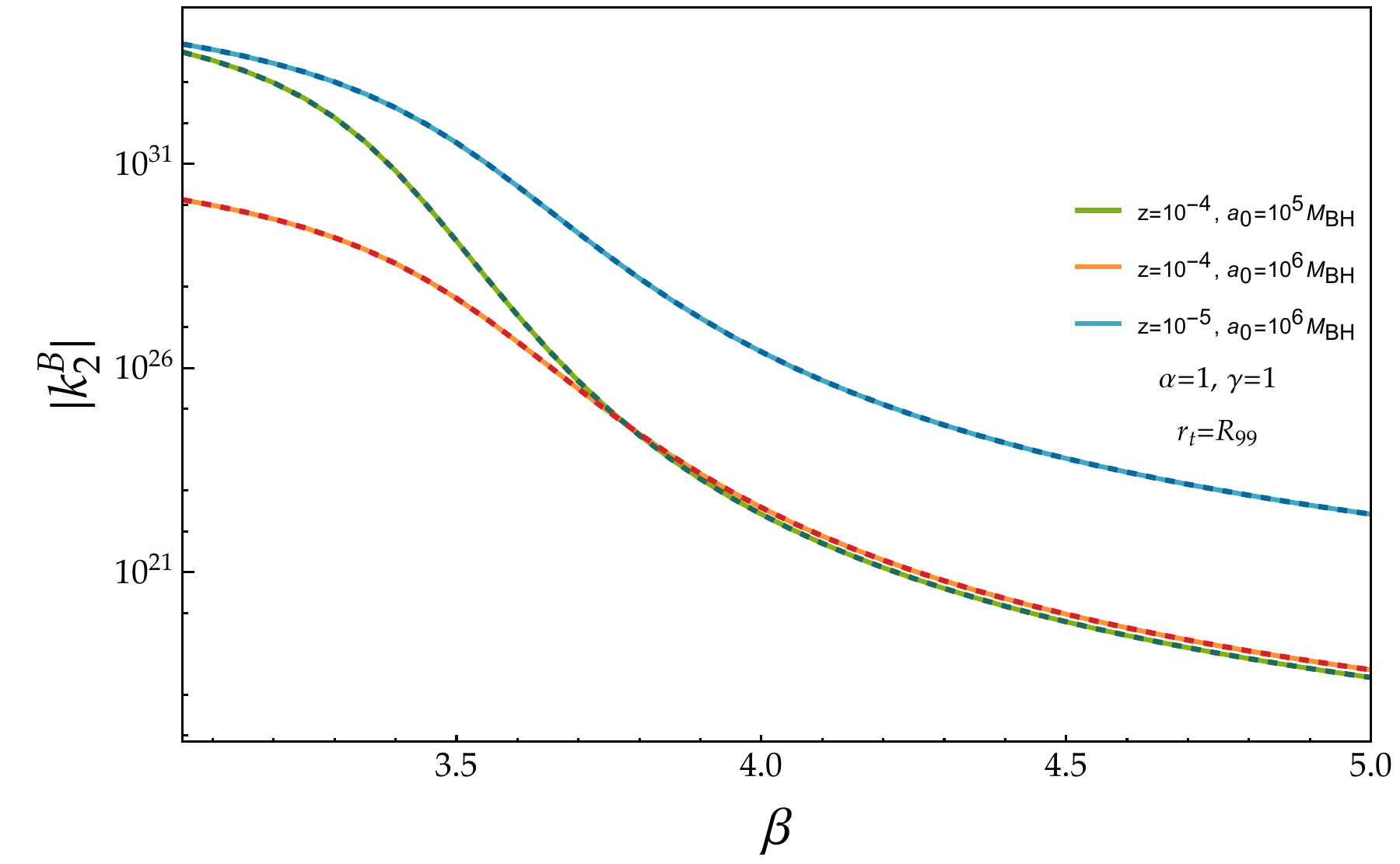}
\caption{ Variation of $|k_2^{B}|$ with the outer logarithmic slope $\beta$, at fixed $\alpha=\gamma=1$.The left panel shows the variation for the mass divergent profiles ($\beta<3$) with fixed truncation radius at $r_t=5a_0$. In contrast, the right panel plots the same with $\beta>3$, and truncation radius $r_t=R_{99}$. Dashed curves. in both panels, represent numerically obtained values of $\lt|k_2^B\rt|$, whereas, solid curves represent analytic results for different $(z,a_0)$ combinations.}
\label{fig:k2Bb}
\end{figure*}
\begin{figure*}
\centering
\includegraphics[width=0.495\linewidth]{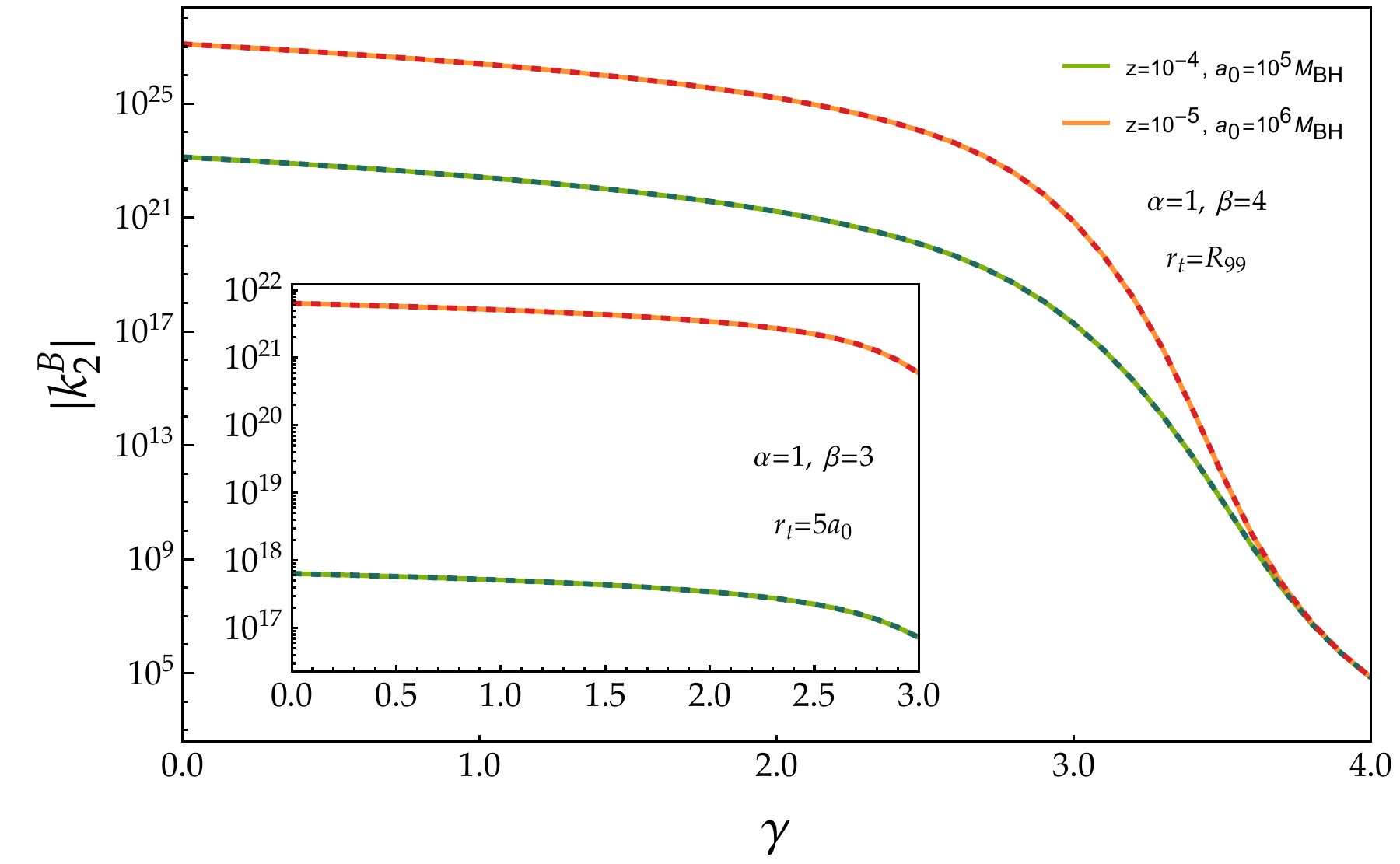}
\includegraphics[width=0.495\linewidth]{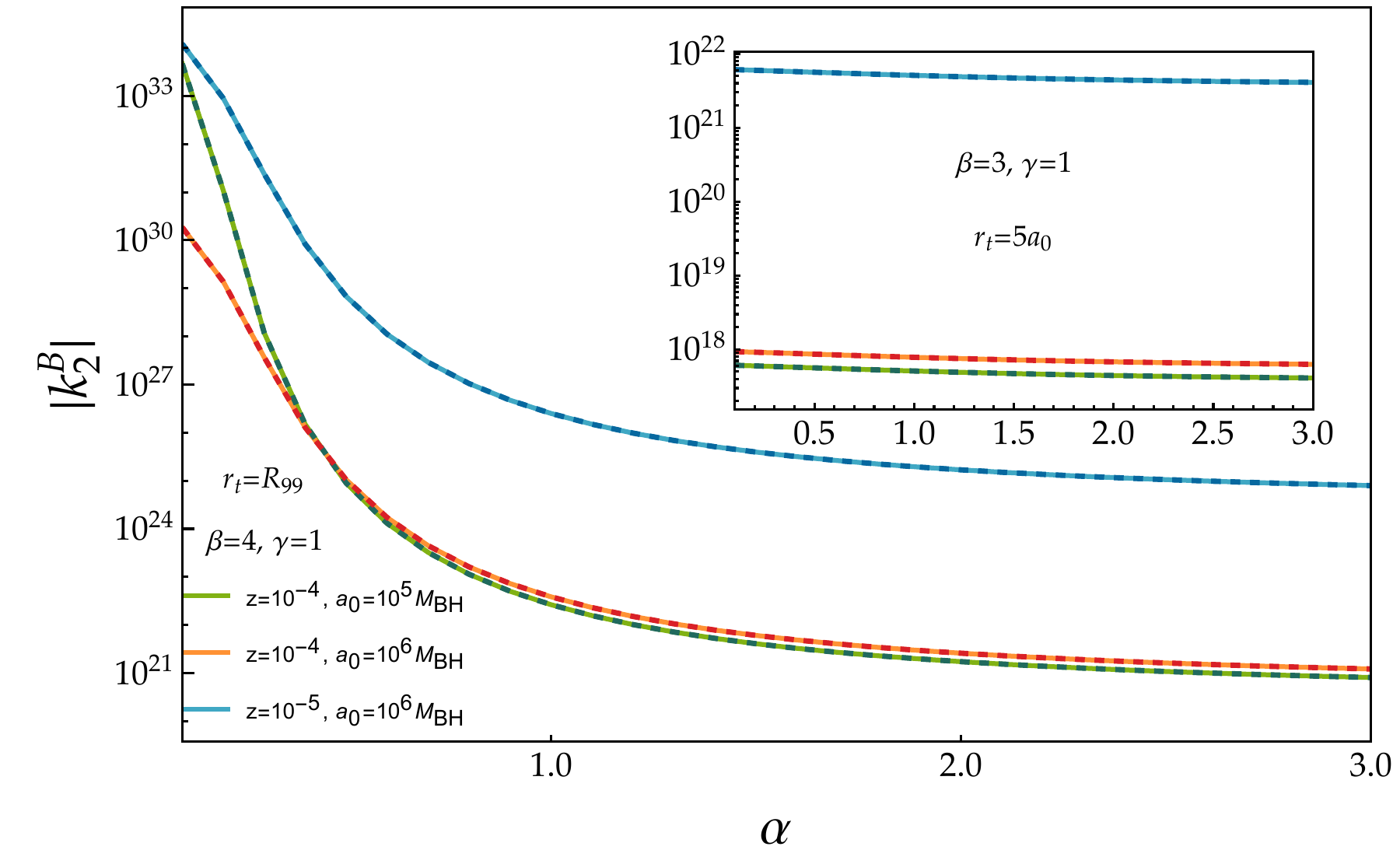}
\caption{Variation of $|k_2^{B}|$ with the inner logarithmic slope $\gamma$ (left) and the transition sharpness $\alpha$ (right). In the left panel, $\alpha=1$ and $\beta=4$, while in the right panel $\beta=4$ and $\gamma=1$. In both panels, the truncation radius is chosen as $r_t=R_{99}$ for different combinations of $a_0$ and $z$. The insets show the corresponding results for $\beta=3$ with a fixed truncation radius $r_t=5a_0$. Dashed curves denote the numerical results, while solid curves show the analytic predictions.}
\label{fig:k2Bga}
\end{figure*}
Figure~\ref{fig:k2Bb} shows the variation of $\lt|k_2^B\rt|$ with the outer slope parameter $\beta$, in the two separate regimes of the outer same. The split at $\beta=3$ is where the halo mass ceases to converge, so the truncation radius is defined differently on either side. In the left panel $\beta<3$ and $r_t=5a_0$ is imposed externally. The extent is then common to all curves, and the panel isolates the effect of redistributing matter within a fixed radius. That effect is weak: $|k_2^{B}|$ falls by a factor of almost $1.9$ across the $\beta$ -range. The weak dependence is specific to fixed truncation radius, increasing the fixed truncation radius only mildly increases the contrast ($\sim 10.95$ for $r_t=10^4 a_0$). 

In the right panel the outer logarithmic slope is steep enough that the halo mass converges, so the truncation radius is set by the halo itself as $R_{99}$, the radius enclosing $99\%$ of the mass. The tidal response is governed by this radius and grows approximately as its fourth power, as can be seen from ~\cref{kappal}, (see also \cite{Cardoso:2019upw}). Therefore, the steep fall of $\lt|k_2^{B}\rt|$ across the panel simply follows the shrinking of $R_{99}$ as $\beta$ increases in ~\cref{fig:R99abg}. The dependence is far from gentle. As $\beta$ approaches 3 the outer density decays so slowly that an appreciable fraction of the mass sits at ever larger radii, and $R_{99}$ must expand rapidly to contain $99\%$ of it. Because $R_{99}$ diverges as the halo mass stops converging at $\beta=3$, the response rises sharply toward $\beta \rightarrow 3^+$. For $\beta>3$, and fixed compactness as $M_{\rm halo}$ is increased the dimensionless response balances two competing effects of this extra mass, a suppression through the larger normalizing mass $M_{\rm ADM}$ and an enhancement through the larger truncation radius. For steep outer logarithmic slopes the truncation radius scales with $a_0$, so the enhancement wins. However, as $\beta\rightarrow3^+$ the slow convergence of the mass causes the enhancement to disappear and only the suppression remains, enhancing the response of the lighter halo. The curves intersect where the two effects balance. This crossing is a consequence of using $r_t=R_{99}$.  With a truncation radius tied to the halo's own scale the two curves remain parallel.

Steepening the inner logarithmic slope $\gamma$ transfers mass toward the inner region, so $\lt|k_2^{B}\rt|$ decreases with $\gamma$ as well (right panel of \cref{fig:k2Bga}). For $\beta>3$ the decrease of $\lt|k_2^B\rt|$ with $\gamma$ is further aided by the contraction of $R_{99}$ (see \cref{fig:R99abg}). As $\gamma$ approaches $\beta$ the $\kappa_2$ become independent of $a_0$ (see \cref{eq:Qlpowerlaw}), this is coupled to the slow near independence of $R_{99}$ on $a_0$ results in the converging value of $\lt|k_2^B\rt|$. However for constant truncation radius $r_t=5a_0$, the $a_0$ dependence persists even as $\gamma$ approaches $\beta$ ~\footnote{Note that halo distribution with larger mass and steeper inner logarithmic slope results in a the appearance of additional light ring pairs (see~\cref{sec: BG-density}).}. 

The parameter $\alpha$ controls how sharply the profile turns from its inner logarithmic slope $\gamma$ to its outer logarithmic slope $\beta$ near $a_0$, leaving both slope parameters unchanged. The right panel of~\cref{fig:k2Bga}, shows the variation of $\lt|k_2^B\rt|$ with $\alpha$. In the main panel the response drops steeply with $\alpha$. A broad transition (small $\alpha$) keeps the density shallow well past $a_0$, so the halo is extended and $R_{99}$ large, whereas a sharp transition confines the mass near $a_0$ and shrinks $R_{99}$ (see \cref{fig:R99abg}). Since the response scales as $R_{99}^{4}$, this alone produces the decline, which steepens at small $\alpha$ and flattens once the transition is already sharp. The inset, with the truncation fixed at $5a_0$, shows that the effect of  $\alpha$ on $|k_2^{B}|$ is small. Thus, the effect of $\alpha$ on $\lt|k_2^B\rt|$ in the main panel arises dominantly from the  change in $R_{99}$.

The three curves share a common shape and differ only through the prefactor set by $(z,a_0)$. The two curves at equal $z$ cross once, near $\alpha\simeq0.45$.
A small $\alpha$ leaves the halo very extended, so $R_{99}$ grows until it
approaches the fixed outer radius used to normalise the total mass. The
larger-$a_0$ halo reaches that limit first. It then loses the $R_{99}^{4}$ enhancement while keeping the penalty of its larger normalising mass, and so falls below the lighter curve. The crossing is therefore set by the fixed cutoff rather than by $\alpha$ itself. Unlike $\beta$, $\alpha$ does not change how fast the halo mass converges, so with a
cutoff scaled to $a_0$ the curves would remain parallel.
\section{Complementarity of axial TLN and QNM}\label{sec:complemenarity}
\begin{figure}
    \centering
    \includegraphics[width=\linewidth]{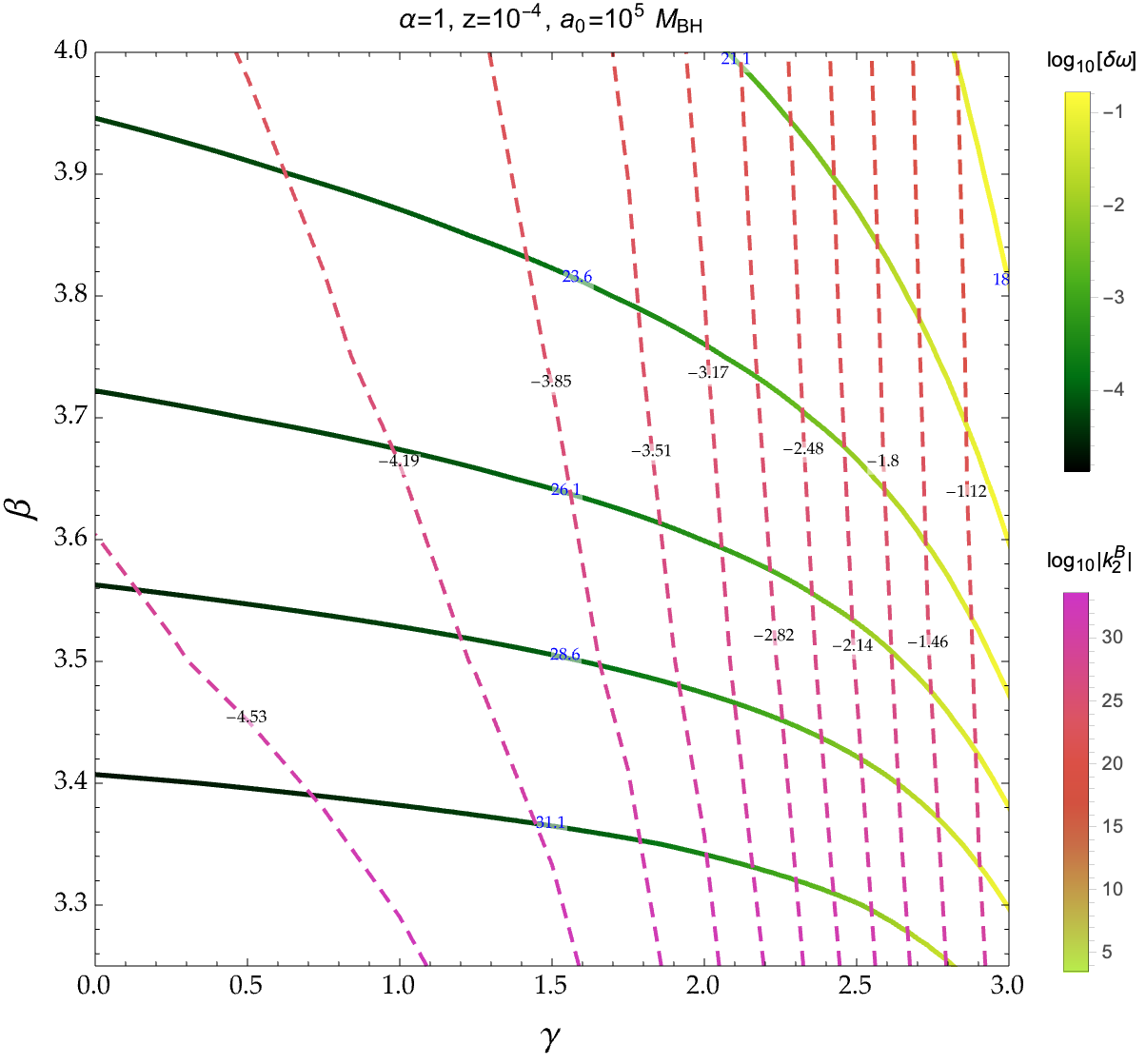}
    \caption{Representative contour map illustrating the complementary dependence of the axial tidal Love number and the QNM redshift on the halo profile. The plot shows the $(\gamma,\beta)$ plane for fixed $\alpha=1$, $z=10^{-4}$, and $a_0=10^5M_{\rm BH}$. Solid curves are contours of constant $\log_{10}|k_2^{B}|$, while dashed curves are contours of constant $\log_{10}\delta\omega$. Each contour family is coloured by the complementary observable: solid contours by $\log_{10}\delta\omega$ and dashed contours by $\log_{10}|k_2^{B}|$. 
    }
    \label{fig:qnm-tln}
\end{figure}
The axial TLN exhibits the opposite dependence to the QNM shift $\delta\omega$, which increases with increasing inner halo density and steeper inner density profiles (\cref{sec:QNMhalo}). This complementarity arises from the distinct radial weighting of the two observables: the axial TLN is primarily sensitive to the outer halo density, whereas the QNM shift is primarily sensitive to the inner halo density. The two observables therefore probe complementary regions of the same halo distribution. \footnote{An independent confirmation of this behaviour for an alternative halo model will be presented in Ref.~\cite{Ghosh-etal}}.

Figure~\ref{fig:qnm-tln} illustrates this complementarity on a representative $(\gamma,\beta)$ slice at fixed $(\alpha,z,a_0)$. Contours of constant $|k_2^{B}|$ and $\delta\omega$ intersect over most of the parameter space and are coloured by the complementary observable. The variation of the color along each contour demonstrates that models degenerate with respect to one observable generally predict distinct values of the other. Consequently, a joint measurement of $(|k_2^{B}|,\delta\omega)$ can substantially reduce the parameter degeneracies associated with either observable alone.
\section{Summary and Discussion}\label{sec: disc}
Astrophysical BHs are expected to reside in complex environments rather than in vacuum. From stellar-mass remnants embedded in gas-rich disks to supermassive BHs at galactic centres, the surrounding matter modifies the spacetime and can therefore influence GW observables~\cite{Ferrarese:2000se,Gebhardt:2000fk,Kormendy:2013dxa,Gondolo:1999ef,Sadeghian:2013laa,Chowdhury:2026cjv}. As ringdown measurements become increasingly precise, understanding these environmental effects is essential both for interpreting observations and for using GWs as probes of the strong-field environment. The central question addressed in this work is whether GW observables are sensitive only to the total amount of surrounding matter, or whether they also encode information about the \emph{shape} of the DM distribution. This distinction is crucial because different halo profiles can produce similar shifts in a single observable, leading to degeneracies between the halo compactness, density profile, and possible deviations from the vacuum spacetime.

In this work we investigated the axial perturbations of a Schwarzschild BH surrounded by a generalized relativistic $(\alpha,\beta,\gamma)$ DM halo (\cref{fig:density}) described within the Einstein-cluster formalism~\cite{Einstein:1939ms,Geralico:2012jt,Cardoso:2021wlq}. For this spacetime we derived the axial master equation, computed the quasinormal spectrum using sixth-order WKB and verified the results through time-domain evolution. We further derived the static axial tidal Love numbers analytically to leading order in the halo compactness and confirmed the analytic expressions by solving the full perturbation equations numerically.

Our analysis shows that the ringdown spectrum depends not only on the halo compactness but also on the detailed shape of the density profile . Increasing the halo compactness systematically redshifts both the oscillation frequency and the damping rate (\cref{fig:zplotnew}), while the profile parameters produce additional corrections that become particularly significant for centrally concentrated halos. Among the shape parameters, the inner logarithmic slope $\gamma$ has the strongest influence because the QNM redshift is governed by the halo redshift integral, whose kernel is concentrated near the inner cut-off. This behaviour is reflected consistently in the light ring frequency, the principal Lyapunov exponent, the WKB spectrum, and the time-domain evolution. The time-domain analysis also reproduces the expected Price-law decay at late times together with the intermediate tail determined by the outer density profile for slowly decaying halos (\cref{fig:TD}).

The static tidal response behaves differently. The axial Love number vanishes identically for an isolated Schwarzschild BH. The surrounding halo renders it finite. To leading order, it is a high-order radial moment of the halo mass, weighted towards the outer halo. Its direct sensitivity is therefore to the outer logarithmic slope $\beta$ and to the truncation radius, not to the inner cusp (\cref{fig:k2Bb,fig:k2Bga}). At fixed compactness its magnitude scales as the fourth power of the truncation radius $r_t$. Crucially, for halos that results in a convergent integrated mass the radius $R_{99}$ enclosing 99\% of the halo mass, contracts as \emph{either} logarithmic slope steepens (\cref{fig:R99abg}). The Love number therefore falls steeply with both $\beta$ and $\gamma$. The inner logarithmic slope acts only indirectly, by fixing the extent of the responding matter, not through the moment itself. Both trends oppose that of the ringdown redshift. The opposite dependence of the tidal response and the ringdown shift on the halo parameters follows directly from the distinct radial kernels governing the two observables.

The principal result of this work is that the ringdown and the static tidal response probe complementary regions of the same matter distribution. The QNM shift is governed by the redshift integral, which receives its dominant contribution from the vicinity of the inner cut-off, whereas the tidal Love number depends on an outer radial moment that is dominated by the extended halo. Consequently, halo configurations that are nearly degenerate in the ringdown generally predict different tidal deformabilities, and vice versa (\cref{fig:qnm-tln}). More generally, our results demonstrate that environmental effects cannot be characterized solely by the total halo mass or compactness; different GW observables respond to different moments of the surrounding matter distribution. Interpreting environmental effects therefore requires multiple observables rather than a single ringdown measurement.

To assess the potential observational relevance of these effects, we compared the predicted QNM shifts with the representative precision inferred from GW250114~\cite{LIGOScientific:2025wao,LIGOScientific:2025rid}. For halo compactness $z=10^{-4}$, a range of centrally concentrated profiles produces frequency shifts of the same order as the current observational precision, whereas for $z=10^{-6}$ only the most concentrated halos remain potentially distinguishable (\cref{fig:abgc4,fig:abgc6}). This comparison is intended only as an indication of the size of the effect rather than as an observational constraint. The GW250114 analysis assumes a Kerr remnant formed in a stellar-mass binary, whereas the present calculations describe a non-rotating BH embedded in a static DM environment. A quantitative constraint will require waveform models that consistently incorporate environmental effects throughout the inspiral, merger, and ringdown.

 The present analysis relies on several key assumptions. The tidal response computed here rests on a choice of closure for the axial fluid velocity. Imposing irrotationality, $\omega^{\mu}=0$, fixes the velocity through \cref{eq:Ulm} and gives a configuration with $\delta u_{\mu}=0$, although the fluid itself moves. The perturbed auxiliary vector $\delta w_{\mu}$, and with it the function $W(t,r)$, is not an independent choice: it follows from the conservation of the perturbed energy--momentum tensor, \cref{eq:cons}. In the static sector, since $h_1=0$, the closure affects only through the term $4f(r) U$ in \cref{eq:rphi}. Since that term is linear in $U$, a different closure changes the induced multipole only by an overall factor in the departure from irrotationality and not suppressed by the halo compactness. The magnetic-type Love numbers of matter-dressed BHs therefore depend on the closure adopted (see, for e.g. \cite{Chakraborty:2026qru}, where a comoving closure $\delta u^{\mu}=0$, has been used to study the static axial TLNs of a Schwarzschild-Hernquist black hole). The radial weighting, on the other hand, remains unchanged, since the departure enters through the same moment of the halo density that determines the response (\cref{eq:kappal}) itself. The complementarity with the ringdown redshift is thus robust against the choice of closure, and only the overall normalisation is convention dependent.

We note that neither of the two dominant mechanisms, that could drive the closure away from irrotationality, namely, viscosity and frame dragging operates here. Viscosity does not arise in the present model at all: the Einstein-cluster construction describes collisionless particles moving on geodesics, so there is no viscous stress to include, and a collisional environment such as baryonic gas would require a different matter model and with it a different closure. Frame dragging is already contained in the treatment, since $h_0$ is itself a dragging potential and \cref{eq:Ulm} carries the fluid along with it at the rate for which $\delta u_\mu$ vanishes. Dragging by the background is a separate matter and requires BH spin, which the present static geometry does not have.
 
 The truncation radius is treated as a parameter of the finite-halo model and provides the natural normalization scale for the tidal response, with its influence becoming increasingly important as $\beta\rightarrow3$. We further restrict the halo parameter space to configurations that do not admit additional light rings, thereby preserving the single-barrier potential structure underlying the WKB calculation of the QNM spectrum. Correlations among the halo parameters $(\alpha,\beta,\gamma,z,r_t)$ may also introduce residual degeneracies in observational inference. Finally, we restrict attention to axial perturbations of a non-spinning BH embedded in an Einstein-cluster halo with vanishing radial pressure. Extending the present framework to the polar sector, rotating BHs, and more general phase-space descriptions of the surrounding matter would broaden its applicability. 

Looking ahead, the most promising observational target is likely to be an extreme-mass-ratio inspiral around a massive BH embedded in a dense DM environment~\citep{Duque:2023seg}. Such systems naturally carry both observables considered here: the tidal response is encoded in the inspiral phasing, while the halo-induced QNM redshift appears in the subsequent ringdown. A direct observational implementation of this complementarity requires extending the present analysis to finite-radius perturbers, which we leave for future work. Combined with independent probes such as galactic rotation curves, which constrain the halo on much larger scales, these complementary GW signatures offer a promising route towards reconstructing the DM distribution in the immediate vicinity of BHs.
\acknowledgements
The authors thank Vitor Cardoso for useful suggestions and comments. The authors also thank Sovan Chakraborty for insightful discussions and for providing computational resources. AC thanks Nils Andersson, Sanved Kolekar, and Sumanta Chakraborty for valuable discussions. AC also thanks Rajes Ghosh and Chiranjeeb Singha for useful comments. The work of AC was partly supported by the National Postdoctoral Fellowship of the Anusandhan National Research Foundation (ANRF), Govt. of India (File No.: PDF/2023/000550) at IIT Guwahati.  AC is grateful to the Raman Research Institute, Bangalore, for the hospitality extended during the conference {\it Symphony of Spacetime} where part of this work was carried out. SC acknowledges support of MATRICS research grant awarded by Science and Engineering Research Board (SERB) and ANRF, Govt. of India through grant no. MTR/2022/000318. 
\appendix
\section{Analytic basis for the light ring results}
\label{App:eikonal}

We derive the results quoted in Sec.~\ref{sec:lightring} and provide
the analytic basis for the trends seen in
\cref{fig:c4c6g,fig:c4c6b,fig:c4c6a}.  

\subsection{Redshift function at the light ring}
\label{app:metric}

The DM density vanishes for $r \leq 4M_{\rm BH}$.  The mass
function therefore satisfies $m(r_c) = M_{\rm BH}$,
$\bar{\rho}(r_c) = 0$, and $m'(r_c) = 0$ at the light ring
$r_c = 3M_{\rm BH}$.  The local geometry at $r_c$ is identical to the
Schwarzschild case.  The halo affects only the global value of
$f(r_c)$.

Integrating \cref{eq:fprime} outward from $r_c$ and imposing
asymptotic flatness gives
\begin{align}
   \ln f(r_c) = -\int_{r_c}^{\infty}
     \frac{2m(r)}{r\bigl[r-2m(r)\bigr]}\,dr.
   \label{app:fexact}
\end{align}
We write $m = M_{\rm BH} + m_h$ and use the relation
\begin{align}
   \frac{2m}{r(r-2m)} - \frac{2M_{\rm BH}}{r(r-2M_{\rm BH})}
   = \frac{2m_h}{(r-2m)(r-2M_{\rm BH})}.
   \label{app:identity}
\end{align}
Subtracting the vacuum integral gives
$f(r_c) = \frac{1}{3}\,e^{-\mathcal{I}}$, which is \cref{eq:frc}, where $\mathcal{I}$ is the halo integral defined in \cref{eq:haloint},
\begin{align}\label{app:haloint}
   \mathcal{I} \equiv \int_{4M_{\rm BH}}^{\infty}
     \frac{2\,m_h(r)}{[r-2m(r)]\,[r-2M_{\rm BH}]}\,dr 
\end{align}
An exterior spacetime requires $r > 2m(r)$ (note that the absence of light ring the matter region invokes a stricter condition $r>3 m(r)$)
and $m_h \geq 0$, so the integrand is non-negative and
$\mathcal{I} > 0$ for any non-vanishing halo.  Equation
\cref{eq:Omega_exact} then follows from
$\Omega_c = \sqrt{f(r_c)}/r_c$ with $r_c = 3M_{\rm BH}$.

\subsection{Universal identity $\lambda_c = \Omega_c$}
\label{app:lyap}

We now derive the identity quoted as \cref{eq:lambdaOmega}.
We define $V_l(r) = f(r)/r^2$ and $g(r) = f(r)(1-2m/r)$, so
$dr_*/dr = g^{-1/2}$.  Since $V_l'(r_c) = 0$,
\begin{align}
   \left.\frac{d^2 V_l}{dr_*^2}\right|_{r_c}
   = g(r_c)\,V_l''(r_c).
\end{align}
At $r_c = 3M_{\rm BH}$, using $m(r_c) = M_{\rm BH}$, $m'(r_c) = 0$,
and $f(r_c) = \frac{1}{3}e^{-\mathcal{I}}$,
\begin{align}
   g(r_c) = f(r_c)\!\left(1 - \frac{2M_{\rm BH}}{r_c}\right)
           = \frac{f(r_c)}{3}.
   \label{app:grc}
\end{align}
Differentiating $f'/f = 2m/[r(r-2m)]$ at $r_c$ with $m = M_{\rm BH}$
and $m' = 0$ gives
\begin{align}
   \frac{f'(r_c)}{f(r_c)} &= \frac{2}{3M_{\rm BH}},
   \label{app:fpf}\\[4pt]
   \frac{f''(r_c)}{f(r_c)}
   &= \left(\frac{f'}{f}\right)^{\!2}\!\bigg|_{r_c}
     + \left.\frac{d}{dr}\frac{2M_{\rm BH}}{r(r-2M_{\rm BH})}\right|_{r_c}
   = -\frac{4}{9M_{\rm BH}^2}.
   \label{app:fppf}
\end{align}
The quotient rule for $V_l = f/r^2$ then gives
\begin{align}
   V_l''(r_c)
   &= \frac{f(r_c)}{r_c^2}
     \!\left[-\frac{4}{9M_{\rm BH}^2}
            - \frac{8}{9M_{\rm BH}^2}
            + \frac{6}{9M_{\rm BH}^2}\right]\nonumber\\
   &= -\frac{2f(r_c)}{3r_c^2 M_{\rm BH}^2} < 0,
   \label{app:Vpp}
\end{align}
consistent with a potential maximum.  Inserting into
\cref{eq:lyapunov} gives
\begin{align}
   \lambda_c^2
   = -\frac{1}{2}\frac{r_c^2}{f(r_c)}\,g(r_c)\,V_l''(r_c)
   = \frac{f(r_c)}{r_c^2}
   = \Omega_c^2.
   \label{app:lyap2}
\end{align}
Hence $\lambda_c = \Omega_c$ exactly.  The key point is that $f(r_c)$
cancels in $\lambda_c^2/\Omega_c^2$.  The local derivatives of $f$ at
$r_c$ depend only on $m(r_c) = M_{\rm BH}$ and $m'(r_c) = 0$, which
take their Schwarzschild values for any halo profile.  The halo
changes only the overall scale of $f(r_c)$ through $\mathcal{I}$,
and both $\Omega_c$ and $\lambda_c$ pick up the same factor
$e^{-\mathcal{I}/2}$.  This confirms
\cref{eq:lambdaOmega} and \cref{eq:deltasEqual} of the main text.

Setting $M_{\rm halo} = 0$ gives $m_h = 0$ and hence
$\mathcal{I} = 0$.  All results reduce to their Schwarzschild
counterparts,
\begin{align}
   \Omega_c^{\rm Sch} = \lambda_c^{\rm Sch}
   = \frac{1}{3\sqrt{3}\,M_{\rm BH}},
   \quad
   \omega_{l n}^{\rm Sch}
   = \frac{l - i(n + \frac{1}{2})}{3\sqrt{3}\,M_{\rm BH}},
   \label{app:Schlimit}
\end{align}
confirming internal consistency.
\subsection{Profile dependence of the halo integral}
\label{app:Idep}

We now make precise the mechanisms described qualitatively in
Sec.~\ref{sec:lightring}.  

\subsubsection{Inner logarithmic slope $\gamma$}
\label{app:gamma_dep}

\paragraph*{Near the inner cut-off.}  Setting $s \equiv r - 4M_{\rm BH}$ and
expanding  the right hand side of \cref{eq:mprime} for  $s \ll M_{\rm BH}$:
\begin{align}
  4 \pi r^{2-\gamma}\!\left(1 - \frac{4M_{\rm BH}}{r}\right)
   = 4 \pi(4M_{\rm BH})^{1-\gamma}\,s + O(s^2/M_{\rm BH}).
\end{align}
The constant term cancels, so $m_h$ vanishes quadratically, 
\begin{align}
   m_h(r) = 2\pi\bar{\rho}_0\,a_0^\gamma\,
             (4M_{\rm BH})^{1-\gamma}
             (r - 4M_{\rm BH})^2\nonumber\\
             + O\!\left[(r-4M_{\rm BH})^3\right],
   \label{app:mhCutoff}
\end{align}
for any $\gamma$.  The integrand of $\mathcal{I}$ therefore vanishes
at the lower limit for all $\gamma \geq 0$, so $\mathcal{I}$ is
finite.

\paragraph*{Power-law regime $4M_{\rm BH} \ll r \ll a_0$.}  Here the
cut-off and outer transition factors are both negligible.  Direct
integration of $\bar\rho \approx \bar\rho_0(r/a_0)^{-\gamma}$ gives,
for $\gamma \neq 3$,
\begin{align}
   m_h(r) \approx
   \frac{4\pi\bar{\rho}_0\,a_0^\gamma}{3-\gamma}
   \left[r^{3-\gamma} - (4M_{\rm BH})^{3-\gamma}\right].
   \label{app:mhUnified}
\end{align}
Both limits are finite.  The sign of $3 - \gamma$ sets the
behaviour.  For $\gamma < 3$, the exponent $3 - \gamma > 0$ and
$m_h \sim r^{3-\gamma}$ grows from zero.  The enclosed fraction
\begin{align}
   \frac{m_h(r)}{M_{\rm halo}} \sim
   \left(\frac{r}{a_0}\right)^{3-\gamma} \ll 1
   \quad (r \ll a_0)
   \label{app:fracLt3}
\end{align}
stays small throughout the inner region, so most of $M_{\rm halo}$
sits at large radii $r \sim a_0$.  For $\gamma > 3$, the exponent
$3 - \gamma < 0$ and $r^{3-\gamma}$ decreases with $r$.  Numerator
and denominator of \cref{app:mhUnified} are both negative, but
their ratio is positive and increases with $r$, giving
\begin{align}
   \frac{m_h(r)}{M_{\rm halo}} \approx
   1 - \left(\frac{4M_{\rm BH}}{r}\right)^{\gamma-3}
   \quad (r \gg 4M_{\rm BH}).
   \label{app:fracGt3}
\end{align}
This rises rapidly to unity just beyond $4M_{\rm BH}$, so most of
$M_{\rm halo}$ sits close to the inner cut-off.  At $\gamma = 3$,
\cref{app:mhUnified} is replaced by
$m_h \approx 4\pi\bar\rho_0 a_0^3 \ln(r/4M_{\rm BH})$.

\paragraph*{Effect on $\mathcal{I}$.}  As $\gamma$ increases, the dominant
radial contribution to $\mathcal{I}$ moves from $r \sim a_0$, where
$m_h \sim M_{\rm halo}$ but $(r - 2M_{\rm BH})^{-1}$ is small, toward
$r \sim 10$--$100\,M_{\rm BH}$, where $m_h \sim M_{\rm halo}$ and
$(r - 2M_{\rm BH})^{-1}$ is large.  This inward shift steadily
amplifies $\mathcal{I}$ and hence $\delta\Omega_c$, which is why
\cref{fig:c4c6g} shows such a strong and rapid growth with $\gamma$.
For $\gamma > 3$, the integrand is $O(M_{\rm halo}/M_{\rm BH}^2)$
over a radial extent $\sim M_{\rm BH}$ near the cut-off, so
$\mathcal{I} \sim M_{\rm halo}/M_{\rm BH}$.  This is parametrically
larger than the $\gamma < 3$ contribution when $a_0 \gg M_{\rm BH}$.

\subsubsection{Outer logarithmic slope $\beta$}

At fixed $M_{\rm halo}$ and $a_0$, a larger $\beta$ steepens the
outer density fall-off and requires a larger normalisation
$\bar\rho_0$ to keep the same total mass.  This raises $m_h(r)$ at
all radii, increasing $\mathcal{I}$ and hence $\delta\Omega_c$, as
seen in \cref{fig:c4c6b}.  The effect is comparable to or weaker
than the $\gamma$ dependence over most of the range shown in the
figures, but for small $\gamma$ or a small gap $\beta-\gamma$, the
sensitivity to $\beta$ can match or exceed that of $\gamma$.  This
is most apparent as $\beta \to \gamma^+$, where the profile
approaches a single power law, concentrating mass further inward
and amplifying $\mathcal{I}$.  The constraint $\beta > \gamma$ links
the two slope parameters throughout.

\subsubsection{Transition parameter $\alpha$}

A smaller $\alpha$ gives a smoother transition near $r \sim a_0$ and keeps more mass at intermediate radii, modestly raising $\bar\rho_0$
and hence $m_h(r)$ in the inner region.  For larger $\gamma$, where
the integrand of $\mathcal{I}$ carries appreciable weight at
intermediate $r$, this raises $\mathcal{I}$, so $\delta\Omega_c$
decreases with $\alpha$, as seen in \cref{fig:c4c6a}.  For small$\gamma$, ($\gamma\lesssim1$), the dependence on $\alpha$ is not weak and can even reverse sign: $\mathcal{I}$ increases with $\alpha$ the opposite of the large-$\gamma$ trend above, with a non-monotonic crossover near $\gamma\approx1$. In both cases the $\alpha$ dependence remains subdominant.

\subsection{Compactness scaling and degeneracy}
\label{app:compact}

At fixed $(\alpha, \beta, \gamma)$, rescaling
$M_{\rm halo} \to \lambda M_{\rm halo}$ at fixed $a_0$ maps
$m_h(r) \to \lambda\,m_h(r)$ at every fixed $r$, turning the single
number $\mathcal{I}$, defined for one fixed halo, into a function
$\mathcal{I}(\lambda)$ that can be differentiated. This is a
bookkeeping device rather than a new physical parameter: the actual
halo always has $\lambda=1$, and every result below is evaluated
there.

The compactness $z=M_{\rm halo}/a_0$ can equally be changed by
holding $M_{\rm halo}$ fixed and varying $a_0$ instead, and the two
routes are not interchangeable in general. For
$4M_{\rm BH}\ll r\ll a_0$, where $m_h(r)/M_{\rm halo}\sim(r/a_0)^{3-\gamma}$
(\cref{app:gamma_dep}), sending $a_0\to a_0/\mu$ at fixed
$M_{\rm halo}$ multiplies $z$ by $\mu$, exactly as $\lambda\to\mu$
would, but it rescales $m_h(r)$ at fixed $r$ by $\mu^{3-\gamma}$
rather than by $\mu$, since $\mu$ enters inside the profile's
argument $r/a_0$ rather than as an overall amplitude. The two
routes therefore agree only where this distinction does not yet
matter: in the linear regime established below, both give the same
leading coefficient $C(\alpha,\beta,\gamma)$, because the dominant
radial contribution to $\mathcal{I}$ sits at $r\ll a_0$ regardless
of route, where the $\mu^{3-\gamma}$ versus $\mu^1$ difference is a
subleading correction. The two routes diverge once $\mathcal{I}$
is no longer small, as the saturation analysis of
\cref{app:saturation} shows explicitly, where holding $a_0$ fixed
and varying $M_{\rm halo}$ leads to a saturation value that depends
on $M_{\rm halo}$, while holding $M_{\rm halo}$ fixed and varying
$a_0$ does not change the saturation value at all. With this scope
in mind, sweeping $\lambda$ gives
\begin{align}
   \mathcal{I}(\lambda z)
   = \int_{4M_{\rm BH}}^{\infty}
     \frac{2\lambda\,m_h(r)}
          {[r - 2M_{\rm BH} - 2\lambda m_h(r)](r - 2M_{\rm BH})}\,dr.
   \label{app:Ilambda}
\end{align}
This is a monotonically increasing function of $\lambda$.
Differentiating under the integral gives
\begin{align}
   \frac{d\mathcal{I}}{d\lambda}
   = \int_{4M_{\rm BH}}^{\infty}
     \frac{2\,m_h(r)}{\bigl[r-2M_{\rm BH}-2\lambda m_h(r)\bigr]^2}\,dr
   > 0.
   \label{app:dIdlambda}
\end{align}
The integrand is non-negative because $m_h \geq 0$ and
$r > 2M_{\rm BH} + 2\lambda m_h$ in the exterior.  A second
derivative gives $d^2\mathcal{I}/d\lambda^2 > 0$, so $\mathcal{I}$
is convex in $\lambda$. The integrand $8m_h^2/[r-2m(r)]^3$ is
manifestly positive in the exterior, where $r > 3m(r)$
(\cref{app:metric}) ensures $r-2m(r) > m(r) > 0$.  As $\lambda$
grows, the denominator $r - 2M_{\rm BH} - 2\lambda m_h$ shrinks and
the integrand grows faster than linear.  The eikonal shift
$\delta\Omega_c = 1 - e^{-\mathcal{I}/2}$ is monotonically increasing
in $\lambda$ (since $\mathcal{I}' > 0$).\footnote{It is interesting to note that for most of the DM distributions studied in the present work, we observe $\delta\Omega_c$ to be a concave function of $\mathcal{I}$, implying $(\mathcal{I}')^2 > 2\mathcal{I}''$.}

\paragraph*{Linear regime.}
The linear approximation $\mathcal{I} \approx C\,z$ holds when the
term $2\lambda m_h(r)$ in the denominator of \cref{app:Ilambda}
stays small compared with $r - 2M_{\rm BH}$ at every radius where
the integrand matters.  The dominant radial scale, and so the
condition for linearity, depends on $\gamma$.

For $\gamma \lesssim 2$, the integrand of $\mathcal{I}$ is dominated
by $r \sim a_0$ (\cref{app:gamma_dep}), where
$m_h(r) \sim M_{\rm halo}$ and $r - 2M_{\rm BH} \sim a_0$.  The
smallness condition then reads $2M_{\rm halo} \ll a_0$, that is
$z \ll \tfrac{1}{2}$, which holds throughout the range $z \ll 1$
shown in the figures.  The denominator reduces to
$(r - 2M_{\rm BH})^2$, giving
\begin{align}
   \mathcal{I} \approx
   C(\alpha,\beta,\gamma)\,z,
   \quad
   C(\alpha,\beta,\gamma) \equiv
   \int_{4M_{\rm BH}}^{\infty}
     \frac{2\hat{m}_h(r)\,a_0}{(r-2M_{\rm BH})^2}\,dr,
   \label{app:Ilinear}
\end{align}
where $\hat{m}_h(r) = m_h(r)/M_{\rm halo}$ is the normalised
enclosed mass, independent of $z$.  Hence $\delta\Omega_c \approx
\tfrac{1}{2}C(\alpha,\beta,\gamma)\,z$ for $\gamma \lesssim 2$.

As $\gamma$ approaches 3, the radial scale that dominates the integrand of $\mathcal{I}$ moves inward gradually rather than switching abruptly at $\gamma=3$.  Numerically, the fraction of
$\mathcal{I}$ coming from $r < 100\,M_{\rm BH}$ rises from below 1\%
at $\gamma=1$ to about 70\% at
$\gamma=2.9$, for $a_0=10^5M_{\rm BH}$, and this fraction barely
changes with $z$ over the range shown.  Equation~(\ref{app:Ilinear})
therefore becomes a progressively cruder approximation for
$2 \lesssim \gamma < 3$, even though $\mathcal{I}$ stays
approximately linear in $z$ across this range.  The dominant radial
scale shifts, but at fixed $\gamma$ the integral remains linear in
$\lambda$ to leading order, since the inner contribution is governed
by \cref{app:mhscaling} of ~\Cref{app:potential}, which
is itself linear in $M_{\rm halo}$ wherever its validity condition
holds.

For $\gamma \geq 3$, the integrand of $\mathcal{I}$ is dominated by
$r \sim \text{few}\,M_{\rm BH}$, where $m_h(r) \sim M_{\rm halo}$ and
$r - 2M_{\rm BH} \sim M_{\rm BH}$.  The smallness condition then
becomes $2M_{\rm halo} \ll M_{\rm BH}$, that is $z \ll M_{\rm
BH}/a_0$.  For the astrophysically relevant case $a_0 \gg M_{\rm
BH}$, this threshold sits far below the $z \ll 1$ range of the
figures.  For $a_0 = 10^5\,M_{\rm BH}$ one needs $z \ll 10^{-5}$.  So
for $\gamma \geq 3$, $\mathcal{I}$ is already in the non-linear,
convex regime through most of the plotted compactness range, and
grows faster than linear in $z$.  The same threshold, generalised to
$z \ll (M_{\rm BH}/a_0)^{\gamma-2}$, governs the onset of
non-linearity already for $2<\gamma<3$
(Appendix~\ref{app:potential}).

The two compactnesses $z = (10^{-6}, 10^{-4})$ in the right panels
of \cref{fig:c4c6g,fig:c4c6b,fig:c4c6a} are obtained by holding
$M_{\rm halo}$ fixed and varying $a_0$. Under this route, $z$ still increases by the
same factor of $10^2$ between the two curves, but $m_h(r)$ at fixed
$r$ no longer rescales by that same factor: writing $a_0\to a_0/\mu$
at fixed $M_{\rm halo}$ multiplies $z$ by $\mu$ while rescaling
$m_h(r)$ at fixed $r$ by $\mu^{3-\gamma}$ in the power-law regime
(\cref{app:gamma_dep}), not by $\mu^1$. For $\gamma \lesssim 2$,
where the linear regime holds most robustly and $3-\gamma$ is not
too small, $\delta\Omega_c$ still grows by close to the full
$10^2$ between the two curves, since the dominant radial
contribution to $\mathcal{I}$ sits at $r\ll a_0$ for both choices of
$a_0$, where the precise value of $a_0$ matters little to the
integral beyond setting its overall normalisation. As $\gamma$ rises
toward 3, however, $3-\gamma\to0$ and the $\mu^{3-\gamma}$ rescaling
of $m_h(r)$ weakens sharply, so the separation between the two
compactness curves \emph{shrinks}. This is  opposite to what naive linear-in-$z$ scaling would suggest, and the
opposite of what would be seen under the $M_{\rm halo}$-route
instead, where the separation stays close to $10^2$ at every
$\gamma$ because $m_h(r)$ there rescales by the full factor $\mu$
regardless of $\gamma$ (\cref{app:compact}). This is consistent with
the saturation mechanism of \cref{app:saturation}: as
$\gamma\to\beta^-$, $\mathcal{I}$ becomes insensitive to $a_0$
altogether, and the shrinking separation seen here is the
finite-$\gamma$ precursor of that same effect. This matches the
behaviour visible directly in the eikonal panels of
\cref{fig:c4c6g}, where the two compactness curves converge as
$\gamma\to3$; the corresponding contrast with the $M_{\rm
halo}$-route, where the separation does not shrink, is shown for
the non-eikonal quantity $\delta\omega$ in \cref{fig:c-g} and
discussed in \cref{app:saturation}.

\paragraph*{Degeneracy.}
$\delta\Omega_c = 1 - e^{-\mathcal{I}/2}$ depends on
$(\alpha, \beta, \gamma, z)$ only through the single number
$\mathcal{I}$.  Any combination with $\beta > \gamma$ that gives the
same $\mathcal{I}$ gives the same light ring shift.  For $\gamma >
3$, where $\mathcal{I} \sim M_{\rm halo}/M_{\rm BH}$, a given shift
is reached at smaller $z$ than for $\gamma < 3$, so a steep inner
cusp at low compactness can mimic a shallower cusp at higher
compactness.  This produces the degeneracy surfaces in
\cref{fig:c4c6g}.  The boundary
$\beta > \gamma$ marks the edge of the physically allowed region.
Breaking this degeneracy needs either subleading corrections to the uniform potential scaling or independent astrophysical priors on the
halo profile.

\onecolumngrid
\subsection{Eikonal - Correspondence}
\label{app:check}
\begin{table}[!h]
\centering
\renewcommand{\arraystretch}{1.1}
\begin{tabular}{|c|c|c|c|}
\hline
{\bf DM Profile} &
\textbf{$l$} &
\textbf{From LR} &
\textbf{From WKB-6}  \\
\hline

\multirow{3}{*}{0.5, 3.5, 1}
& 2  & $0.3848872-0.0962218\,i$ & $0.3736068-0.0888880\,i$  \\ \cline{2-4}
& 3  & $0.5773308-0.0962218\,i$ & $0.5994232-0.0926994\,i$  \\ \cline{2-4}
& 10 & $1.9244360-0.0962218\,i$ & $1.9967205-0.0958606\,i$  \\
\hline

\multirow{3}{*}{0.5, 3.5, 2.5}
& 2  & $0.3805829-0.0951457\,i$ & $0.3694286-0.0878939\,i$ \\ \cline{2-4}
& 3  & $0.5708744-0.0951457\,i$ & $0.5927197-0.0916627\,i$  \\ \cline{2-4}
& 10 & $1.9029146-0.0951457\,i$ & $1.9743907-0.0947886\,i$ \\
\hline

\multirow{3}{*}{1, 3, 1}
& 2  & $0.3848667-0.0962167\,i$ & $0.3735869-0.0888832\,i$  \\ \cline{2-4}
& 3  & $0.5773001-0.0962167\,i$ & $0.5993913-0.0926945\,i$  \\ \cline{2-4}
& 10 & $1.9243336-0.0962167\,i$ & $1.9966142-0.0958555\,i$  \\
\hline

\multirow{3}{*}{1, 3, 2.5}
& 2  & $0.3819387-0.0954847\,i$ & $0.3707447-0.0882070\,i$ \\ \cline{2-4}
& 3  & $0.5729081-0.0954847\,i$ & $0.5948312-0.0919893\,i$  \\ \cline{2-4}
& 10 & $1.9096936-0.0954847\,i$ & $1.9814243-0.0951263\,i$  \\
\hline

\multirow{3}{*}{1, 4, 1}
& 2  & $0.3848617-0.0962154\,i$ & $0.3735820-0.0888821\,i$  \\ \cline{2-4}
& 3  & $0.5772926-0.0962154\,i$ & $0.5993834-0.0926932\,i$ \\ \cline{2-4}
& 10 & $1.9243085-0.0962154\,i$ & $1.9965882-0.0958543\,i$\\
\hline

\multirow{3}{*}{1, 4, 2.5}
& 2  & $0.3803444-0.0950861\,i$ & $0.3691971-0.0878388\,i$  \\ \cline{2-4}
& 3  & $0.5705166-0.0950861\,i$ & $0.5923482-0.0916053\,i$  \\ \cline{2-4}
& 10 & $1.9017221-0.0950861\,i$ & $1.9731534-0.0947292\,i$ \\
\hline

\multirow{3}{*}{Einasto}
& 2  & $0.3847691-0.0961923\,i$ & $0.3734921-0.0888607\,i$  \\ \cline{2-4}
& 3  & $0.5771536-0.0961923\,i$ & $0.5992392-0.0926710\,i$ \\ \cline{2-4}
& 10 & $1.9238454-0.0961923\,i$ & $1.9961077-0.0958312\,i$  \\
\hline
\multirow{3}{*}{Schwarzschild}
& 2  & $0.3849002-0.0962250\,i$ & $0.3736194-0.0888910\,i$  \\ \cline{2-4}
& 3  & $0.5773503-0.0962250\,i$ & $0.5994434-0.0927025\,i$ 
 \\ \cline{2-4}
& 10 & $1.9245009-0.0962250\,i$ & $1.9967878-0.0958639\,i$  \\
\hline

\end{tabular}
\caption{Comparison of the angular frequency and the principal Lyapunov exponent at the light ring with the fundamental QNM frequencies obtained using sixth order WKB for different multipoles for different halo configurations with compactness $z=10^{-4}$ and scale radius $a_0=10^{5}$. For the Einasto profile $r_e=a_0$. The Schwarzshild values are also shown for reference.}
\label{tab:qnm_grouped_dm}
\end{table}
\twocolumngrid
\section{Analytic basis for the non-eikonal QNM results}
\label{App:QNM}
 
We derive the results and trends described in \cref{sec:QNMhalo}
and provide the analytic basis for the observations in
\cref{fig:zplotnew,fig:c4c6g,fig:c4c6b,fig:c4c6a,fig:c-g}.
 
\subsection{Potential scaling and exact frequency relation}
\label{app:potential}
 
The Regge--Wheeler potential governing axial perturbations,
\cref{eq:RWpot}, reads
\begin{align}
   V(r) = f(r)\!\left[\frac{l(l+1)}{r^2}
           - \frac{6m(r)}{r^3}
           + \frac{m'(r)}{r^2}\right].
   \label{app:Vdef}
\end{align}
This potential is defined on the full domain $r \in [2M_{\rm BH},
\infty)$, with or without the halo. Its functional form changes at
$r = 4M_{\rm BH}$ because the halo density $\bar\rho(r)$, and hence
$m(r)$ and $f(r)$, is defined piecewise. $\bar\rho(r) = 0$ for $r
\leq 4M_{\rm BH}$, the assumed DM-free cavity, and $\bar\rho(r) > 0$
for $r > 4M_{\rm BH}$. We treat $V_{\rm DM}(r)$ on these two
sub-domains in turn and ask in each case how closely it tracks
$e^{-\mathcal{I}}\,V_{\rm Sch}$.
 
\paragraph*{Sub-domain $r \leq 4M_{\rm BH}$.}
Here $\bar\rho = 0$ and $m_h = 0$, so $m = M_{\rm BH}$ and $m' = 0$.
Both $f_{\rm DM}$ and $f_{\rm Sch}$ obey the same first-order ODE,
\cref{eq:fprime},
\begin{align}
   \frac{f'}{f} = \frac{2M_{\rm BH}}{r(r-2M_{\rm BH})},
\end{align}
and so differ only by a multiplicative constant fixed by their
boundary conditions at large $r$. From \cref{App:eikonal} this
constant is $e^{-\mathcal{I}}$, where $\mathcal{I}$ is the halo
integral of \cref{eq:haloint}, a fixed number rather than a function
of $r$. Therefore
\begin{align}
   V_{\rm DM}(r) = e^{-\mathcal{I}}\,V_{\rm Sch}(r)
   \quad (r \leq 4M_{\rm BH})
   \label{app:Vscale}
\end{align}
exactly. $V_{\rm DM}$ is simply $V_{\rm Sch}$ rescaled by the
constant $e^{-\mathcal{I}}$, so it has the same shape and the same
extrema, lowered uniformly in amplitude. A positive rescaling does
not move an extremum, so any local maximum of
$e^{-\mathcal{I}}V_{\rm Sch}(r)$ on this sub-domain sits at the same
$r$ as the corresponding extremum of $V_{\rm Sch}$. We show below
that for $l \geq 2$ this extremum, the global potential maximum,
falls at $r_{\rm max} < 4M_{\rm BH}$, inside this sub-domain.
Whether it is also the global maximum of $V_{\rm DM}(r)$ over the
full domain depends on the other sub-domain, treated next.
 
\paragraph*{Sub-domain $r > 4M_{\rm BH}$.}
Here $m_h(r) > 0$ and $\bar\rho(r) > 0$, so \cref{app:Vscale} no
longer holds exactly. Even so, $V_{\rm DM}(r) \approx
e^{-\mathcal{I}}V_{\rm Sch}(r)$ remains an excellent approximation,
for reasons that depend on $\gamma$.
 
The size of $m_h(r)$ at $r \sim \text{few}\,M_{\rm BH}$, well inside
$a_0$, refines the simple split between $\gamma<3$ and $\gamma>3$.
From the enclosed-mass fraction of \cref{app:gamma_dep},
$m_h(r)/M_{\rm halo} \sim (r/a_0)^{3-\gamma}$ for $4M_{\rm BH} \ll r
\ll a_0$ and $\gamma<3$. Writing $M_{\rm halo}/M_{\rm BH} =
z\,a_0/M_{\rm BH}$, the enclosed mass at $r \sim c\,M_{\rm BH}$ with
$c = O(1)$ scales as
\begin{align}
   \frac{m_h(cM_{\rm BH})}{M_{\rm BH}}
   \sim z\,c^{3-\gamma}\left(\frac{M_{\rm BH}}{a_0}\right)^{2-\gamma}.
   \label{app:mhscaling}
\end{align}
For $\gamma \leq 2$ the exponent $2-\gamma \geq 0$, so
$(M_{\rm BH}/a_0)^{2-\gamma} \leq 1$, and \cref{app:mhscaling} holds
trivially. $m_h(r) \ll M_{\rm BH}$ for any $z \ll 1$. For $2 <
\gamma < 3$ the exponent flips sign, $(M_{\rm BH}/a_0)^{2-\gamma} =
(a_0/M_{\rm BH})^{\gamma-2} \gg 1$, and the bound needs the stronger
condition
\begin{align}
   z \ll \left(\frac{M_{\rm BH}}{a_0}\right)^{\gamma-2}
   \qquad (2 < \gamma < 3),
   \label{app:gammaThreshold}
\end{align}
which tightens as $\gamma \to 3^-$ and in that limit matches onto
the threshold $z \ll M_{\rm BH}/a_0$ derived below for $\gamma > 3$.
For $a_0 = 10^5 M_{\rm BH}$, \cref{app:gammaThreshold} requires
$z \ll 3\times10^{-5}$ at $\gamma=2.9$, so $m_h(r)\ll M_{\rm BH}$ can
fail at moderate compactness even when $\gamma<3$.  Where
\cref{app:gammaThreshold} holds, the ratio $V_{\rm DM}/V_{\rm Sch}
\approx e^{-\mathcal{I}}$ is preserved through the inner region, up
to corrections of order $m_h(r)/M_{\rm BH} \ll 1$. Near $r \sim a_0$,
where $m_h(r) \sim M_{\rm halo}$, this approximation breaks down
regardless of $\gamma$. There, $V(r) \sim l(l+1)/r^2 \sim
l(l+1)/a_0^2$ is negligible against $\omega^2$, so the
contribution to the QNM condition is suppressed by
$(M_{\rm BH}/a_0)^2 \ll 1$ either way. This is a magnitude bound,
not the exact value of $V$: the other two terms in
\cref{app:Vdef} stay subdominant to the centrifugal term, since
$-6m(r)/r^3$ is suppressed by a factor $\sim M_{\rm BH}/r$, and
$m'(r)/r^2$ carries an overall factor $\sim z/l(l+1) \ll 1$
from the small compactness. Both corrections stay below the leading
term throughout $r > 4M_{\rm BH}$.
 
For $\gamma\geq3$ (and for $\gamma$ near $3$ at the compactnesses
used in the figures, where \cref{app:gammaThreshold} fails), $m_h(r)
\sim M_{\rm BH}$ already at $r\sim\text{few}\,M_{\rm BH}$, so the
$O(m_h/M_{\rm BH})$ correction to $V_{\rm DM}$ is not small in the
DM-occupied zone.  Nevertheless, the potential barrier governing
the $l\geq2$ fundamental mode is concentrated at $r<4M_{\rm BH}$
(shown below), where \cref{app:Vscale} holds exactly.  The
modification to $V_{\rm DM}$ for $r>4M_{\rm BH}$ enters only the
classically allowed propagating region and contributes a only a small
correction to the outgoing wave phase.
 
In either case, $V(r)$ at $r > 4M_{\rm BH}$ stays bounded by $\sim
l(l+1)/r^2$, for the same reason given above, and this never
reaches the value $V_{\rm max} = e^{-\mathcal{I}}V_{\rm Sch}(r_{\rm
max})$ attained at $r_{\rm max} < 4M_{\rm BH}$ on the first
sub-domain. No competing maximum appears for $r > 4M_{\rm BH}$, so
$r_{\rm max}$, computed below from the exact rescaling
\cref{app:Vscale}, is the global maximum of $V_{\rm DM}(r)$, not
merely a local one confined to $r \leq 4M_{\rm BH}$.
 
The uniform scaling $V_{\rm DM} \approx e^{-\mathcal{I}}\,V_{\rm
Sch}$ therefore holds across the QNM-relevant region in both cases.
To estimate the residual correction from $r > 4M_{\rm BH}$, we use
the standard real-axis WKB heuristic of the numerical QNM literature~\cite{Schutz:1985km,Iyer:1986np}, in which the potential barrier extends
between the points where $V(r) = \omega_{\rm Re}^2$ on the real
$r$-axis.\footnote{This is a leading-order approximation, valid when
$|\omega_{\rm Im}| \ll \omega_{\rm Re}$. For $l=2$, $n=0$,
$|\omega_{\rm Im}/\omega_{\rm Re}| \approx 0.24$, so it is indicative
rather than exact. The rigorous treatment locates turning points as
zeros of $Q_0(r) = \omega^2 - V(r)$ in the complex $r$-plane, as in
the exact WKB formalism of Miyachi {\it et al.}~\cite{Miyachi:2025ptm}. For $r <
4M_{\rm BH}$, $Q_0$ inherits the meromorphic structure of the
Schwarzschild problem exactly under $\omega \to
e^{\mathcal{I}/2}\omega_{\rm DM}$, so an exact-WKB treatment of this
region alone should be achievable in principle. For $r > 4M_{\rm
BH}$, the profile factor $[1+(r/a_0)^\alpha]^{(\gamma-\beta)/\alpha}$
has branch points in the complex plane for generic non-integer
$\alpha$, so $Q_0(r)$ is not meromorphic there and the
Borel-summation machinery behind exact WKB does not directly apply.
A full treatment would also need $r=4M_{\rm BH}$ as an interior matching point joining the WKB bases of the two regions, a connection problem not covered by the existing literature, which treats only the two asymptotic ends. We regard a full exact-WKB treatment of the generic $(\alpha,\beta,\gamma)$ profile as out of reach
with current methods, though tractable for restricted cases such as integer $\alpha$, and leave it for future work.}
We quantify this correction below, after deriving the Schwarzschild potential maximum and its value at the cut-off.
 
The tortoise coordinate satisfies $dr_*/dr = [f(r)(1-2m/r)]^{-1/2}$.
For $r \leq 4M_{\rm BH}$, using $m = M_{\rm BH}$ and $f_{\rm DM} =
e^{-\mathcal{I}}\,f_{\rm Sch}$,
\begin{align}
   \frac{dr_*^{\rm DM}}{dr}
   = \frac{1}{\sqrt{e^{-\mathcal{I}}\,f_{\rm Sch}(r)(1-2M_{\rm BH}/r)}}
   = e^{\mathcal{I}/2}\,\frac{dr_*^{\rm Sch}}{dr}.
   \label{app:tortoise}
\end{align}
Since $\mathcal{I}$ is a constant, integrating gives
$r_*^{\rm DM} = e^{\mathcal{I}/2}\,r_*^{\rm Sch} + C$ with $C$ a
constant of integration, and no $d\mathcal{I}/dr_*$ term appears.
Applying the chain rule $d/dr_*^{\rm DM} =
e^{-\mathcal{I}/2}\,d/dr_*^{\rm Sch}$ to the DM Regge--Wheeler
equation and using \cref{app:Vscale} gives
\begin{align}
   e^{-\mathcal{I}}\frac{d^2\psi}{dr_*^{{\rm Sch}\,2}}
   + \left[\omega_{\rm DM}^2
           - e^{-\mathcal{I}}V_{\rm Sch}\right]\psi &= 0.
   \label{app:RWintermediate}
\end{align}
Multiplying through by $e^{\mathcal{I}}$ gives
\begin{align}
   \frac{d^2\psi}{dr_*^{{\rm Sch}\,2}}
   + \left[e^{\mathcal{I}}\omega_{\rm DM}^2
           - V_{\rm Sch}(r_*^{\rm Sch})\right]\psi = 0
   \quad (r \leq 4M_{\rm BH}),
   \label{app:RWrescaled}
\end{align}
the Schwarzschild Regge--Wheeler equation with $\omega$ replaced by
$e^{\mathcal{I}/2}\omega_{\rm DM}$. At the horizon ($r_*^{\rm Sch}
\to -\infty$), the ingoing boundary condition $\psi \sim
e^{-i\omega_{\rm DM}r_*^{\rm DM}} = e^{-i\omega_{\rm
DM}e^{\mathcal{I}/2}r_*^{\rm Sch}} \times \text{phase}$ matches the
Schwarzschild form with $\omega_{\rm Sch} =
e^{\mathcal{I}/2}\omega_{\rm DM}$. The integration constant $C$
adds only an irrelevant overall phase. The QNM resonance condition
then gives
\begin{align}
   \omega_{\rm DM} = e^{-\mathcal{I}/2}\,\omega_{\rm Sch},
   \label{app:omegaDM}
\end{align}
and hence
\begin{align}
   \delta\omega_{\rm Re} = \delta\omega_{\rm Im}
   = 1 - e^{-\mathcal{I}/2}
   = \delta\Omega_c.
   \label{app:deltasQNM}
\end{align}
The fractional redshifts of the real and imaginary parts of the QNM
frequency match each other and the light ring shift $\delta\Omega_c$
for every multipole $l \geq 2$, every overtone, and every halo
profile. This is why $\delta\omega_{\rm Re}$ and $\delta\omega_{\rm
Im}$ overlap so closely in \cref{fig:c4c6a}, and why both track the
eikonal shifts there.
 
The derivation rests on \cref{app:Vscale} holding through the
region that sets the QNM eigenvalue, bounded within the real-axis
heuristic by the points where $V(r) = \omega_{\rm Re}^2$. Setting
$dV/dr = 0$ for the Schwarzschild potential with $V = (1-2M/r)
[l(l+1)/r^2 - 6M/r^3]$ gives the quadratic $2L\,r^2 -
6M(3+L)\,r + 48M^2 = 0$, where $L = l(l+1)$, whose outer root
gives the potential maximum
\begin{align}
   r_{\rm max}(l) = \frac{3(3+L)
   + \sqrt{9(3+L)^2 - 96L}}{2L}\,M_{\rm BH}~,\nonumber\\
   r_{\rm max}(l)\xrightarrow{\,l\to\infty\,} 3M_{\rm BH}.
   \label{app:rmax}
\end{align}
For $l = 2$ ($L = 6$) this evaluates analytically to $r_{\rm max}
= \tfrac{1}{4}(9+\sqrt{17})\,M_{\rm BH} \approx 3.28\,M_{\rm BH}$,
inside the DM-free zone. The potential at the inner cut-off
evaluates analytically to
\begin{align}
   V(4M_{\rm BH}) = \frac{9}{64\,M_{\rm BH}^2}
   \quad (l = 2).
   \label{app:V4M}
\end{align}
$V$ falls monotonically beyond $r_{\rm max}$, so the heuristic outer
boundary satisfies $V(r_{\rm outer}) = \omega_{\rm Re}^2$, and
$r_{\rm outer}$ falls outside $4M_{\rm BH}$ only if $\omega_{\rm
Re}^2 < V(4M_{\rm BH})$. Numerically, $\omega_{\rm Re}^2 \approx
0.9929\,V(4M_{\rm BH})$ for the $l = 2$, $n = 0$ mode, so $r_{\rm
outer}$ sits just outside $4M_{\rm BH}$, with $r_{\rm outer} -
4M_{\rm BH} \approx [V(4M_{\rm BH}) - \omega_{\rm Re}^2]/
|V'(4M_{\rm BH})| \approx 0.04\,M_{\rm BH}$.  The WKB phase
accumulated in the strip $[4M_{\rm BH},\,r_{\rm outer}]$ is
\begin{align}
   \phi_{\rm strip}
   \approx \frac{1}{f(4M_{\rm BH})}
   \sqrt{|V'(4M_{\rm BH})|}\,
   \tfrac{2}{3}\,(r_{\rm outer}-4M_{\rm BH})^{3/2}
   \label{app:phi_strip}
\end{align}
The Schutz--Will quantisation condition~\cite{Schutz:1985km} requires the full barrier integral $\phi_{\rm total}$ to equal
$(n+\frac{1}{2})\pi = \pi/2$ for the fundamental mode; a fractional shift $\phi_{\rm strip}/(\pi/2)$ in that integral from the DM strip translates directly into the same fractional shift in $\omega$, giving $\delta\omega/\omega \sim \phi_{\rm strip}/(\pi/2)$
It is the uniform rescaling of the entire barrier by $e^{-\mathcal{I}}$, not any special sensitivity
near $r = 4M_{\rm BH}$, that produces \cref{app:omegaDM}, with the DM-occupied region adding only this small correction.  For
$l \geq 3$ both the potential maximum and the turning points lie strictly inside $4M_{\rm BH}$, verified numerically, so the
correction is smaller still.
 
\subsection{Compactness dependence}
\label{app:compactness_QNM}
 
From \cref{app:omegaDM,app:Ilambda}, $\delta\omega = 1 -
e^{-\mathcal{I}/2}$. For fixed halo shape at small compactness $z =
M_{\rm halo}/a_0$, the halo integral satisfies $\mathcal{I} \propto
z$ (\cref{app:compact}), so
\begin{align}
   \delta\omega \approx \tfrac{1}{2}\mathcal{I} \propto z
   \quad (z \ll 1),
   \label{app:linearZ}
\end{align}
predicting a linear growth of $\delta\omega$ with compactness on a log-log plot, with slope close to unity, matching the initial rise in \cref{fig:zplotnew}. As $M_{\rm halo}$ grows, $\mathcal{I}$ grows faster than linear in $z$ since it is convex (\cref{app:dIdlambda}).
The observable $\delta\omega = 1-e^{-\mathcal{I}/2}$ is also a striclty growing function in $z$ and is dominantly concave in $z$ (for most of the DM distributions studied in this work), hence the gradual flattening at the right end
of \cref{fig:zplotnew}. Both the convexity of $\mathcal{I}$ and the ,omotonic increase of $\delta\omega$ are independent of the specific
halo profile, which is why all the curves share the same
qualitative shape.
 
\subsection{Shape-parameter dependence of $\delta\omega$}
\label{app:shape_QNM}
 
Since $\delta\omega_{\rm Re} = \delta\omega_{\rm Im} =
1-e^{-\mathcal{I}/2}$, and $\mathcal{I}$ depends on
$(\alpha,\beta,\gamma)$ through the mechanisms of
\cref{app:Idep}, the shape-parameter trends in \cref{fig:c4c6a}
follow directly.
 
The strong growth of $\delta\omega$ with $\gamma$ mirrors the growth
of $\mathcal{I}$ established in \cref{app:gamma_dep}. As $\gamma$
increases, the dominant radial contribution to $\mathcal{I}$ moves
inward from $r \sim a_0$ toward $r \sim 10$--$100\,M_{\rm BH}$,
where the weight $(r-2M_{\rm BH})^{-1}$ is large. For small
$\mathcal{I}$, $\delta\omega \approx \mathcal{I}/2$, and the
polynomial behaviour seen in the data,
\begin{align}
   \delta\omega \approx \sum_n a_n\,\gamma^n,\qquad n\geq0
   \label{app:poly}
\end{align}
follows from a Taylor expansion of $\mathcal{I}(\gamma)$ about
$\gamma = 0$. The coefficients $a_n$ encode the sensitivity of the
enclosed halo mass to the inner logarithmic slope at each order. At larger
$\gamma$ the growth speeds up because $\mathcal{I}$ increases
non-linearly.
 
The rise of $\delta\omega$ with $\beta$, and its drop with $\alpha$
at larger $\gamma$, follow the same mechanisms described for
$\delta\Omega_c$ in \cref{app:Idep}, including the same caveats,
$\beta$'s effect is comparable to or weaker than $\gamma$'s over
most of the range shown, but can match or exceed it for small
$\gamma$ or a small gap $\beta-\gamma$, and the $\alpha$ dependence
reverses sign below $\gamma\approx1$.

\subsection{Degeneracy and the iso-$\mathcal{I}$ surface}
\label{app:degen_QNM}
 
Equation~\eqref{app:deltasQNM} shows that $\delta\omega$ depends on
$(\alpha,\beta,\gamma,z)$ only through the single number
$\mathcal{I}$. Any combination of halo parameters that gives the
same $\mathcal{I}$ gives the same $\delta\omega$. In the
$(\gamma,z)$ plane at fixed $(\alpha,\beta)$, the curves of constant
$\delta\omega$ are curves of constant $\mathcal{I}$. Since
$\mathcal{I}$ ( hence $\delta\omega$) grows strongly with $\gamma$, and $\delta\omega$ is an monotonically increasing function of $z$ (numerically observed to be growing concavely with $z$), a
larger $\gamma$ can offset a smaller $z$ and still give the same
total shift, producing the degeneracy curves in \cref{fig:c-g}.
 
This degeneracy is exact within the approximation
\cref{app:omegaDM}. Because $\omega_{\rm DM} =
e^{-\mathcal{I}/2}\,\omega_{\rm Sch}$ for every $(l,n)$ mode, all
frequency ratios between modes keep their Schwarzschild values and
carry no information about $\mathcal{I}$. Every mode gives the same
equation $\delta\omega = 1-e^{-\mathcal{I}/2}$ independently, so
measuring several modes gives redundant determinations of
$\mathcal{I}$ but does not pin down which halo parameters
$(\alpha,\beta,\gamma,z)$ produce it. The quality factor $Q =
\omega_{\rm Re}/|\omega_{\rm Im}|$ is likewise unchanged: $Q_{\rm
DM} = Q_{\rm Sch}$ for every mode.
 
Subleading corrections come from the small portion of the WKB
phase integral that extends past $r = 4M_{\rm BH}$, where $m_h(r)$
and $\bar\rho(r)$ modify the potential beyond $f(r)$ alone. These
corrections are of order $(r_{\rm tp}-4M_{\rm BH})^{3/2}/\phi_{\rm
WKB} \lesssim 10^{-3}$, where $\phi_{\rm WKB}$ is the total WKB
phase, and introduce weak mode-dependent contributions to
$\delta\omega$ that can differ between modes with different
turning-point structures. Such corrections could in principle break
the mode degeneracy, but they sit well below current measurement
precision.
 
\subsection{Saturation at large $\gamma$}
\label{app:saturation}
 
Figures~\ref{fig:c4c6a} and \ref{fig:c-g} show $\delta\omega$ growing rapidly with
$\gamma$ but flattening as $\gamma$ approaches its maximum allowed
value $\beta$. We now show that this saturation is exact, shared by
the eikonal and non-eikonal regimes alike, and set by a single
computable quantity.
 
\paragraph*{Origin of the saturation.}
From \cref{app:deltasQNM}, both $\delta\Omega_c$ and
$\delta\omega_{\rm Re/Im}$ equal $1-e^{-\mathcal{I}/2}$, a
monotonically increasing function of $\mathcal{I}$. Since
$\mathcal{I}$ is itself increasing in $\gamma$, the shifts grow with
$\gamma$. The physical constraint $\beta > \gamma \geq 0$ sets an
upper bound $\gamma < \beta$, and as $\gamma \to \beta^-$ the
profile approaches the pure power-law limit
\begin{align}
   \bar\rho(r) \to \bar\rho_0
   \left(\frac{r}{a_0}\right)^{-\beta}
   \!\left(1-\frac{4M_{\rm BH}}{r}\right)
   \quad (\gamma\to\beta^-),
   \label{app:purepowerlaw}
\end{align}
where the inner-to-outer transition disappears entirely. This is the
most centrally concentrated profile allowed for a given $\beta$, so
no further inward shift of halo mass is possible. The halo integral
then approaches a finite saturation value,
\begin{align}
   \mathcal{I}_{\rm sat}(\beta,M_{\rm halo})
   \equiv \mathcal{I}\bigl[\bar\rho\to\cref{app:purepowerlaw},
          \,M_{\rm halo}\;{\rm fixed}\bigr],
   \label{app:Isat}
\end{align}
and both the eikonal and non-eikonal shifts saturate at the same
value,
\begin{align}
   \delta_{\rm sat}(\beta, M_{\rm halo})
   = 1 - e^{-\mathcal{I}_{\rm sat}(\beta,M_{\rm halo})/2}.
   \label{app:deltaSat}
\end{align}
We write $\mathcal{I}_{\rm sat}$ as a function of $M_{\rm halo}$
rather than of the compactness $z=M_{\rm halo}/a_0$ used elsewhere,
because, as shown next, $a_0$ drops out of the limiting profile
entirely and $\mathcal{I}_{\rm sat}$ depends on $M_{\rm halo}$ and
$M_{\rm BH}$ alone.
 
\paragraph*{Normalization of the limiting profile.}
 For the pure power law \cref{app:purepowerlaw} with $\beta > 3$, the
total halo mass integral evaluates analytically to
\begin{align}
   M_{\rm halo}
   = \frac{4\pi\bar\rho_0\,a_0^\beta\,(4M_{\rm BH})^{3-\beta}}
          {(\beta-3)(\beta-2)},
   \label{app:Mhalo_sat}
\end{align}
where the factor $[(\beta-3)(\beta-2)]^{-1}$ comes from the combined
effect of the power-law fall-off and the inner cut-off
$(1-4M_{\rm BH}/r)$. Solving for $\bar\rho_0$ and substituting back
into the enclosed mass $m_h(r)=\int_{4M_{\rm BH}}^r
4\pi r'^2\bar\rho(r')\,dr'$ shows that every factor of $a_0$
cancels, leaving
\begin{align}
   \frac{m_h(r)}{M_{\rm halo}}
   \approx 1 - \left(\frac{4M_{\rm BH}}{r}\right)^{\beta-3}
   \qquad (r \gg 4M_{\rm BH}),
   \label{app:fracSat}
\end{align}
a function of $r/M_{\rm BH}$ and $\beta$ alone, with no remaining
$a_0$ dependence. In the $\gamma\to\beta^-$ limit, $a_0$ no longer marks a transition scale in the density profile, since the profile
is a single power law all the way from $4M_{\rm BH}$ to infinity. The scale factor $a_0$ enters only through normalising $\bar\rho_0$ against $M_{\rm halo}$, and that dependence cancels exactly once the
normalisation is carried through. Consequently $\mathcal{I}_{\rm
sat}$ and $\delta_{\rm sat}$, built entirely from $m_h(r)$ and $r$,
depend on $M_{\rm halo}$ and $\beta$ only, not on $a_0$
independently.
 
\paragraph*{Saturation value and its dependence on $\beta$ and $M_{\rm
halo}$.} 
The density $\bar\rho\propto r^{-\beta}(1-4M_{\rm BH}/r)$ peaks at $r_{\rm peak}=4M_{\rm BH}(1+\beta)/\beta$, which approaches $4M_{\rm BH}$ monotonically as $\beta$ grows.  A steeper outer profile therefore concentrates the density maximum more tightly near the inner edge of the halo, and the enclosed mass $m_h(r)$ rises more steeply from $r=4M_{\rm BH}$.  Since the integrand of $\mathcal{I}_{\rm sat}$ carries the weight $[r-2m(r)]^{-1}$, which is largest near $r=2M_{\rm BH}$, contributions from the inner halo are amplified. A larger $\beta$ thus raises $\mathcal{I}_{\rm sat}$ and $\delta_{\rm sat}$.  For small $M_{\rm halo}/M_{\rm BH}$, $\mathcal{I}_{\rm sat}\propto M_{\rm halo}$ and  $\delta_{\rm
sat}\approx\mathcal{I}_{\rm sat}/2\propto M_{\rm halo}$. At larger $M_{\rm halo}$ the non-linear denominator in \cref{app:Ilambda} slows the growth.  That $a_0$ plays no independent role once $\gamma=\beta$ follows directly from the $a_0$-independence of $m_h(r)/M_{\rm halo}$ in the pure power-law limit established above.

This $a_0$-independence has a direct consequence for how the saturation appears in figures that hold different quantities fixed. Figure~\ref{fig:c-g} plots $\delta\omega_{\rm Re}$ and $\delta\omega_{\rm
Im}$ against $\gamma$ at fixed $\alpha=0.5$, $\beta=3.5$, for three compactnesses $z=10^{-5}$, $10^{-4.5}$, $10^{-4}$ obtained by holding $a_0$ fixed and varying $M_{\rm halo}$. The three curves remain separated by roughly an order of magnitude across the full range, including as $\gamma\to\beta^-$: they do not converge at the right edge of the plot. This is expected, since $\delta_{\rm sat}$ depends directly on $M_{\rm halo}$, which differs between the three curves. The successive terminal values increase faster than the linear-in-$M_{\rm
halo}$ scaling that holds only once $\mathcal{I}_{\rm sat}$ is small, consistent with the convex growth of $\mathcal{I}$ with compactness established in \cref{app:compact}.
 
Figure~\ref{fig:c4c6g}, by contrast, obtains its different compactnesses by holding $M_{\rm halo}$ fixed and varying $a_0$ instead. Because $\mathcal{I}_{\rm sat}$ does not depend on $a_0$, the curves shown
there for different $a_0$ at the same $M_{\rm halo}$ converge to a common value as $\gamma\to\beta^-$, in contrast with \cref{fig:c-g}. The two figures are therefore not in tension: each varies a different one of the two independent quantities entering $z=M_{\rm halo}/a_0$, and only $M_{\rm halo}$ survives in the saturated limit.

\paragraph*{Observational implication.}
A finite saturation means that for a given $(\beta, M_{\rm halo})$
there is a maximum frequency redshift reachable by varying $\gamma$.
A measured $\delta\omega$ above $\delta_{\rm sat}(\beta,M_{\rm
halo})$ would be inconsistent with the $(\alpha, \beta, \gamma)$ family at
that halo mass, giving a direct constraint on the halo parameters. A
measured shift below $\delta_{\rm sat}$ is consistent with a range
of $\gamma$ values, feeding into the degeneracy discussed in
\cref{app:compact,app:degen_QNM}. Because $\delta_{\rm sat}$ depends
on $M_{\rm halo}$ rather than on $a_0$, as \cref{fig:c-g,fig:c4c6g}
together show, this saturation bound constrains $\gamma$ once
$(\beta, M_{\rm halo})$ are otherwise fixed or independently known,
regardless of $a_0$.
\twocolumngrid
\bibliographystyle{./utphys1}
\bibliography{ref}
\end{document}